%% file: LTDLFN.tex
\documentclass[%
 reprint,
superscriptaddress,
nofootinbib,
amsmath,amssymb,
aps,
prx,
floatfix,
]{revtex4-2}

\PassOptionsToPackage{dvipsnames}{xcolor}
\usepackage{graphicx}
\usepackage{dcolumn}
\usepackage{booktabs}
\usepackage{bm}
\usepackage{hyperref}
\usepackage[nameinlink, capitalise]{cleveref}
\usepackage{physics}
\usepackage{amssymb,mathtools}
\usepackage{subcaption}
\usepackage{placeins}
\usepackage{tikz}
\usetikzlibrary{external}
\usepackage{soul}
\usepackage{pgfplots}
\pgfplotsset{compat=1.18}
\usepgfplotslibrary{fillbetween}
\usepgfplotslibrary{groupplots}
\input{results/result_commands.tex}
\usepackage{tikzit}
\input{quantum.tikzdefs}
\input{gcp.tikzstyles}
\newtheorem{definition}{Definition}

\renewcommand{\selectlanguage}[1]{}
\begin{document}
\title{Loss-tolerant distributed lattice surgery using fusion networks}

\author{Felix Burt}
  \email{f.burt23@imperial.ac.uk}
  \affiliation{Electrical and Electronic Engineering, Imperial College London, SW7 2AZ, United Kingdom}
\author{Richard Meister}
  \affiliation{Department of Computing, Imperial College London, SW7 2AZ, United Kingdom}
\author{Sheng-Ku Lin}
  \affiliation{Department of Computing, Imperial College London, SW7 2AZ, United Kingdom}
\author{Kuan-Cheng Chen}
  \affiliation{Electrical and Electronic Engineering, Imperial College London, SW7 2AZ, United Kingdom}
\author{Michael Hanks}
  \affiliation{Blackett Laboratory, Imperial College London, SW7 2AZ, United Kingdom}
\author{Roberto Bondesan}
  \affiliation{Department of Computing, Imperial College London, SW7 2AZ, United Kingdom}
\author{M.S. Kim}
  \affiliation{Blackett Laboratory, Imperial College London, SW7 2AZ, United Kingdom}
\author{Kin K. Leung}
  \affiliation{Electrical and Electronic Engineering, Imperial College London, SW7 2AZ, United Kingdom}

\begin{abstract}
    \input{sections/0_abstract.tex}
\end{abstract}    
\maketitle

\input{sections/1_introduction.tex}
\input{sections/2_preliminaries_zx.tex}
\input{sections/3_syndrome_extraction.tex}
\input{sections/4_hybrid_dqec.tex}
\input{sections/5_protocols.tex}
\input{sections/6_simulations.tex}
\input{sections/7_results.tex}
\input{sections/8_discussion.tex}
\input{sections/9_conclusion.tex}

\section*{Data and code availability}
The data and simulation code supporting this work will be made publicly available in a future release, and are available from the corresponding author upon reasonable request.

\section*{Acknowledgements}
The authors acknowledge funding from the Engineering and Physical Sciences Research Council (EPSRC)-funded Distributed Quantum Computing project, grant number EP/W032643/1.

\bibliography{references}

\end{document}

%% file: results/result_commands.tex
\pgfplotstableread[col sep=comma]{results/prep_and_erasure_correctable_regime_pm_corr_mergeZZ_obsZZ_data.csv}\ResourceZZData
\pgfplotstableread[col sep=comma]{results/prep_and_erasure_correctable_regime_pm_corr_mergeZZ_obsXX_data.csv}\ResourceXXData
\pgfplotstableread[col sep=comma]{results/circuit_and_erasure_correctable_regime_pm_corr_mergeZZ_obsZZ_data.csv}\CircuitZZData
\pgfplotstableread[col sep=comma]{results/circuit_and_erasure_correctable_regime_pm_corr_mergeZZ_obsXX_data.csv}\CircuitXXData

\newif\ifresultfound
\newif\iflossphysical
\newif\iflossdatafound

\newcommand{\ResultFindValue}[5]{%
  \global\resultfoundfalse
  \gdef\resultvalue{}%
  \pgfplotstablegetrowsof{#1}%
  \pgfmathtruncatemacro{\resultlastrow}{\pgfplotsretval-1}%
  \pgfplotsforeachungrouped \resultrow in {0,...,\resultlastrow}{%
    \pgfplotstablegetelem{\resultrow}{series}\of{#1}%
    \edef\resultseries{\pgfplotsretval}%
    \pgfplotstablegetelem{\resultrow}{#3}\of{#1}%
    \edef\resultselector{\pgfplotsretval}%
    \ifnum\pdfstrcmp{\resultseries}{#2}=0\relax
      \ifnum\pdfstrcmp{\resultselector}{#4}=0\relax
        \pgfplotstablegetelem{\resultrow}{#5}\of{#1}%
        \xdef\resultvalue{\pgfplotsretval}%
        \global\resultfoundtrue
      \fi
    \fi
  }%
}

\newcommand{\ResultThresholdPercent}[4]{%
  \ResultFindValue{#1}{#2}{local_prob}{#3}{threshold}%
  \ifresultfound
    \ensuremath{\pgfmathprintnumber[fixed,fixed zerofill,precision=#4]{\fpeval{100*\resultvalue}}\%}%
  \else
    \textemdash
  \fi
}

\newcommand{\ResultThresholdWithErrorPercent}[4]{%
  \ResultFindValue{#1}{#2}{local_prob}{#3}{threshold}%
  \ifresultfound
    \xdef\resultcentral{\resultvalue}%
    \ResultFindValue{#1}{#2}{local_prob}{#3}{stderr}%
    \xdef\resulterror{\resultvalue}%
    \ensuremath{%
      \pgfmathprintnumber[fixed,fixed zerofill,precision=#4]{\fpeval{100*\resultcentral}}%
      (\pgfmathprintnumber[fixed,precision=0]{\fpeval{round(100*\resulterror*10^(#4),0)}})\%%
    }%
  \else
    \textemdash
  \fi
}

\newcommand{\ResultThresholdWithError}[4]{%
  \ResultFindValue{#1}{#2}{local_prob}{#3}{threshold}%
  \ifresultfound
    \xdef\resultcentral{\resultvalue}%
    \ResultFindValue{#1}{#2}{local_prob}{#3}{stderr}%
    \xdef\resulterror{\resultvalue}%
    \ensuremath{%
      \pgfmathprintnumber[fixed,fixed zerofill,precision=#4]{\resultcentral}%
      (\pgfmathprintnumber[fixed,precision=0]{\fpeval{round(\resulterror*10^(#4),0)}})%
    }%
  \else
    \textemdash
  \fi
}

\newcommand{\ErasureToLossPercent}[3]{%
  \ensuremath{\pgfmathprintnumber[fixed,fixed zerofill,precision=#3]{\fpeval{100*(1-((1-(#1))/(1-2^(-((#2)+1))))^(2^(-(#2))))}}\%}%
}

\newcommand{\CalculateLossThreshold}[4]{%
  \ResultFindValue{#1}{#2}{local_prob}{#3}{threshold}%
  \global\lossphysicalfalse
  \ifresultfound
    \pgfmathparse{\resultvalue >= pow(2,-((#4)+1)) ? 1 : 0}%
    \ifnum\pgfmathresult=1\relax
      \xdef\lossvalue{\fpeval{100*(1-((1-\resultvalue)/(1-2^(-((#4)+1))))^(2^(-(#4))) )}}%
      \global\lossphysicaltrue
    \fi
  \fi
}

\newcommand{\CalculateLossThresholdWithError}[4]{%
  \global\lossdatafoundfalse
  \CalculateLossThreshold{#1}{#2}{#3}{#4}%
  \ifresultfound
    \global\lossdatafoundtrue
    \xdef\lossthresholdpe{\resultvalue}%
    \iflossphysical
      \ResultFindValue{#1}{#2}{local_prob}{#3}{stderr}%
      \xdef\lossthresholderror{\resultvalue}%
      \xdef\losserror{\fpeval{%
        100*2^(-(#4))/(1-2^(-((#4)+1)))%
        *((1-\lossthresholdpe)/(1-2^(-((#4)+1))))^(2^(-(#4))-1)%
        *\lossthresholderror
      }}%
    \fi
  \fi
}

\newcommand{\LossThresholdPercent}[5]{%
  \CalculateLossThreshold{#1}{#2}{#3}{#4}%
  \ifresultfound
    \iflossphysical
      \ensuremath{\pgfmathprintnumber[fixed,fixed zerofill,precision=#5]{\lossvalue}\%}%
    \fi
  \else
    \textemdash
  \fi
}

\newcommand{\BestLossThresholdPercent}[5]{%
  \CalculateLossThreshold{#1}{#2}{#3}{#4}%
  \ifresultfound
    \iflossphysical
      \ensuremath{\mathbf{\pgfmathprintnumber[fixed,fixed zerofill,precision=#5]{\lossvalue}\%}}%
    \fi
  \else
    \textemdash
  \fi
}

\newcommand{\LossThresholdWithErrorPercent}[5]{%
  \CalculateLossThresholdWithError{#1}{#2}{#3}{#4}%
  \iflossdatafound
    \iflossphysical
      \ensuremath{%
        \pgfmathprintnumber[fixed,fixed zerofill,precision=#5]{\lossvalue}%
        (\pgfmathprintnumber[fixed,precision=0]{\fpeval{round(\losserror*10^(#5),0)}})\%%
      }%
    \fi
  \else
    \textemdash
  \fi
}

\newcommand{\BestLossThresholdWithErrorPercent}[5]{%
  \CalculateLossThresholdWithError{#1}{#2}{#3}{#4}%
  \iflossdatafound
    \iflossphysical
      \ensuremath{\mathbf{%
        \pgfmathprintnumber[fixed,fixed zerofill,precision=#5]{\lossvalue}%
        (\pgfmathprintnumber[fixed,precision=0]{\fpeval{round(\losserror*10^(#5),0)}})\%%
      }}%
    \fi
  \else
    \textemdash
  \fi
}

%% file: gcp.tikzstyles
\tikzstyle{white}=[inner sep=0mm, minimum size=3mm, shape=circle, draw=black, fill=white]
\tikzstyle{white dashed}=[inner sep=0mm, minimum size=3mm, shape=circle, draw=black, fill=white, dashed, dash pattern=on 1pt off 1pt]
\tikzstyle{black node}=[inner sep=0mm, minimum size=3mm, shape=circle, draw=black, fill=black]
\tikzstyle{Z dot}=[inner sep=0mm, minimum size=2.5mm, shape=circle, draw=black, fill=zxgreen, tikzit fill={rgb,255: red,221; green,255; blue,221}, tikzit category=zx]
\tikzstyle{Z dot fade}=[inner sep=0mm, minimum size=2.5mm, shape=circle, draw=black!20, fill=zxgreen!20, tikzit fill={rgb,255: red,231; green,255; blue,231}, tikzit category=zx]
\tikzstyle{Z dot dark}=[Z dot, fill=zxgreendark, tikzit fill={rgb,255: red,136; green,204; blue,136}]
\tikzstyle{Z dot g}=[Z dot, shape=circle, draw=black, fill=zxgreen, tikzit fill={rgb,255: red,221; green,255; blue,221}, tikzit category=zx]
\tikzstyle{Z capsule dot}=[shape=rectangle, rounded corners=1.6mm, minimum height=3mm, minimum width=3mm, inner xsep=0.9mm, inner ysep=0mm, draw=black, fill=zxgreen, tikzit fill={rgb,255: red,221; green,255; blue,221}, tikzit category=zx]
\tikzstyle{X capsule dot}=[shape=rectangle, rounded corners=1.6mm, minimum height=3mm, minimum width=3mm, inner xsep=0.9mm, inner ysep=0mm, draw=black, fill=zxred, tikzit fill={rgb,255: red,255; green,136; blue,136}, tikzit category=zx]
\tikzstyle{Z bold dot}=[inner sep=0mm, minimum size=2mm, shape=circle, draw=black, fill=zxgreen, tikzit fill={rgb,255: red,221; green,255; blue,221}, line width=1.2pt, tikzit category=zx]
\tikzstyle{Z meas}=[Z dot, tikzit fill={rgb,255: red,166; green,255; blue,166}, append after command={\pgfextra{\draw[meas tick] ($(\tikzlastnode.center)$) -- ($(\tikzlastnode.center)+(0mm,2.8mm)$);}}, tikzit category=zx]
\tikzstyle{Z meas graph}=[Z dot, tikzit fill={rgb,255: red,166; green,255; blue,166}, append after command={\pgfextra{\draw[meas tick] ($(\tikzlastnode.center)$) -- ($(\tikzlastnode.center)+(1.5mm,1.5mm)$);}}, tikzit category=zx]
\tikzstyle{Z init}=[Z dot, tikzit fill={rgb,255: red,166; green,255; blue,166}, tikzit category=zx]
\tikzstyle{Z phase dot}=[minimum size=5mm, font={\footnotesize\boldmath}, shape=rectangle, rounded corners=2mm, inner sep=0.2mm, outer sep=-2mm, scale=0.8, tikzit shape=circle, draw=black, fill=zxgreen, tikzit fill={rgb,255: red,221; green,255; blue,221}, tikzit draw=blue, tikzit category=zx]
\tikzstyle{Z tiny dot}=[inner sep=0mm, minimum size=1mm, shape=circle, draw=black, fill=zxgreen, tikzit fill={rgb,255: red,221; green,255; blue,221}]
\tikzstyle{X dot}=[Z dot, shape=circle, draw=black, fill=zxred, tikzit fill={rgb,255: red,255; green,136; blue,136}, tikzit category=zx]
\tikzstyle{X meas}=[Z dot, fill=zxred, tikzit fill={rgb,255: red,255; green,93; blue,93}, append after command={\pgfextra{\draw[meas tick] ($(\tikzlastnode.center)$) -- ($(\tikzlastnode.center)+(0mm,2.8mm)$);}}, tikzit category=zx]
\tikzstyle{X dot g}=[Z dot, shape=circle, draw=black, fill=zxred, tikzit fill={rgb,255: red,255; green,136; blue,136}, tikzit category=zx]
\tikzstyle{X bold dot}=[inner sep=0mm, minimum size=2mm, shape=circle, draw=black, fill=zxred, tikzit fill={rgb,255: red,255; green,136; blue,136}, line width=1.2pt, tikzit category=zx]
\tikzstyle{X phase dot}=[Z phase dot, tikzit shape=circle, tikzit draw=blue, fill=zxred, tikzit fill={rgb,255: red,255; green,136; blue,136}, font={\footnotesize\boldmath}, tikzit category=zx]
\tikzstyle{X meas graph}=[Z dot, fill=zxred, tikzit fill={rgb,255: red,255; green,93; blue,93}, append after command={\pgfextra{\draw[meas tick] ($(\tikzlastnode.center)$) -- ($(\tikzlastnode.center)+(1.5mm,1.5mm)$);}}, tikzit category=zx]
\tikzstyle{X init}=[X dot, tikzit fill={rgb,255: red,255; green,93; blue,93}, tikzit category=zx]
\tikzstyle{X tiny dot}=[inner sep=0mm, minimum size=1mm, shape=circle, draw=black, fill=zxred, tikzit fill={rgb,255: red,255; green,136; blue,136}]
\tikzstyle{hadamard}=[fill=yellow, draw=black, shape=rectangle, inner sep=0.6mm, minimum height=1.5mm, minimum width=1.5mm, tikzit category=zx]
\tikzstyle{whitebox}=[fill=white, draw=black, shape=rectangle, inner sep=0.6mm, minimum height=2mm, minimum width=2mm, tikzit category=zx]
\tikzstyle{greenbox}=[whitebox, tikzit shape=rectangle, minimum height=3mm, minimum width=3mm, fill=zxgreen, tikzit fill={rgb,255: red,221; green,255; blue,221}, tikzit category=zx]
\tikzstyle{redbox}=[whitebox, tikzit shape=rectangle, minimum height=3mm, minimum width=3mm, fill=zxred, tikzit fill={rgb,255: red,255; green,136; blue,136}, tikzit category=zx]
\tikzstyle{greenredbox}=[whitebox, tikzit shape=rectangle, minimum height=3mm, minimum width=3mm, path picture={\fill[zxgreen] (path picture bounding box.north west) rectangle (path picture bounding box.east); \fill[zxred] (path picture bounding box.west) rectangle (path picture bounding box.south east); \draw[black] (path picture bounding box.south west) rectangle (path picture bounding box.north east);}, tikzit fill={rgb,255: red,221; green,255; blue,221}, tikzit category=zx]
\tikzstyle{redgreenbox}=[whitebox, tikzit shape=rectangle, minimum height=3mm, minimum width=3mm, path picture={\fill[zxred] (path picture bounding box.north west) rectangle (path picture bounding box.east); \fill[zxgreen] (path picture bounding box.west) rectangle (path picture bounding box.south east); \draw[black] (path picture bounding box.south west) rectangle (path picture bounding box.north east);}, tikzit fill={rgb,255: red,255; green,136; blue,136}, tikzit category=zx]
\tikzstyle{H}=[draw=black, fill=yellow, shape=rectangle]
\tikzstyle{greyTriangleE}=[draw=black, fill=gray!30, regular polygon, regular polygon sides=3, shape border rotate=30, minimum size=7mm, inner sep=0pt, tikzit shape=triangle]
\tikzstyle{greyTriangleN}=[draw=black, fill=gray!30, regular polygon, regular polygon sides=3, shape border rotate=0, minimum size=7mm, inner sep=0pt, tikzit shape=triangle]
\tikzstyle{gate}=[shape=rectangle, text height=1.5ex, text depth=0.25ex, yshift=-0.5mm, fill=white, draw=black, minimum height=5mm, minimum width=5mm, font={\small}, tikzit category=circuit]
\tikzstyle{big gate}=[shape=rectangle, text height=1.5ex, text depth=0.25ex, yshift=-0.5mm, fill=white, draw=black, minimum height=10mm, minimum width=5mm, font={\small}, tikzit category=circuit]
\tikzstyle{paulibox}=[fill={rgb,255: red,221; green,221; blue,255}, draw=black, shape=rectangle, inner sep=0.6mm, minimum height=6mm, minimum width=6mm, font={\footnotesize}, text height=1.5ex, text depth=0.25ex, tikzit category=zx]
\tikzstyle{paul meas box}=[paulibox, fill=zxpaulibox, tikzit fill={rgb,255: red,221; green,221; blue,255}, append after command={\pgfextra{\draw[fill=zxpaulibox, draw=black] ($(\tikzlastnode.east)+(-0.1mm,1.6mm)$) arc[start angle=90,end angle=-90,radius=1.6mm] -- ($(\tikzlastnode.east)+(-0.1mm,-1.4mm)$) -- cycle; \draw[->] ($(\tikzlastnode.east)+(-0.1mm,-0.2mm)$) -- ($(\tikzlastnode.east)+(2.2mm,1.5mm)$);}}]
\tikzstyle{pauli meas box}=[paul meas box]
\tikzstyle{small black dot}=[fill=black, draw=black, shape=circle, inner sep=0pt, minimum width=1.2mm, tikzit category=circuit]
\tikzstyle{small white dot}=[fill=white, draw=black, shape=circle, inner sep=0pt, minimum width=1.2mm, tikzit category=circuit]
\tikzstyle{small red dot}=[fill=red, draw=black, shape=circle, inner sep=0pt, minimum width=1.2mm, tikzit category=circuit]
\tikzstyle{small blue dot}=[fill=blue, draw=black, shape=circle, inner sep=0pt, minimum width=1.2mm, tikzit category=circuit]
\tikzstyle{data dot}=[fill=black, draw=black, shape=circle, inner sep=0pt, minimum width=1.7mm, tikzit category=circuit]
\tikzstyle{cnot ctrl}=[fill=black, draw=black, shape=circle, inner sep=0pt, minimum width=1.2mm, tikzit category=circuit]
\tikzstyle{cnot targ}=[fill=white, draw=white, shape=circle, tikzit category=circuit, label={center:$\oplus$}, inner sep=0pt, minimum width=2.1mm, tikzit fill={rgb,255: red,102; green,204; blue,255}, tikzit draw=black]
\tikzstyle{cnot targ rot}=[fill=white, draw=white, shape=circle, tikzit category=circuit, label={center:\rotatebox{45}{$\oplus$}}, inner sep=0pt, minimum width=2.1mm, tikzit fill={rgb,255: red,255; green,170; blue,102}, tikzit draw=black]
\tikzstyle{clabel}=[fill=none, draw=none, shape=rectangle, tikzit fill={rgb,255: red,56; green,255; blue,242}, inner sep=1pt, tikzit category=labels]
\tikzstyle{clabel west}=[clabel, anchor=west, tikzit category=labels]
\tikzstyle{box}=[shape=rectangle, text height=1.5ex, text depth=0.25ex, yshift=-0.5mm, fill=white, draw=black, minimum height=5mm, minimum width=5mm, font={\small}]
\tikzstyle{medium box}=[shape=rectangle, text height=1.5ex, text depth=0.25ex, yshift=-0.5mm, fill=white, draw=black, minimum height=10mm, minimum width=5mm, font={\small}]
\tikzstyle{time}=[simple]
\tikzstyle{simple}=[-, line cap=round, shorten <=-0.17mm, shorten >=-0.17mm]
\tikzstyle{blue}=[-, draw=blue, line cap=round, shorten <=-0.17mm, shorten >=-0.17mm]
\tikzstyle{hadamard edge}=[-, thick, draw=black, tikzit draw={rgb,255: red,68; green,136; blue,255}, postaction={decorate, decoration={markings, mark=at position 0.5 with {\node[hadamard] {};}}}]
\tikzstyle{boldedge}=[-, line width=2pt, shorten <=-0.5mm, shorten >=-0.5mm, line cap=round, tikzit line width=2]
\tikzstyle{meas tick}=[-, line width=1.2pt, shorten <=-0.5mm, shorten >=-0.5mm, line cap=round, tikzit line width=1.5]
\tikzstyle{H boldedge}=[-, line width=2pt, shorten <=-0.5mm, shorten >=-0.5mm, line cap=round, tikzit line width=2, fill=none, tikzit draw={rgb,255: red,68; green,136; blue,255}, draw=black, postaction={decorate, decoration={markings, mark=at position 0.5 with {\node[hadamard] {};}}}]
\tikzstyle{webgreen}=[-, draw=black, preaction={draw, zxgreenweb, line width=4pt, line cap = round,  shorten <=-0.17mm, shorten >=-0.17mm}, tikzit draw=black, tikzit halo={rgb,255: red,102; green,255; blue,102}]
\tikzstyle{webgreennoline}=[-, draw=none, preaction={draw, zxgreenweb, line width=2pt, line cap = round,  shorten <=-0.17mm, shorten >=-0.17mm}, tikzit draw=black, tikzit halo={rgb,255: red,102; green,255; blue,102}]
\tikzstyle{boldwebgreen}=[-, line width=2pt, draw=black, line cap=round, preaction={draw, zxgreenweb, line width=4pt, line cap = round,  shorten <=-0.17mm, shorten >=-0.17mm}, fill=none, tikzit draw=black, tikzit line width=2, tikzit halo={rgb,255: red,102; green,255; blue,102}]
\tikzstyle{meastickweb green}=[-, line width=1.2pt, draw=black, line cap=round, preaction={draw, zxgreenweb, line width=3.4pt, line cap=round, shorten <=-0.0mm, shorten >=-0.4mm}, fill=none, tikzit draw=black, tikzit line width=1.5, tikzit halo={rgb,255: red,102; green,255; blue,102}]
\tikzstyle{webred}=[-, draw=black, preaction={draw, zxredweb, line width=4pt, line cap = round,  shorten <=-0.17mm, shorten >=-0.17mm}, tikzit draw=black, tikzit halo={rgb,255: red,255; green,102; blue,102}]
\tikzstyle{webrednoline}=[-,draw=none, preaction={draw, zxredweb, line width=2pt, line cap = round,  shorten <=-0.17mm, shorten >=-0.17mm}, tikzit draw=black, tikzit halo={rgb,255: red,255; green,102; blue,102}]
\tikzstyle{boldwebred}=[-, line width=2pt, line cap=round, draw=black, preaction={draw, zxredweb, line width=4pt, line cap = round,  shorten <=-0.17mm, shorten >=-0.17mm}, tikzit draw=black, tikzit line width=2, tikzit halo={rgb,255: red,255; green,102; blue,102}]
\tikzstyle{meastickweb red}=[-, line width=1.2pt, line cap=round, draw=black, preaction={draw, zxredweb, line width=3pt, line cap=round, shorten <=-0.0mm, shorten >=-0.4mm}, fill=none, tikzit draw=black, tikzit line width=1.5, tikzit halo={rgb,255: red,255; green,102; blue,102}]
\tikzstyle{webredgreen}=[-, draw=black, preaction={decorate, decoration={show path construction, lineto code={\draw[zxredweb, line width=4pt, line cap=round, shorten <=-0.17mm] (\tikzinputsegmentfirst) -- ($(\tikzinputsegmentfirst)!0.5!(\tikzinputsegmentlast)$); \draw[zxgreenweb, line width=4pt, line cap=round, shorten >=-0.17mm] ($(\tikzinputsegmentfirst)!0.5!(\tikzinputsegmentlast)$) -- (\tikzinputsegmentlast);}, curveto code={\coordinate (wrgA) at ($(\tikzinputsegmentfirst)!0.5!(\tikzinputsegmentsupporta)$); \coordinate (wrgB) at ($(\tikzinputsegmentsupporta)!0.5!(\tikzinputsegmentsupportb)$); \coordinate (wrgC) at ($(\tikzinputsegmentsupportb)!0.5!(\tikzinputsegmentlast)$); \coordinate (wrgD) at ($(wrgA)!0.5!(wrgB)$); \coordinate (wrgE) at ($(wrgB)!0.5!(wrgC)$); \coordinate (wrgM) at ($(wrgD)!0.5!(wrgE)$); \draw[zxredweb, line width=4pt, line cap=round, shorten <=-0.17mm] (\tikzinputsegmentfirst) .. controls (wrgA) and (wrgD) .. (wrgM); \draw[zxgreenweb, line width=4pt, line cap=round, shorten >=-0.17mm] (wrgM) .. controls (wrgE) and (wrgC) .. (\tikzinputsegmentlast);}}}, postaction={decorate, decoration={markings, mark=at position 0.5 with {\node[hadamard] {};}}}, tikzit draw=black, tikzit halo={rgb,255: red,255; green,102; blue,102}, tikzit halo2={rgb,255: red,102; green,255; blue,102}]
\tikzstyle{webgreenred}=[-, draw=black, preaction={decorate, decoration={show path construction, lineto code={\draw[zxgreenweb, line width=4pt, line cap=round, shorten <=-0.17mm] (\tikzinputsegmentfirst) -- ($(\tikzinputsegmentfirst)!0.5!(\tikzinputsegmentlast)$); \draw[zxredweb, line width=4pt, line cap=round, shorten >=-0.17mm] ($(\tikzinputsegmentfirst)!0.5!(\tikzinputsegmentlast)$) -- (\tikzinputsegmentlast);}, curveto code={\coordinate (wgrA) at ($(\tikzinputsegmentfirst)!0.5!(\tikzinputsegmentsupporta)$); \coordinate (wgrB) at ($(\tikzinputsegmentsupporta)!0.5!(\tikzinputsegmentsupportb)$); \coordinate (wgrC) at ($(\tikzinputsegmentsupportb)!0.5!(\tikzinputsegmentlast)$); \coordinate (wgrD) at ($(wgrA)!0.5!(wgrB)$); \coordinate (wgrE) at ($(wgrB)!0.5!(wgrC)$); \coordinate (wgrM) at ($(wgrD)!0.5!(wgrE)$); \draw[zxgreenweb, line width=4pt, line cap=round, shorten <=-0.17mm] (\tikzinputsegmentfirst) .. controls (wgrA) and (wgrD) .. (wgrM); \draw[zxredweb, line width=4pt, line cap=round, shorten >=-0.17mm] (wgrM) .. controls (wgrE) and (wgrC) .. (\tikzinputsegmentlast);}}}, postaction={decorate, decoration={markings, mark=at position 0.5 with {\node[hadamard] {};}}}, tikzit draw=black, tikzit halo={rgb,255: red,102; green,255; blue,102}, tikzit halo2={rgb,255: red,255; green,102; blue,102}]
\tikzstyle{boldwebredgreen}=[-, line width=2pt, line cap=round, draw=black, preaction={decorate, decoration={show path construction, lineto code={\draw[zxredweb, line width=4pt, line cap=round, shorten <=-0.17mm] (\tikzinputsegmentfirst) -- ($(\tikzinputsegmentfirst)!0.5!(\tikzinputsegmentlast)$); \draw[zxgreenweb, line width=4pt, line cap=round, shorten >=-0.17mm] ($(\tikzinputsegmentfirst)!0.5!(\tikzinputsegmentlast)$) -- (\tikzinputsegmentlast);}, curveto code={\coordinate (bwrgA) at ($(\tikzinputsegmentfirst)!0.5!(\tikzinputsegmentsupporta)$); \coordinate (bwrgB) at ($(\tikzinputsegmentsupporta)!0.5!(\tikzinputsegmentsupportb)$); \coordinate (bwrgC) at ($(\tikzinputsegmentsupportb)!0.5!(\tikzinputsegmentlast)$); \coordinate (bwrgD) at ($(bwrgA)!0.5!(bwrgB)$); \coordinate (bwrgE) at ($(bwrgB)!0.5!(bwrgC)$); \coordinate (bwrgM) at ($(bwrgD)!0.5!(bwrgE)$); \draw[zxredweb, line width=4pt, line cap=round, shorten <=-0.17mm] (\tikzinputsegmentfirst) .. controls (bwrgA) and (bwrgD) .. (bwrgM); \draw[zxgreenweb, line width=4pt, line cap=round, shorten >=-0.17mm] (bwrgM) .. controls (bwrgE) and (bwrgC) .. (\tikzinputsegmentlast);}}}, postaction={decorate, decoration={markings, mark=at position 0.5 with {\node[hadamard] {};}}}, tikzit draw=black, tikzit line width=2, tikzit halo={rgb,255: red,255; green,102; blue,102}, tikzit halo2={rgb,255: red,102; green,255; blue,102}]
\tikzstyle{boldwebgreenred}=[-, line width=2pt, line cap=round, draw=black, preaction={decorate, decoration={show path construction, lineto code={\draw[zxgreenweb, line width=4pt, line cap=round, shorten <=-0.17mm] (\tikzinputsegmentfirst) -- ($(\tikzinputsegmentfirst)!0.5!(\tikzinputsegmentlast)$); \draw[zxredweb, line width=4pt, line cap=round, shorten >=-0.17mm] ($(\tikzinputsegmentfirst)!0.5!(\tikzinputsegmentlast)$) -- (\tikzinputsegmentlast);}, curveto code={\coordinate (bwgrA) at ($(\tikzinputsegmentfirst)!0.5!(\tikzinputsegmentsupporta)$); \coordinate (bwgrB) at ($(\tikzinputsegmentsupporta)!0.5!(\tikzinputsegmentsupportb)$); \coordinate (bwgrC) at ($(\tikzinputsegmentsupportb)!0.5!(\tikzinputsegmentlast)$); \coordinate (bwgrD) at ($(bwgrA)!0.5!(bwgrB)$); \coordinate (bwgrE) at ($(bwgrB)!0.5!(bwgrC)$); \coordinate (bwgrM) at ($(bwgrD)!0.5!(bwgrE)$); \draw[zxgreenweb, line width=4pt, line cap=round, shorten <=-0.17mm] (\tikzinputsegmentfirst) .. controls (bwgrA) and (bwgrD) .. (bwgrM); \draw[zxredweb, line width=4pt, line cap=round, shorten >=-0.17mm] (bwgrM) .. controls (bwgrE) and (bwgrC) .. (\tikzinputsegmentlast);}}}, postaction={decorate, decoration={markings, mark=at position 0.5 with {\node[hadamard] {};}}}, tikzit draw=black, tikzit line width=2, tikzit halo={rgb,255: red,102; green,255; blue,102}, tikzit halo2={rgb,255: red,255; green,102; blue,102}]
\tikzstyle{box edge}=[-, dashed, dash pattern=on 2pt off 0.5pt, thick, draw={rgb,255: red,203; green,192; blue,225}]
\tikzstyle{brace edge}=[-, tikzit draw=blue, decorate, decoration={brace,amplitude=1mm,raise=-1mm}]
\tikzstyle{diredge}=[->]
\tikzstyle{double edge}=[-, double, shorten <=-1mm, shorten >=-1mm, double distance=2pt]
\tikzstyle{gray edge}=[-, gray!60!white]
\tikzstyle{pointer edge}=[->, very thick, gray]
\tikzstyle{class}=[-, line width=2pt, shorten <=-0.17mm, shorten >=1.5mm, line cap=round]
\tikzstyle{bidir edge}=[<->, very thick, draw=black]
\tikzstyle{surface}=[-, draw=none, tikzit fill={rgb,255: red,172; green,173; blue,220}, fill={rgb,255: red,172; green,173; blue,220}]
\tikzstyle{surface X red}=[-, draw=none, tikzit fill={rgb,255: red,255; green,145; blue,146}, fill=zxred, fill opacity=0.8]
\tikzstyle{surface Z green}=[-, draw=none, tikzit fill={rgb,255: red,221; green,255; blue,221}, fill=zxgreen, fill opacity=0.9]
\tikzstyle{paulifill}=[-, draw=black, fill={rgb,255: red,191; green,193; blue,255}]
\tikzstyle{gray}=[-, draw=gray]
\tikzstyle{sparse dotted}=[-, draw=black, line cap=round, dash pattern=on 0pt off 2pt, dotted]
\tikzstyle{ZZ}=[-, draw=blue, tikzit draw=blue, dashed]
\tikzstyle{cg fusion}=[-, dashed, line width=2pt, line cap=round, draw=black]
\tikzstyle{snake style}=[-, decorate, decoration={snake,amplitude=.4mm,segment length=2mm,pre length = 0.5mm, post length=0.5mm}, draw=black, tikzit snake]
\tikzstyle{edgeStyle}=[-, black]
\tikzstyle{dashedStyle}=[-, dashed]
\tikzstyle{arrowStyle}=[->]
\tikzstyle{snakeArrowStyle}=[->, decorate, decoration={snake,amplitude=.4mm,segment length=1.5mm,post length=1mm}, tikzit snake]
\tikzstyle{bluefill}=[-, draw=none, tikzit fill={rgb,255: red,30; green,199; blue,255}, fill=blue, fill opacity=0.2]
\tikzstyle{greyfill}=[-, draw=none, tikzit fill={rgb,255: red,169; green,171; blue,218}, fill={rgb,255: red,169; green,171; blue,218}, fill opacity=0.8]
\tikzstyle{lightbrownfill}=[-, draw=black, tikzit fill={rgb,255: red,221; green,205; blue,148}, fill={rgb,255: red,221; green,205; blue,148}, fill opacity=0.8, dashed]
\tikzstyle{redfill}=[-, dashed, draw=black, tikzit fill={rgb,255: red,255; green,133; blue,113}, fill=red!60, fill opacity=0.5]
\tikzstyle{yellowfill}=[-, draw=black, tikzit fill={rgb,255: red,255; green,255; blue,155}, fill=yellow!60, fill opacity=0.5]
\tikzstyle{qpu fill}=[-, draw=none, tikzit fill={rgb,255: red,224; green,240; blue,240}, fill=teal!12]
\tikzstyle{link fill}=[-, draw=none, tikzit fill={rgb,255: red,255; green,241; blue,214}, fill=orange!16]
\tikzstyle{black}=[-, thick, draw=black]
\tikzstyle{xlog}=[-, draw=magenta, dotted]
\tikzstyle{zlog}=[-, dotted, draw={rgb,255: red,0; green,179; blue,255}]
\tikzstyle{resource}=[-, draw={rgb,255: red,0; green,0; blue,81}]
\tikzstyle{xz}=[-, dashed, draw=orange]
\tikzstyle{webgreen arrow}=[->, draw=black, preaction={draw, -, zxgreenweb, line width=4pt, line cap=round, shorten <=-0.17mm, shorten >=-0.17mm}, tikzit draw=black, tikzit halo={rgb,255: red,102; green,255; blue,102}]
\tikzstyle{webred arrow}=[->, draw=black, preaction={draw, -, zxredweb, line width=4pt, line cap=round, shorten <=-0.17mm, shorten >=-0.17mm}, tikzit draw=black, tikzit halo={rgb,255: red,255; green,102; blue,102}]

%% file: sections/0_abstract.tex
Networking matter-based quantum processing units (QPUs) offers a promising route to scaling fault-tolerant quantum computers. This requires distributed logical operations to be performed across photonic links, where noise is characteristically different from and stronger than in local QPUs owing to photon loss and probabilistic linear-optical operations. Measurement- and fusion-based quantum computing are designed to be robust against loss and probabilistic photonic operations, suggesting they could complement circuit-based error correction at the interface between networked QPUs. We use ZX calculus transformations to construct \textit{hybrid} syndrome extraction protocols and apply them to distributed rotated surface code lattice surgery, demonstrating that several protocols attain a $50\%$ interface-erasure threshold when local noise is absent, including circuit-based lattice surgery using fused Bell pairs and hybrid protocols using linear cluster states. Relative to a straight Bell-pair interface geometry, the hybrid protocols increase the merge-observable interface distance from $d+1$ to $2d+1$ and restore the perpendicular-observable interface distance from $\lfloor(d+1)/2\rfloor$ to $d$. We calculate how the threshold decreases when using truncated resource states and map correctable regions under resource-state errors, local circuit noise, and fusion erasure. We then convert these erasure thresholds into photon-loss thresholds and probe subthreshold performance using fusion boosting. At circuit and resource-state error rates of $10^{-3}$, local errors largely mask the interface-distance advantage. As local noise is reduced to $10^{-4}$ and below, the linear-chain protocols improve more rapidly with decreasing erasure, suggesting that interface distance enhancements are effective in low local noise regimes.

%% file: sections/1_introduction.tex
\section{Introduction}

Scaling quantum computers to practically relevant logical qubit counts at low error rates requires more qubits than are typically expected to fit on a single quantum processing unit (QPU)~\cite{vanmeterPathScalableDistributed2016,saffmanQuantumComputingAtomic2016,cetinaControlTransverseMotion2022,angARQUINArchitecturesMultinode2024}. This has motivated interest in distributed quantum computing (DQC) as a model for scalable fault-tolerant quantum computation (FTQC), in which multiple QPUs containing matter-based qubits are connected via a photonic network~\cite{monroeLargescaleModularQuantumcomputer2014,mainDistributedQuantumComputing2025,caleffiDistributedQuantumComputing2024}. Combining matter-based qubits with photons leads to noise asymmetry. Noise in matter-based processors is dominated by decoherence and imperfect gates~\cite{laddQuantumComputers2010} -- typically modelled as random Pauli errors in a circuit-level noise model, while noise in photonic networks is dominated by loss of photons due to absorption in fibre and probabilistic linear-optical gates~\cite{knillSchemeEfficientQuantum2001}.

\begin{figure*}[tb]
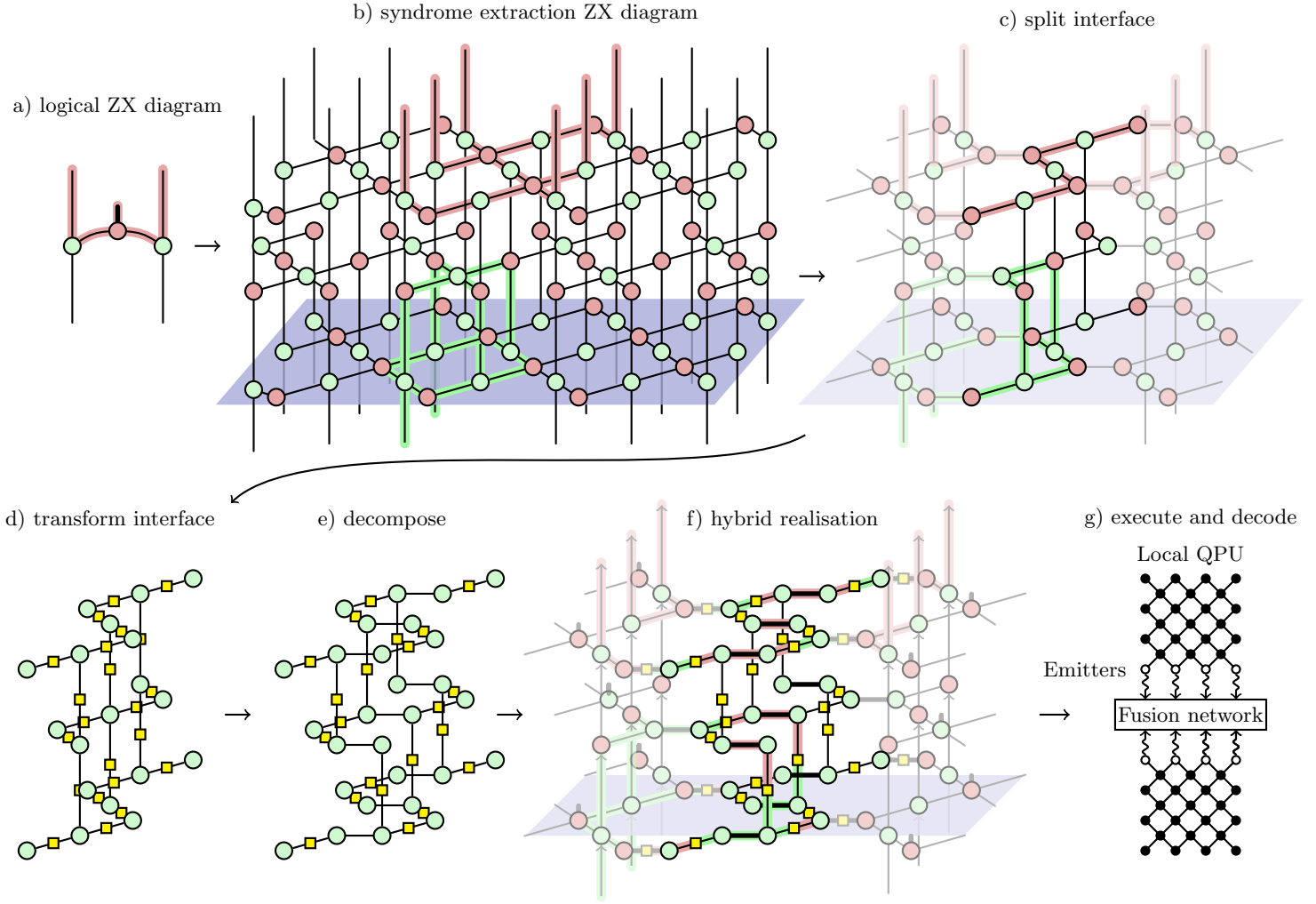

    \centering
    \tikzfig{1_introduction/01_overview}[scale=0.9]
    \caption{\textbf{Overview of the construction.} A lattice surgery merge, carried from the logical
    level down to a hybrid protocol designed for heterogeneous hardware. At the logical level the merge
    implements a joint measurement, whose outcome carries the joint observable (red web).
    This maps to an abstract \textit{syndrome extraction diagram} with Pauli webs corresponding to logical observables and detecting regions (\cref{sec:sed_framework}). Using web-preserving transformations (\cref{sec:transformations}) the seam is split and transformed into a graph-like sub-diagram (\cref{sec:clusterisation}) which can be split into resource state components. Measurement and world-line annotations fix a hybrid realisation with fusion networks at the interface and circuit-based syndrome extraction elsewhere. The diagram is then further mapped to heterogeneous hardware fixing an error model and decoder.}
    \label{fig:overview_pipeline}
\end{figure*}

In photonics, loss of a qubit can be \textit{heralded} by measurements, making it an instance of an \textit{erasure}~\cite{silvaThresholdsLinearOptics2005}. Erasures provide more information than computational errors alone and can therefore be tolerated at much higher rates. The quantum capacity of the erasure channel vanishes at a rate of $50\%$, setting an upper bound on protecting a qubit in transmission~\cite{bennettCapacitiesQuantumErasure1997,woottersSingleQuantumCannot1982}. Varnava et al. proposed a measurement-based quantum computing (MBQC)~\cite{raussendorfOneWayQuantumComputer2001} scheme that reaches the $50\%$ limit using adaptive measurements on tree-like cluster states~\cite{varnavaLossToleranceOneway2006} and provided a procedure for constructing the necessary cluster states using linear optics~\cite{varnavaLossTolerantLinear2007,nielsenOpticalQuantumComputation2004}. Stace et al. showed that two-dimensional planar surface codes~\cite{bravyiQuantumCodesLattice1998,kitaevFaulttolerantQuantumComputation2003} exhibit erasure thresholds reaching up to $50\%$ as predicted by bond percolation on square lattices~\cite{kestenPercolationTheoryMathematicians1982} and investigated the threshold degradation in the presence of computational errors~\cite{staceThresholdsTopologicalCodes2009}. Barrett and Stace extended this to the three-dimensional topological RHG cluster state~\cite{raussendorfFaulttolerantOnewayQuantum2006,raussendorfTopologicalFaulttoleranceCluster2007}, obtaining a $24.9\%$ erasure threshold from the bond-percolation threshold of the corresponding cubic lattice~\cite{barrettFaultTolerantQuantum2010a}. Alternative cluster state geometries can push thresholds up to $50\%$, and can even exceed $50\%$ if the cluster topology is modified to favour a particular logical observable (though this sacrifices protection on a complementary observable)~\cite{nickersonMeasurementBasedFault2018}.

Cluster state erasure thresholds are highly idealised and do not account for cluster state preparation. Recently, fusion-based quantum computing (FBQC)~\cite{bartolucciFusionbasedQuantumComputation2023} has emerged as a variant of MBQC tailored to photonics that decomposes cluster states into smaller resource states connected by linear-optical fusion measurements~\cite{browneResourceefficientLinearOptical2005}. In this way, it implements the cluster state model while also partially handling errors in preparation. Erasure thresholds in FBQC are dependent on the exact decomposition of the cluster into resource states -- for example, the rotated surface code can be decomposed into a network of six-qubit ring resource states, leading to a threshold of $12\%$ for independently erased fusion outcomes~\cite{bartolucciFusionbasedQuantumComputation2023}. Higher thresholds can be achieved using pre-encoded resource states~\cite{songEncodedFusionBasedQuantumComputation2024}, alternative codes such as foliated Floquet codes and graph codes~\cite{paesaniHighthresholdQuantumComputing2023, bellOptimizingGraphCodes2023} or by dynamically exploiting error biases~\cite{bombinIncreasingErrorTolerance2023,bartolucciComparisonSchemesHighly2025}. Since linear-optical fusions have an inherent failure probability, the corresponding loss thresholds are lower, though the exact translation depends on the implementation of the fusion measurement.

Decomposing cluster states into smaller resource states enables distributed computation, in which systems are scaled via large linear-optical networks of resource state generators~\cite{bombinInterleavingModularArchitectures2021,montfortFaulttolerantStructuresMeasurementbased2025}. In contrast, matter-based DQC systems reserve photonic networks for communication between QPUs that process information primarily locally~\footnote{This is a common model for DQC, though there are other models in which most operations are performed non-locally between few-qubit modules~\cite{nickersonTopologicalQuantumComputing2013,singhModularArchitecturesEntanglement2024,deboneThresholdsDistributedSurface2024,sutcliffeDistributedQuantumError2025,vakninFaulttolerantDistributedQuantum2026}.}. When QPUs are only large enough to store a small number of logical qubits, non-local gates must be performed via the network, typically via teleported transversal gates~\cite{ryan-andersonHighfidelityTeleportationLogical2024,stackTransversalFaultTolerant2026}, lattice surgery~\cite{horsmanSurfaceCodeQuantum2012,rametteFaulttolerantConnectionErrorcorrected2024,martonLatticeSurgerybasedLogical2025a,jacintoNetworkRequirementsDistributed2025,bodekerLatticeSurgeryNearterm2026,chelluriFaultTolerantLogicalOperations2026,schmidtResourceadaptiveDistributedFault2026} or logical entanglement distillation~\cite{sunamiEntanglementBoostingLowvolume2025,marqversenFaulttolerantInterfacesModular2025,pattisonFastQuantumInterconnects2024,ataidesConstantOverheadFaultTolerantBellPair2025}. Performing distributed lattice surgery requires some boundary stabiliser measurements to be performed non-locally. This can be done using distributed stabiliser measurement gadgets that either teleport non-local two-qubit gates~\cite{eisertOptimalLocalImplementation2000}, prepare multi-qubit entangled ancillas~\cite{nickersonFreelyScalableQuantum2014,jacintoNetworkRequirementsDistributed2025,ainleyErrorCorrectionDistributed2026a}, or jointly measure distributed ancillas using fusions~\cite{haugLatticeSurgeryBell2025}. Ramette et al. demonstrated that, in surface-code lattice surgery, error rates of $10\%$ can be tolerated at the interface with minimal effect on the local threshold~\cite{rametteFaulttolerantConnectionErrorcorrected2024}. Shalby et al. constructed precise circuits for interface stabiliser measurements that avoid distance-reducing hook errors, verifying that local circuit-noise thresholds are maintained with interface noise an order of magnitude stronger than the local noise~\cite{shalbyOptimizedNoiseresilientSurface2025}. Similar methods were proposed by Jacinto et al. in the context of large-scale DQC resource estimation~\cite{jacintoNetworkRequirementsDistributed2025}. Alternatively, Haug et al. proposed \textit{Bell measurements} for interface stabiliser measurements and found a $17.4\%$ link-error threshold when the local error rate was fixed at $0.1\%$~\cite{haugLatticeSurgeryBell2025}. Additional techniques to preserve lattice surgery distance have recently been proposed by Schmidt et al.~\cite{schmidtResourceadaptiveDistributedFault2026} and Thakur et al.~\cite{thakurSyndromeExtractionFramework2026} based on alternative decompositions of weight-$4$ boundary stabilisers. These works all follow the paradigm of a ``black-box'' network, accounting for network noise using augmented phenomenological noise~\cite{rametteFaulttolerantConnectionErrorcorrected2024}, elevated depolarisation on inter-QPU gates~\cite{shalbyOptimizedNoiseresilientSurface2025}, or depolarisation on entangled resource states~\cite{jacintoNetworkRequirementsDistributed2025,haugLatticeSurgeryBell2025,schmidtResourceadaptiveDistributedFault2026}. None of these models explicitly treat loss and probabilistic operations in the network. By abstracting away network details, heralding information is either discarded or assumed to be used for post-selection. This contrasts with FBQC, in which loss and fusion failures are treated as erasures to be decoded in real time.

A natural question to ask is whether the advantages of loss tolerance in photonic MBQC can be exploited alongside circuit-based DQC. A number of related proposals combine circuit-, measurement- and fusion-based methods at different architectural levels. These include spin--optical hardware for circuit-based syndrome extraction~\cite{gliniastySpinOpticalQuantumComputing2024}, deterministic gates combined with fusions to construct cluster states~\cite{sahayHighThresholdCodesNeutralAtom2023}, MBQC-like teleportation through networks of small fault-tolerant nodes~\cite{naitoNetworkBasedQuantumComputing2026}, and encoded-fusion subsystems that supply logical Bell pairs or execute subroutines for CBQC modules~\cite{larasatiHybridCircuitMeasurementBased2026}. These approaches combine paradigms across hardware components, computational stages or logical subsystems. While techniques for converting full protocols between circuit- and measurement-based error correction have been developed~\cite{boltFoliatedQuantumErrorCorrecting2016,brownUniversalFaulttolerantMeasurementbased2020a,bombinUnifyingFlavorsFault2024,pesahFaulttolerantTransformationsSpacetime2025b}, the design space between the two paradigms can be expanded by considering protocols which mix the two at the level of \textit{syndrome extraction}.

In this work, we construct \textit{hybrid} syndrome extraction protocols using the ZX calculus~\cite{coeckeInteractingQuantumObservables2011} and the Pauli web formalism introduced by Bombin et al.~\cite{bombinUnifyingFlavorsFault2024}. We extend the constructions of Ref.~\cite{bombinUnifyingFlavorsFault2024} to build hybrid protocols for lattice surgery between distributed logical qubits, replacing the sequential measurement of merge stabilisers with equivalent cluster states that locally interact with ancilla qubits. We provide decompositions of cluster state interfaces into \textit{fusion networks} with varying \textit{resource state} sizes. Our constructions are illustrated in \cref{fig:overview_pipeline}. Additionally, we decompose standard circuit-based protocols for distributed lattice surgery into variants using fusions of auxiliary Bell pairs. We analytically predict the fusion-erasure thresholds for our circuit-based protocols and hybrid protocols with varying resource state sizes. Using a heterogeneous error model that combines circuit-level depolarising noise on matter qubits with resource-state errors, photon loss and fusion failures, we analyse the thresholds achievable for both local and non-local noise and explore subthreshold performance. Through this, we show:

\begin{itemize}
    \item Under negligible local noise, independent fusion outcome erasure thresholds of $50\%$ are achievable in the network for distributed lattice surgery, both with an interface mediated by fused Bell pairs and with a cluster-state interface decomposed into a fusion network of linear-chain resource states.
    \item Relative to a straight surgery interface geometry using Bell pairs, the fusion-network interface approximately doubles the interface distances on \textit{both} joint observables: $d+1 \to 2d+1$ for the merge observable and $\lfloor(d+1)/2\rfloor \to d$ for the perpendicular observable. The zig-zag geometry~\cite{shalbyOptimizedNoiseresilientSurface2025,jacintoNetworkRequirementsDistributed2025} for Bell pair interfaces restores the perpendicular distance to $d$ though leaves the merge-observable distance at $d+1$.
    \item Truncating the linear resource states to a fixed size leaves both interface distances unchanged while lowering the erasure threshold predictably, to $38\%$ for eight-qubit chains and $32\%$ for five-qubit chains.
    \item Subthreshold performance exhibits a crossover between local-error and interface-erasure-dominated regimes. With local noise rates of $10^{-3}$, the increased interface distance has only a small effect. As local noise decreases to $10^{-4}$ and below, the effect becomes more pronounced and hybrid protocols provide better suppression of the merge observable than both Bell-pair interfaces, though the perpendicular observable remains best protected by the zig-zag Bell pair interface.
\end{itemize}

This work serves as an initial demonstration of the benefits of hybrid syndrome extraction protocols, highlighting the potential to tailor error correction schemes towards dominant noise sources such as photon loss. The paper is organised as follows. Section~\ref{sec:background} reviews the ZX calculus and Pauli webs. Section~\ref{sec:sed_framework} introduces syndrome extraction diagrams and web-preserving transformations, and Sec.~\ref{sec:hybrid_dqec} develops their hybrid realisations and applies them to distributed lattice surgery. Section~\ref{sec:protocols} describes resource states and interface erasure models. Sections~\ref{sec:numerical_simulations} and \ref{sec:results} present the noise models, simulations, and results, followed by a discussion and conclusion.

%% file: sections/2_preliminaries_zx.tex
\section{Preliminaries}\label{sec:background}
\subsection{ZX calculus}\label{sec:zx}

The ZX calculus~\cite{coeckeInteractingQuantumObservables2011} is a graphical language for describing and transforming quantum circuits~\cite{coeckePicturingQuantumProcesses2017} and has proven to be useful for describing quantum error correction~\cite{kissingerPhasefreeZXDiagrams2022}. We assume basic familiarity with ZX and provide only the definitions relevant to the constructions of \cref{sec:sed_framework,sec:hybrid_dqec}. For introductions to ZX, and the ZX-instrument notation used here, see van de Wetering~\cite{weteringZXcalculusWorkingQuantum2020} and Bombin et al.~\cite{bombinUnifyingFlavorsFault2024}, respectively.

We restrict to the stabiliser fragment of the ZX calculus~\cite{backensZXcalculusCompleteStabilizer2014} and confine further to consider only $0$ and $\pi$ phases. ZX diagrams are used at two levels: the level of stabiliser measurements~\cite{bombinUnifyingFlavorsFault2024}, at which ZX diagrams represent syndrome extraction circuits, and the level of logical operations~\cite{beaudrapZXCalculusLanguage2020}, at which they represent the higher-level primitives of lattice surgery. The constructions of this work are carried out at the first level; the logical-level description needed for lattice surgery is included for completeness in \cref{sec:hybrid_logical_circs}. Throughout, we adopt the extensions introduced by Bombin et al.~\cite{bombinUnifyingFlavorsFault2024}, namely \textit{ZX-instruments} used for representing classical information and \textit{Pauli webs}, though provide our own notation and definitions.

\begin{figure*}[tb]
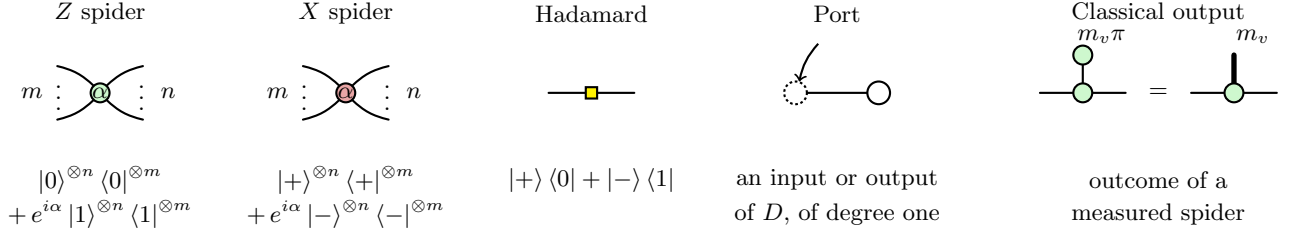

    \centering
    \tikzfig{2_preliminaries/01_alphabet}[scale=2]
    \caption{\textbf{Node types.} The four values of the type map $t$, together with the bold leg of a classical output. Spiders are drawn green ($Z$) and red ($X$) and carry a phase $\alpha \in \{0,\pi\}$; Hadamard nodes have degree two, while ports and classical outputs degree one. For the standard ZX calculus and ZX-instrument conventions, see Refs.~\cite{weteringZXcalculusWorkingQuantum2020,bombinUnifyingFlavorsFault2024}.}
    \label{fig:zx_alphabet}
\end{figure*}

The basic relevant components of ZX diagrams are shown in \cref{fig:zx_alphabet}. Here we fix the notation used throughout, and provide definitions for Pauli webs. This will provide the basis for defining diagrammatic transformations which preserve the structure of syndrome extraction diagrams while changing the physical realisation of the syndrome extraction.

\subsubsection{ZX diagrams}\label{sec:zx_diagrams}

The constructions of \cref{sec:sed_framework} proceed by explicit graphical transformations of ZX diagrams, so we treat a ZX diagram as a decorated graph, similar to the definition of Ref.~\cite{rodatzFloquetifyingStabiliserCodes2024}, though restricted to $0$ and $\pi$ phases.

\begin{definition}[ZX diagram]\label{def:zx_diagram}

A \textit{ZX diagram} is a graph $D = (V, E)$ with a type map $t : V \to \{Z, X, H, P\}$ and a phase map $\alpha(v) = a_v \pi$ on $t^{-1}(\{Z,X\})$, where each $a_v \in \mathbb{F}_2$ is either fixed by the diagram or a measurement outcome, subject to $t(v) = H \implies \deg(v) = 2$ and $t(v) = P \implies \deg(v) = 1$.
\end{definition}

The four node types of \cref{def:zx_diagram} are shown explicitly in \cref{fig:zx_alphabet}. $Z$ and $X$ nodes are green and red spiders respectively, representing tensors in the $Z$ and $X$ bases, $H$ nodes are Hadamard gates and $P$ nodes are ports, corresponding to the quantum inputs and outputs of the linear map represented by $D$. The ports are labelled as $m$ inputs and $n$ outputs. Edges incident to ports form the set of \textit{boundary edges}:

\begin{equation}\label{eq:boundary_edges}
\mathcal{B}(D) := \{e = (u,v) \in E : t(u) = P \text{ or } t(v) = P\}.
\end{equation}

An edge is \textit{internal} if it is not a boundary edge. For a given node $v$, we write $E(v) := \{ e \in E : v \in e \}$ for the \textit{legs} of $v$, so that $\deg(v) = |E(v)|$. Since all Hadamard nodes are degree $2$, we will refer to \textit{Hadamard} \textit{edges} as the combination of the node and its two legs. Distinct from ports, we define a \textit{classical output} as a degree-one spider $v$ with a colour matching that of its sole neighbour and a phase $m_v \pi$ that is parametrised by a measurement outcome $m_v$. Following Ref.~\cite{bombinUnifyingFlavorsFault2024} we draw the classical output as an open bold leg. Classical outputs are used to fix a realisation of a diagram, turning abstract detecting regions into experimental parity checks.

\subsubsection{Pauli webs}\label{sec:zx_webs}
Pauli webs are an additional element introduced by Bombin~et~al.~\cite{bombinUnifyingFlavorsFault2024}, which are termed a ``graphical overlay notation'' to ZX diagrams. Pauli webs display the Clifford properties of a ZX diagram, and can be used to identify stabilisers, detecting regions and logical operators. Pauli webs and analogous notations have been used for various purposes, including construction of Floquet codes~\cite{townsend-teagueFloquetifyingColourCode2023,rodatzFloquetifyingStabiliserCodes2024}, fault-tolerant circuit transformations~\cite{rodatzFaultToleranceConstruction2026,schmidtResourceadaptiveDistributedFault2026} and resource reductions~\cite{mcewenRelaxingHardwareRequirements2023,litinskiBlockletConcatenationLowoverhead2025}, to name a few.

Graphically, a Pauli web is a highlighting of the edges of a ZX diagram obeying the following rules\footnote{Note that many works using Pauli webs use an inverted colour convention~\cite{rodatzFloquetifyingStabiliserCodes2024,rodatzFaultToleranceConstruction2026,townsend-teagueFloquetifyingColourCode2023}, though we retain the convention of Ref.~\cite{bombinUnifyingFlavorsFault2024}.}:

\begin{itemize}
    \item A green (red) spider must have all or none of its legs highlighted green (red).
    \item A green (red) spider must have an \textit{even} number of legs highlighted red (green).
    \item The leg of a classical output is highlighted if and only if all \textit{other} legs of its sole neighbour are highlighted in a colour matching that of the neighbour.
    \item Any Hadamard exchanges the two colours between its input and its output.
\end{itemize}

\begin{figure*}[tb]
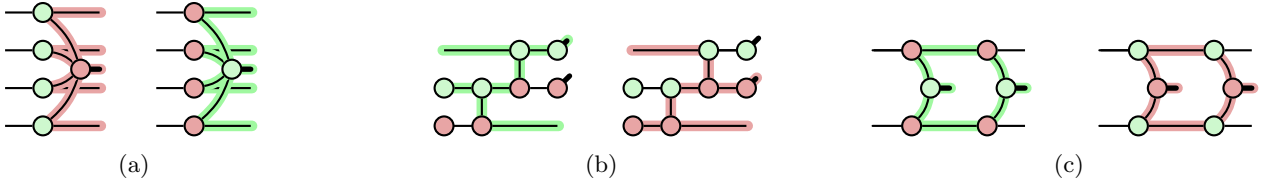

    \centering
    \begin{subfigure}{0.31\linewidth}
        \centering
        \tikzfig{2_preliminaries/02_stabiliser_web}
        \caption{}
        \label{fig:web_stabiliser}
    \end{subfigure}
    \hfill
    \begin{subfigure}{0.31\linewidth}
        \centering
        \tikzfig{2_preliminaries/03_teleportation_web}
        \caption{}
        \label{fig:web_logical}
    \end{subfigure}
    \hfill
    \begin{subfigure}{0.31\linewidth}
        \centering
        \tikzfig{2_preliminaries/04_check_web}
        \caption{}
        \label{fig:web_check}
    \end{subfigure}
    \caption{\textbf{The three classes of Pauli web}, distinguished by the input and output support of the boundary operator of \cref{eq:web_boundary}: (a) a \textit{stabiliser}, highlighted on outputs only; (b) a \textit{logical operator}, highlighted on both inputs and outputs; (c) a \textit{detecting region}, highlighted on neither, whose highlighted measurement outcomes impose a check.}
    \label{fig:web_basics}
\end{figure*}

More formally, we can define a web as an assignment of a Pauli operator to each edge. Specifically, a green edge carries $X$, a red edge carries $Z$, and an edge highlighted in both colours carries $Y$. Writing the Pauli on edge $e$ as $w_e = X^{x_e} Z^{z_e}$ up to phase, with $(x_e, z_e) \in \mathbb{F}_2^2$, we can define a \textit{valid} Pauli web as one that obeys the following constraints at each node.

\begin{definition}[Pauli web]\label{def:pauli_web}
A \textit{Pauli web} on a ZX diagram $D$ is an assignment $w : E \to \{I, X, Y, Z\}$ such that, at every node $v$ with $t(v) = Z$, the $x$-components are constant on the legs of $v$ and the $z$-components have even parity\footnote{These conditions hold for spiders of phase $k\pi$, which by \cref{def:zx_diagram} is every spider considered here. For spiders of arbitrary phase the parity condition acquires a phase-dependent correction, for which we refer the reader to Ref.~\cite{rodatzFaultToleranceConstruction2026}.},
\begin{equation}\label{eq:web_rules}
    x_e = x_{e'} \quad \forall\, e, e' \in E(v) , \qquad \sum_{e \in E(v)} z_e = 0 ,
\end{equation}
with $x$ and $z$ interchanged at nodes with $t(v) = X$, and such that $(x_e, z_e)$ is exchanged across every $H$ node.
\end{definition}

The first condition of \cref{eq:web_rules} gives the all-or-none rule and the second the even-parity rule, while the Hadamard rule is absorbed into the exchange of $(x_e, z_e)$ across an $H$ node. The rule for measurement outcomes falls out of the restriction placed on classical outputs. Since a classical-output spider has degree one, the parity rule forbids its leg from being highlighted in a non-matching colour. The all-or-none rule at its neighbour then requires this leg to be highlighted when all other legs of the neighbour are highlighted in its matching colour. Two webs may be added by multiplying their Pauli operators edgewise, which is an addition of $(x_e, z_e)$ over $\mathbb{F}_2^2$. The set of valid webs on a given diagram, which we write $\mathcal{W}(D)$, is closed under this addition and forms a vector space over $\mathbb{F}_2$. For our purposes, we will be interested only in a generating set of webs corresponding to stabilisers, detecting regions and logical operators in syndrome extraction diagrams.

A Pauli web can be interpreted by iterating through the spiders of the diagram and placing the corresponding Pauli contributions on each of their legs~\cite{rodatzFloquetifyingStabiliserCodes2024}. Internal edges receive two contributions, which cancel automatically, while boundary edges receive only one contribution each, which cancel only after being propagated through the diagram. Splitting the boundary edges as $\mathcal{B}(D)=\mathcal{B}_{\mathrm{in}}(D)\sqcup\mathcal{B}_{\mathrm{out}}(D)$ we can define the input and output Paulis of a web by
\begin{equation*}
    P_{\mathrm{in}}(w):=\bigotimes_{e\in\mathcal{B}_{\mathrm{in}}(D)}w_e,
    \qquad
    P_{\mathrm{out}}(w):=\bigotimes_{e\in\mathcal{B}_{\mathrm{out}}(D)}w_e,
\end{equation*}
where an empty tensor product is the identity. The \textit{boundary operator} of $w$ is the resulting map from its input Pauli to its output Pauli,
\begin{equation}\label{eq:web_boundary}
    O_w:P_{\mathrm{in}}(w)\longmapsto P_{\mathrm{out}}(w),
\end{equation}
defined without its overall sign. The cancellation of webs is only up to a scalar: a node of phase $\pi$ whose legs are all highlighted in its matching colour contributes a factor $-1$. For the phase-free generating webs considered below, these signs arise only from the outcome-dependent phases and are accounted for through the measurement outcomes. The classes of web used throughout this work are distinguished by the input and output support of $O_w$:

\begin{itemize}
    \item $P_{\mathrm{in}}(w)=I$ and $P_{\mathrm{out}}(w)\neq I$: a \textbf{stabiliser} (with input and output exchanged, a \textbf{co-stabiliser}), as shown in \cref{fig:web_stabiliser}.
    \item $P_{\mathrm{in}}(w)\neq I$ and $P_{\mathrm{out}}(w)\neq I$: a \textbf{logical operator}, as shown in \cref{fig:web_logical}.
    \item $P_{\mathrm{in}}(w)=P_{\mathrm{out}}(w)=I$: a \textbf{detecting region}, as shown in \cref{fig:web_check}.
\end{itemize}

Where webs have non-trivial support on classical outputs, the measurement outcomes determine either \textit{frame updates} or \textit{parity constraints}. We will be interested in the restricted cases in which the generating webs have no edge highlighted in both colours and all spiders (except for classical outputs) are phaseless. In this case, the signs come solely from the classical outputs. Calling $\mathcal{C}(w)$ the set of classical output vertices whose incident edges are highlighted by $w$, we define the outcome parity

\begin{equation}
    \mu(w) := \sum_{c \in \mathcal{C}(w)} m_c \mod 2.
\end{equation}

For a detecting region, $O_w:I\mapsto I$, and the outcome parity is $\mu(w)=0$ in the absence of errors. For stabilisers and logical operators, the outcomes supply the sign omitted from \cref{eq:web_boundary}, giving the signed boundary map
\begin{equation}
    O_w^{(m)}:P_{\mathrm{in}}(w)\longmapsto(-1)^{\mu(w)}P_{\mathrm{out}}(w).
\end{equation}

These outcome-dependent signs determine the relevant Pauli frame updates.

%% file: sections/3_syndrome_extraction.tex
\section{Syndrome extraction diagrams and transformations}\label{sec:sed_framework}

Error correction relies on the extraction of error syndromes from measurement outcomes. What changes between circuit-based error correction and measurement-based variations is \textit{how} the syndromes are extracted. A measurement-based implementation of a code built via foliation~\cite{boltFoliatedQuantumErrorCorrecting2016} has the same detecting regions and logical webs as the circuit-based implementation~\cite{bombinUnifyingFlavorsFault2024}, but a detecting region that is reported by two ancilla outcomes in the circuit-based case is reported by many single-qubit measurements in the measurement-based case. Likewise, corresponding logical webs retain the same boundary correlations, while the measurement outcomes determining their signs can differ between implementations.

Since the same detecting region can be reported by different sets of measurement outcomes, syndrome extraction protocols need not commit to one paradigm throughout. Using a set of diagrammatic transformations which preserve the Pauli webs of a syndrome extraction diagram, we can ensure that detecting regions and logical webs are preserved.

\begin{figure*}[tb]
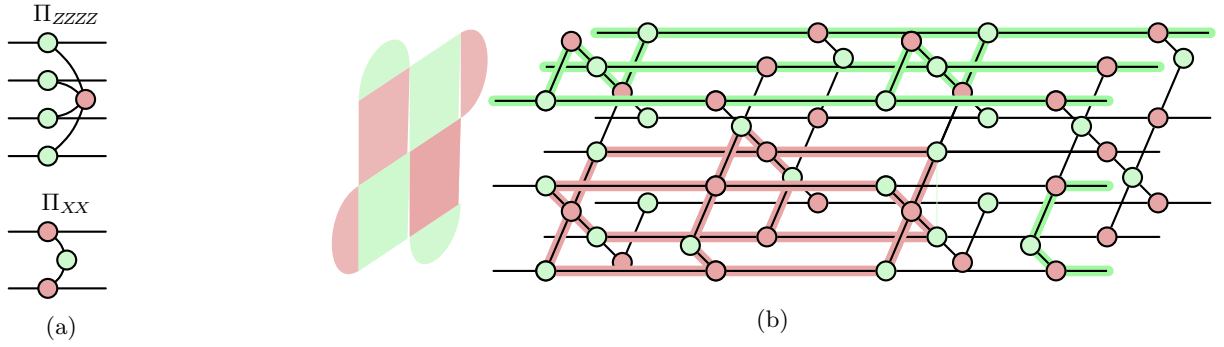

    \centering
    \begin{subfigure}[c]{0.25\linewidth}
        \centering
        \tikzfig{3_syndrome_extraction/01_generator_projectors}[scale=1.0]
        \caption{}
        \label{fig:multi_qubit_pauli_measurements}
    \end{subfigure}
    \hfill
    \begin{subfigure}[c]{0.7\linewidth}
        \centering
        \tikzfig{3_syndrome_extraction/02_surface_code_zx_rot}[scale=0.9]
        \caption{}
        \label{fig:surface}
    \end{subfigure}
    \caption{\textbf{Syndrome extraction diagrams and their Pauli webs.} (a) Phase-free projector gadgets for a four-qubit $Z$-type generator and a two-qubit $X$-type generator, each with a generator spider joined to data spiders. Exchanging spider colours exchanges the generator type. (b) All consecutive measurements of the same Pauli type can be merged into a single layer, and composing layers gives a syndrome extraction diagram (SED), with detecting regions, stabilisers, and logical webs corresponding to the code. A $Z$-type detecting region is shown in red, an $X$-type stabiliser from a two-qubit boundary generator in green, and an $X$ logical web along the top boundary in green.}
    \label{fig:se_diagrams}
\end{figure*}

\subsection{Syndrome extraction diagrams}\label{sec:se_diagrams}

For each chosen CSS stabiliser \textit{generator} $S$, the phase-free ZX gadget in \cref{fig:multi_qubit_pauli_measurements} represents the positive-eigenspace projector

\begin{equation*}
    \Pi_S = \frac{I+S}{2},
\end{equation*}
up to scalar normalisation. A \textit{generator spider} is joined to \textit{data spiders} in the support of $S$: a red generator spider with green data spiders for a $Z$-type generator, with the colours exchanged for an $X$-type generator. Data spiders do not necessarily correspond to data qubits -- this will depend on the physical realisation chosen later.

Composing projector gadgets in alternating $Z$-type and $X$-type layers and fusing all neighbouring spiders of the same colour gives a \textit{syndrome extraction diagram} (SED). An SED alone does not commit to a \textit{realisation} for the syndrome extraction, though it provides the core from which specific implementations can be derived. An SED carries three families of Pauli web that our transformations must preserve: 1) two consecutive occurrences of a generator give a detecting region, 2) a web extending from a generator in the last layer of its type to the outputs gives a stabiliser relation, while the corresponding input-supported web in the first layer gives a co-stabiliser relation, and 3) each chosen logical Pauli generator gives a web carrying its correlation from the input to the output boundary -- the logical membranes, or correlation surfaces, of Refs.~\cite{raussendorfTopologicalFaulttoleranceCluster2007,bartolucciFusionbasedQuantumComputation2023,bombinUnifyingFlavorsFault2024}. \Cref{fig:surface} shows a rotated surface code SED, highlighting a $Z$-type detecting region in red, an $X$-type stabiliser in green, and an $X$ logical web in green.

Without classical outputs, a detecting region does not yet report an experimental parity check -- distinguishing different realisations of a code requires additional diagrammatic annotations which will be specified in \cref{sec:outcome_introduction}. Before fixing these annotations, it may be necessary to transform the base SED. Before defining realisations, we specify a set of transformations that will be used to locally manipulate the SED while preserving its underlying code structure.
 
\subsection{Web-preserving transformations}\label{sec:transformations}

Starting with a phase-free ZX diagram with no classical outputs, the following graphical transformations preserve the represented map up to the scalar normalisation suppressed here, and importantly preserve the Pauli webs of the diagram. Equality of maps alone does not ensure this -- a rewrite can add or remove internal webs~\cite{rodatzFloquetifyingStabiliserCodes2024}. Throughout, input and output ports retain their labels, so their boundary edges can be compared directly.

\begin{definition}[Web-preserving transformation]\label{def:web_preserving}
A transformation $D \mapsto D'$ is \textit{web-preserving} if it induces a linear bijection $\tau : \mathcal W(D) \to \mathcal W(D')$ such that every web retains the same boundary operator,
\begin{equation}
    O_{\tau(w)}=O_w \qquad \forall\,w\in\mathcal W(D).
\end{equation}
\end{definition}

Thus each web in $D$ maps uniquely to a web in $D'$ that specifies the same input-to-output Pauli transformation, and addition of webs is preserved. The result of this is that all detecting regions are preserved, and additionally all stabilisers and logical operators continue to represent correlations between the same operators on their inputs and outputs. This does not by itself establish \textit{distance preservation}. This depends on a lower level decomposition and a noise model. This is investigated in detail in Refs.~\cite{rodatzFloquetifyingStabiliserCodes2024,rodatzFaultToleranceConstruction2026}.

Each rule we consider acts locally on a spider. We establish its web correspondence by fixing the labels on the legs entering the affected region and showing a unique correspondence between the valid internal labellings before and after the rule. Since each rule acts on a spider, we can split the analysis into the cases of \textit{matching} (web colour on spider legs matches spider colour) and \textit{non-matching} (web colour on spider legs does not match spider colour) highlight components, which obey independent binary constraints according to \cref{eq:web_rules} in \cref{def:pauli_web}. All illustrated rules also hold with red and green exchanged.

\subsubsection{Duality}

The \textit{duality} $\mathcal D_v$ exchanges the colour of spider $v$ and inserts a Hadamard on each of its legs.

\ctikzfig{3_syndrome_extraction/transformations/01_duality}[scale=1.4]\label{diag:duality}

The corresponding web exchanges its two highlight colours between the spider and the new Hadamards. Each Hadamard exchanges them back, leaving the external labels unchanged. This correspondence is unique and reversible: applying duality twice restores the diagram after cancelling adjacent Hadamard pairs.

\subsubsection{Identity insertion}

The \textit{identity insertion} $\mathcal I_e^c$ inserts a phase-free degree-two spider of colour $c\in\{Z,X\}$ on an edge $e$.

\ctikzfig{3_syndrome_extraction/transformations/02_identity_insertion}[scale=1.4]\label{diag:identity_insertion}

Any node placed on an edge will have two legs, thus both rules of \cref{eq:web_rules} are automatically satisfied by highlighting both legs with the same colour as the web at the edge.

\subsubsection{Split}

Let $A$ and $B$ partition the legs of spider $v$. The \textit{split} $\mathcal S_v(A,B)$ replaces $v$ by two spiders of the same colour, joined by one new edge and inheriting the legs in $A$ and $B$, respectively.

\ctikzfig{3_syndrome_extraction/transformations/03_split}[scale=1.4]\label{diag:split}

For the matching highlight component, the new edge is forced to carry the common value on the original legs. For the non-matching component, this splits into two distinct cases. Since the number of highlighted legs in $A$ and $B$ together must be even, either both have an even number of highlighted legs or both have an odd number. If both have an odd number, then the new edge is forced to be highlighted to restore even parity at both new spiders. If both have an even number, then parity is obeyed after the split without highlighting the new edge, so it remains unhighlighted. The original even-parity rule gives the same parity for $B$, so both new spiders satisfy the web rules. These values are forced, and merging the two spiders reverses the correspondence.

Note that the split is only web-preserving because the two split nodes are joined by a single edge, even though the spider fusion rewrite is valid for any number of legs. If we allowed two edges between $v_A$ and $v_B$, this would introduce a loop web in the non-matching colour of the node, thus breaking the bijection. 

\ctikzfig{3_syndrome_extraction/transformations/04_split_double}[scale=1.4]\label{diag:split_double}

The above transformation is \textit{not} web-preserving, and its inverse would remove a web. Forbidding these transformations restricts the scope, though ensures that we retain the set of webs corresponding to the original code. Such transformations may be useful in introducing new detecting regions to catch hook errors, particularly in lower level decompositions of local stabiliser measurements~\cite{rodatzFloquetifyingStabiliserCodes2024}, though are not used in our constructions.

Any sequence of the transformations introduced and their inverses is web-preserving, since their linear bijections compose and retain the same boundary labels. Classical outcomes and physical realisations are introduced in the next section.

\subsection{Splitting diagrams and sub-diagram boundaries}\label{sec:hybrid_construction}

Split transformations can be used to isolate subsets of nodes as distinct sub-diagrams, which interact only through sets of \textit{boundary edges} connecting the sub-diagrams. Let $S$ be a non-empty, proper, connected set of spiders in the initial SED, and let $\bar S$ denote the remaining spiders. We write $D[S]$ for the sub-diagram induced by $S$, using the Hadamard-edge shorthand. Define its \textit{boundary edges} and \textit{boundary spiders} by
\begin{align}
    \partial E(S)&=\{(u,v)\in E:u\in S,\ v\in\bar S\},\label{eq:region_boundary_edges}\\
    \partial V(S)&=\{u\in S:\exists v\in\bar S,\ (u,v)\in E\}.\label{eq:region_boundary_nodes}
\end{align}

The sets $\partial V(S)$ and $\partial E(S)$ describe the boundary between $D[S]$ and $D[\bar S]$, while $\mathcal{B}(D)$ continues to describe the full diagram boundary. The edge boundary $\partial E(S)$ is the cut associated with the partition $S\sqcup\bar S$. We say that $D[S]$ has a \textit{matching boundary} if this cut is a matching, so no two boundary edges share an endpoint. Equivalently,

\begin{equation}\label{eq:matching_boundary}
    |E(v)\cap\partial E(S)|=1
    \qquad \forall\,v\in\partial V(S)\cup\partial V(\bar S).
\end{equation}
The physical significance of this condition will become clear when the boundary edges are realised in \cref{sec:clusterisation}. The following procedure enforces it by splitting each original boundary spider into an inside and an outside component joined by a single edge.

\begin{figure}[tb]
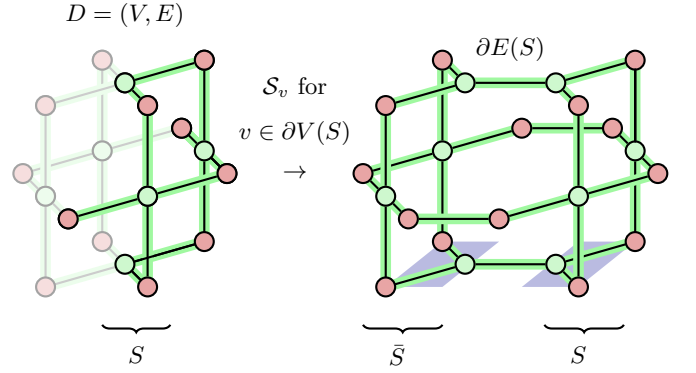

    \centering
    \tikzfig{3_syndrome_extraction/05_split_check}[scale=1.2]
    \caption{\textbf{Splitting a detecting region.} (left) An SED with a selected region $S$ and a highlighted detecting region. (right) Splitting the boundary generator spiders gives $D[S]$ a matching boundary.
    }
    \label{fig:split_check}
\end{figure}

We demonstrate an example splitting of a single surface code detecting region in \cref{fig:split_check}. By choosing $S$ such that every spider in $\partial V(S)$ is a generator spider, each spider $v$ can be split using $\mathcal S_v(A,B)$, where $A$ contains its legs leading into $S$ and $B$ its legs leading into $\bar S$. The two new spiders $v_S$ and $v_{\bar S}$ inherit these legs and are joined by one new edge. $S$ is then updated by retaining $v_S$ in place of $v$, while $v_{\bar S}$ is retained in the complement. We continue to write $D,S,\bar S$ for the updated diagram and regions. The new set of boundary edges $\partial E(S)$ consists of the split edges $(v_S,v_{\bar S})$, one per original boundary spider. These edges have distinct endpoints and hence form a matching, so $D[S]$ has a matching boundary.

%% file: sections/4_hybrid_dqec.tex
\section{Hybrid realisations of syndrome extraction}\label{sec:hybrid_dqec}

The previous section introduced syndrome extraction diagrams (SEDs) as the general representation of an error correction protocol for a CSS code. SEDs capture the structure of stabilisers and input/output correlations, though do not commit to a particular \textit{realisation} of a diagram. We distinguish different realisations of SEDs using two additional features: classical outcomes and world-lines. Classical outcomes convert detecting regions into parity checks on measurement outcomes while world-lines track the temporal evolution of individual qubits. A realisation can still be abstract, and need not tell us the exact decomposition of stabiliser measurements or preparation circuits for resource states, though the specification of world-lines and classical outcomes must provide enough information to commit to either circuit-based, measurement-based or hybrid implementations.

\subsection{World-lines and classical outcomes}\label{sec:outcome_introduction}

\subsubsection{World-lines}

To specify physical qubit identity and succession, we optionally add direction to selected quantum edges, inspired by previous works~\cite{townsend-teagueFloquetifyingColourCode2023,rodatzFloquetifyingStabiliserCodes2024,bombinUnifyingFlavorsFault2024}. Edge direction determines the temporal order of operations on the same physical qubit, with the arrow pointing from earlier to later interactions, thus specifying the \textit{world-line} of the qubit. Formally, we specify world-lines by vertex-disjoint directed paths $P_i=(v_1\to\cdots\to v_k)$, where the vertices are distinct and consecutive vertices are joined by quantum edges. Each $P_i$ corresponds to a single qubit and can be written as a sequence of directed edges $(v_1\to v_2, v_2\to v_3, \ldots, v_{k-1}\to v_k)$, where the directed edge $v_i\to v_{i+1}$ means that the qubit acted on at $v_i$ persists and is subsequently acted on at $v_{i+1}$. For a directed edge $v_i \to v_{i+1}$, if $v_i$ is a port, then the port must be an input, and if $v_{i+1}$ is a port, then the port must be an output. Classical output spiders are excluded from the world-lines. Each spider and each edge is part of at most one path. Edges and spiders which do not correspond to a path represent interactions between different qubits. A \textit{resource state} is a connected sub-diagram with no input ports in which every spider belongs to a distinct world-line containing no other spider. Such a world-line may terminate at that spider or continue to a unique output port.

\begin{figure*}[tb]
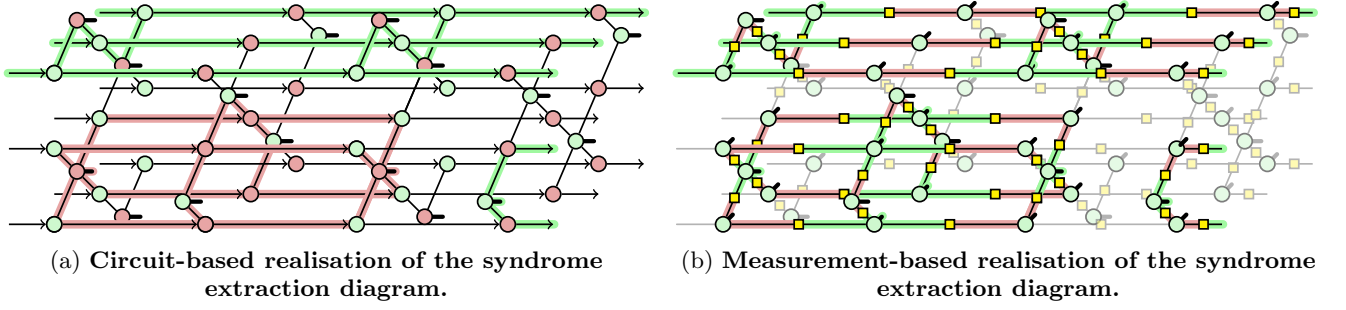

    \begin{subfigure}{0.49\linewidth}
        \scalebox{0.8}{
        \tikzfig{4_hybrid_dqec/01_surface_code_zx_rot_circuit}[scale=1.0]}
        \caption{\textbf{Circuit-based realisation of the syndrome extraction diagram.}}
        \label{fig:sed_circuit_realisation}
    \end{subfigure}
    \begin{subfigure}{0.49\linewidth}
        \scalebox{0.8}{
        \tikzfig{4_hybrid_dqec/02_surface_code_zx_rot_cluster}[scale=1.0]
        }
        \caption{\textbf{Measurement-based realisation of the syndrome extraction diagram.}}
        \label{fig:sed_cluster_realisation}
    \end{subfigure}
    \caption{\textbf{Two realisations of the syndrome extraction diagram: circuit-based (a) and measurement-based (b).} In the circuit-based realisation, directed edges represent data-qubit world-lines, while each generator spider lies on a distinct world-line and has a classical output recording its stabiliser measurement outcome. In the measurement-based realisation, each bulk spider lies on a distinct world-line and represents a qubit initialised in $\ket{+}$, while edges represent $CZ$ gates forming a cluster state and classical outputs record the measurement outcomes.}
    \label{fig:realisations}
\end{figure*}

\subsubsection{Classical outcomes}

For a spider $v$ with no classical output, let $\mathcal C_v$ attach a matching-colour classical-output spider carrying a measurement outcome bit. Adding a classical output is no longer an equality of linear maps -- setting the new bit to zero recovers the original diagram, while setting it to one introduces a $\pi$ phase at $v$.

\ctikzfig{4_hybrid_dqec/transformations/01_measurement}[scale=1.4]\label{diag:classical_outcome}

The new leg cannot carry the non-matching highlight component, by the degree-one output spider's parity rule. Its matching component is fixed by the all-or-none rule at $v$. Thus every web extends uniquely, and deleting the output leg recovers the original web. This extension is linear and leaves quantum boundary labels unchanged, so outcome introduction is web-preserving in the sense of \cref{def:web_preserving}. Measurements change the recorded parity $\mu(w)$ for webs highlighting the new leg. Here the parent has at least one pre-existing leg (adding an output to a disconnected node would create a new trivial web on the classical output), as in the gadgets considered throughout. Attaching a classical output to a generator spider turns its corresponding detecting regions into parity checks. Outcome $m$ selects the generator eigenspace with eigenvalue $(-1)^m$. A detecting region between consecutive occurrences of a generator then reports $a\oplus b=0$ in the absence of errors, where $a,b$ are the two recorded bits.

\subsubsection{Circuit- and measurement-based realisations}

A \textit{circuit-based realisation} of an SED has a data qubit world-line from each input to its corresponding output (defined by the shortest path from a given input to an output), while each ancilla spider lies on a distinct world-line containing no other spider. An ancilla here represents one preparation-and-measurement instance; hardware may be reset and reused in later rounds. The arrows specify the succession of Pauli-measurement gadgets along each data qubit, leaving the gate sequence within each gadget to a further decomposition. The circuit-based realisation of \cref{fig:surface} is shown in \cref{fig:sed_circuit_realisation}.

Alternatively, a \textit{measurement-based realisation} of an SED can be obtained first by applying duality $\mathcal{D}_{v}$ to all red spiders in the base SED. This converts the diagram into \textit{graph-like} form, since it contains only green spiders and Hadamard edges~\cite{duncanGraphtheoreticSimplificationQuantum2020}. Following this, each spider is assigned to a distinct world-line containing no other spider and is given a classical output~\cite{bombinUnifyingFlavorsFault2024}. The resulting ZX diagram is a \textit{cluster state} where each spider is a qubit initialised in $\ket{+}$ and each internal edge is a $CZ$ gate, or cluster bond~\cite{heinEntanglementGraphStates2006}. We use the term \textit{cluster state} to refer to a graph-like resource state diagram. The addition of classical outputs completes the measurement-based realisation -- bulk detecting regions acquire an additional four measurement outcomes in their parity checks, and $d$ outcomes are added to the frame updates of each logical operator per round. The measurement-based realisation is shown in \cref{fig:sed_cluster_realisation}.

\subsection{Realisations of transformed SEDs}\label{sec:bell_fusion_realisations}

While the base SED provides a starting point for both circuit-based and measurement-based realisations, architectural constraints may necessitate modifications. For example, an SED that has been transformed using splits could correspond to a circuit-based realisation using distributed entangled ancillas, or a measurement-based realisation where cluster states are decomposed into smaller resource states connected by fusions. Here we define specific realisation procedures for these transformed diagrams, which include the necessary world-line annotations and classical outputs while specifying additional implementation details.

\subsubsection{Bell preparation}

An SED could be realised across two processors by splitting ancilla nodes along an interface, and preparing them in a Bell state $\ket{\Phi^+}=(\ket{00}+\ket{11})/\sqrt{2}$.

\ctikzfig{4_hybrid_dqec/transformations/02_bell_preparation_walkthrough}[scale=1.3]\label{diag:bell_preparation}

The realisation uses the web-preserving identity insertion and split transformations to decompose the Bell preparation, and fixes the paths such that two qubit world-lines connected by a $CX$ gate are established. The world-lines enforce that the Bell state is prepared before any subsequent operations on the qubits. Adding classical outputs to the two endpoints represents local readouts of the ancillas, with outcome bits $a$ and $b$. Matching Pauli webs are preserved by including both outcomes in their parity, while non-matching webs remain unaffected.

\subsubsection{Fusion}

Alternatively, a split edge can be realised as a joint \textit{fusion measurement}. A \textit{Bell fusion} jointly measures the $X\!X$ and $Z\!Z$ parity of the two ancillas after the local gates have been performed~\cite{bartolucciFusionbasedQuantumComputation2023, browneResourceefficientLinearOptical2005}. The measurement consumes the two incoming qubits and records the eigenvalues of the commuting operators $X\!X$ and $Z\!Z$ as $(-1)^{m_{X\!X}}$ and $(-1)^{m_{Z\!Z}}$.

\ctikzfig{4_hybrid_dqec/transformations/03_fusion_walkthrough}[scale=1.3]\label{diag:fusion}

The fusion realisation is structurally identical to the Bell preparation, though it inverts the world-lines, such that fusion occurs after other operations at the two endpoints, and adds classical outcomes at the end of the world-lines. The resulting webs include either the $m_{X\!X}$ or $m_{Z\!Z}$ outcomes, depending on the colour of the web at the split edge. Fusions thus contribute outcomes to Pauli webs in both matching and non-matching cases. Fusions can be used to measure stabilisers in the circuit-based realisation, as in Ref.~\cite{haugLatticeSurgeryBell2025}, though can also be used to decompose cluster states into smaller resource-state components~\cite{bartolucciFusionbasedQuantumComputation2023}.

The fusion takes a different form when the split edge carries a Hadamard, which is the case when duality has been applied to one of the nodes. In this case, the fusion measures the two-qubit operators $X\!Z$ and $Z\!X$ instead of $X\!X$ and $Z\!Z$.

\ctikzfig{4_hybrid_dqec/transformations/04_hadamard_fusion_split}[scale=1.3]\label{diag:hadamard_fusion}

A Hadamard fusion preserves webs in the same way as a Bell fusion, though measures different Pauli operators in the process.

Bell preparations and fusions can be combined to create a lower-level realisation of remote entanglement generation. A Bell pair can be created by fusing two auxiliary Bell pairs in a process known as an \textit{entanglement swap}.

\ctikzfig{4_hybrid_dqec/transformations/05_bell_prep_fusion}[scale=1.3]\label{diag:bell_prep_fusion}

This will allow us to treat circuit-based protocols using Bell pairs under noise models incorporating loss and fusion failure as in FBQC. We will describe this further in \cref{sec:protocols}.

\subsection{Hybrid diagrams and hybrid checks}\label{sec:hybrid_diagrams}

The circuit-based and measurement-based pictures are two extreme cases of the hybrid realisations of syndrome extraction. Informally, a hybrid SED has a circuit-based region coupled to a measurement-based region, with some detecting regions or logical webs supported across both. To make this definition more formal, we first introduce the notion of a cluster region and the process of clusterisation, which allows us to identify and transform parts of the SED into a graph-like form suitable for measurement-based operations.

\subsubsection{Cluster regions and clusterisation}\label{sec:clusterisation}

\begin{definition}[Cluster region]\label{def:cluster_region}\label{def:cluster}
In a chosen physical realisation, a connected subset of nodes $S$ with $|S|>1$ and with an induced sub-diagram $D[S]$ is a \textit{cluster region} if $D[S]$ is a graph-like resource state (a cluster state) and each spider in $S$ is connected to an output port, classical output, or fusion.
\end{definition}

A fusion consumes each participating resource qubit, so a boundary resource qubit can participate in at most one fusion. Since the boundary-splitting procedure of \cref{sec:hybrid_construction} ensures that boundary edges form a matching, every boundary spider is incident on exactly one boundary edge by \cref{eq:matching_boundary}. After each spider in the cluster region is assigned to a distinct world-line, every boundary spider represents a distinct resource qubit participating in exactly one boundary fusion. In order to obtain a cluster region in a graph, we define a \textit{clusterisation} procedure.

\begin{definition}[Clusterisation]\label{def:clusterisation}
A \textit{clusterisation} of a sub-diagram $D[S]$ consists of:

\begin{enumerate}
    \item converting $D[S]$ into a graph-like form;
    \item assigning each spider in $S$ to a distinct world-line containing no other spider;
    \item adding classical outputs to spiders in $S \setminus \partial V(S)$, and realising edges in $\partial E(S)$ as fusions.
\end{enumerate}
\end{definition}

A measurement-based realisation is thus obtained by choosing $S$ as the full set of nodes. In this case, inputs and outputs become ports for composing together larger diagrams, rather than holding output qubit states. If the purpose of the SED is to prepare a particular state, qubits connected to output ports may be optionally left unmeasured. Additionally, a fusion based decomposition can be obtained by first a applying a series of splits to the SED, then applying clusterisation to each sub-diagram~\footnote{Alternatively the full diagram could be clusterised then subsequently split into sub-diagrams. We choose to transform first such that all transformations precede realisation.}.

\subsubsection{Hybrid realisation}

We can then define a hybrid realisation of an SED as one that contains both a cluster region and a circuit-based extraction region, connected by a matching boundary.

\begin{definition}[Hybrid diagram]\label{def:hybrid_diagram}
An SED with a specified physical realisation is \textit{hybrid} if it contains a cluster region $D[S]$ coupled through a non-empty set of boundary edges forming a matching $\partial E(S)$ to a circuit-based extraction region $D[\bar{S}]$ containing at least one data-qubit world-line with more than one spider.
\end{definition}

\begin{figure*}[tb]
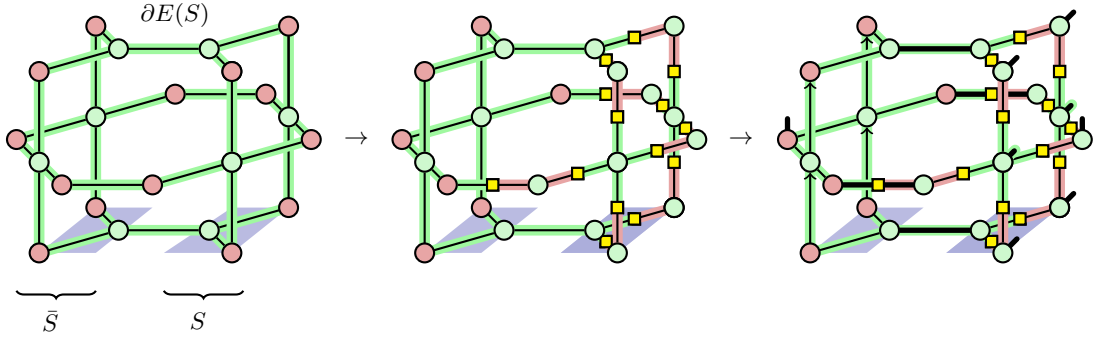

    \centering
    \tikzfig{4_hybrid_dqec/03_hybrid_check}[scale=1.2]
    \caption{\textbf{Hybrid realisation of a split detecting region.} (left) the SED from \cref{fig:split_check} after splitting the boundary generator spiders. $S$ is clusterised in two stages: (centre) duality makes $D[S]$ graph-like and converts select boundary edges to Hadamard edges; (right) nodes in $S$ are fixed with single spider world-lines, classical outcomes and fusions, while data qubit world-lines fix $D[\bar{S}]$ as a circuit-based diagram. The corresponding check contains outcomes from local ancilla readouts in boundary fusions and resource measurements in the cluster region, making it a hybrid check.}
    \label{fig:hybrid_check}
\end{figure*}

We can build a simple example of a hybrid diagram using the single-detecting-region SED from \cref{fig:split_check}. Once $D[S]$ has a matching boundary, we fix the hybrid realisation by applying clusterisation to $D[S]$ and realising $D[\bar{S}]$ as a circuit-based diagram by adding world-lines on data nodes and classical outcomes on all generator spiders (except those on the boundary). The clusterisation definition forces the boundary edges to be realised as fusions, which we refer to as \textit{boundary fusions}: fusion measurements coupling qubits across the boundary between the circuit-based and measurement-based regions. Regular boundary edges are realised as $X\!X$--$Z\!Z$ fusions, while Hadamard boundary edges arising from clusterisation are realised as $X\!Z$--$Z\!X$ fusions. Since each spider in the cluster region lies on a distinct world-line containing no other spider, there is no inherent ordering to measurements in the bulk of the cluster. Boundary fusions, however, depend on operations in the circuit-based regions and therefore inherit an ordering from the data-qubit world-lines. The bulk of the cluster can be prepared and measured freely, while fusions on boundary qubits must respect the causal ordering of the circuit-based operations.

Transforming to a hybrid SED results in detecting regions which cross the boundary between the cluster and circuit-based extraction regions. This gives rise to parity checks which contain contributions from circuit-based ancilla measurements and cluster-based resource measurements -- we refer to these as \textit{hybrid checks}. The example in \cref{fig:hybrid_check} follows one detecting region through the construction. Its external support remains trivial through the splits and dualities. The final annotations extend it to boundary-fusion outcomes and local resource readouts, yielding a hybrid parity constraint. The same graphical transformations and outcome-extension rules act on all webs, retaining the stabiliser and logical correlations as well as the detecting regions.

\subsection{Constructing hybrid lattice surgery diagrams}\label{sec:hybrid_logical_circs}

\begin{figure*}[tb]
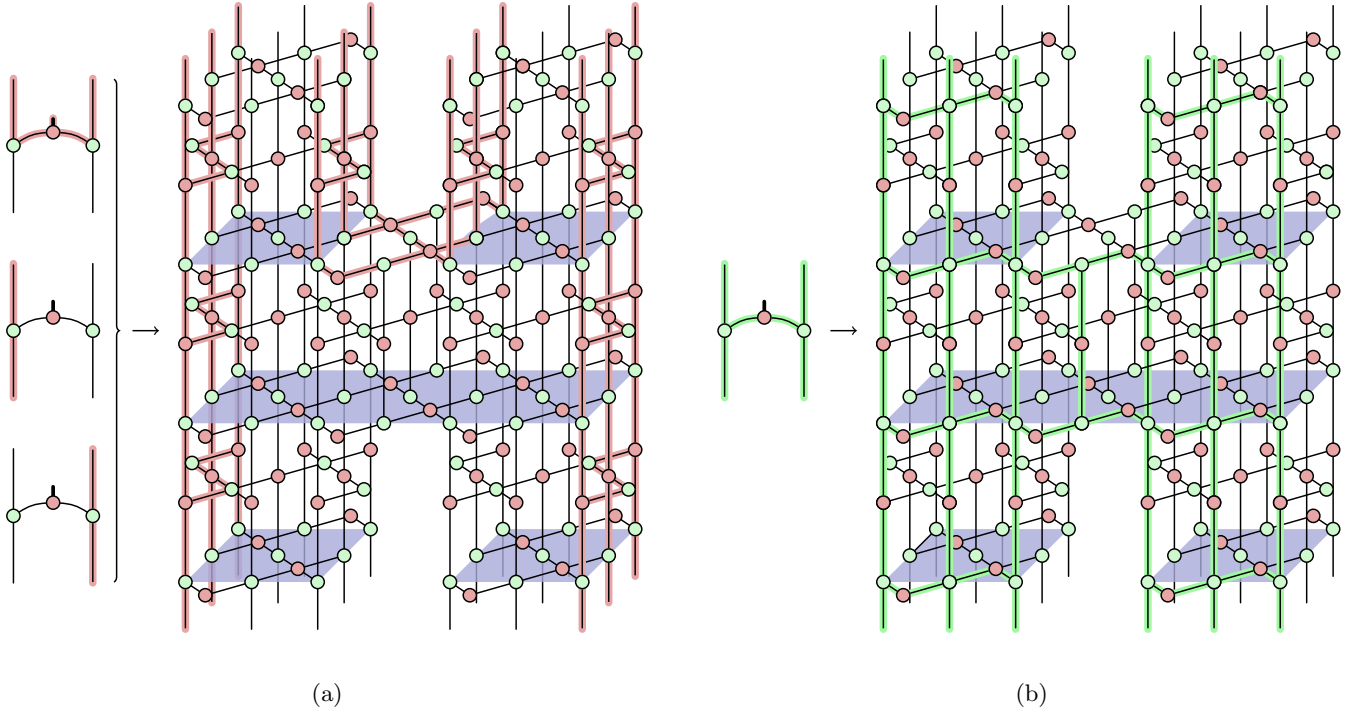

    \centering
    \begin{subfigure}{0.48\textwidth}
        \centering
        \scalebox{0.7}{\tikzfig{4_hybrid_dqec/04_phys_to_log_surgery_zz}[scale=1.0]}
        \caption{}
             \label{fig:zx_surgery_phys_to_log}
    \end{subfigure}
    \hfill
    \begin{subfigure}{0.48\textwidth}
        \centering
        \scalebox{0.7}{\tikzfig{4_hybrid_dqec/05_phys_to_log_surgery_other_obs_zz}[scale=1.0]}
        \caption{}
        \label{fig:zx_surgery_other_obs}
    \end{subfigure}
    \caption{\textbf{Correlation surfaces in a lattice surgery merge-and-split SED.} The logical-level Pauli webs map to correlation surfaces in the SED. (a) The two individual $Z$ correlations and the joint $Z\!Z$ correlation reporting the logical measurement. (b) The joint $X\!X$ correlation, whose outcome support includes a seam readout at the split. Small diagrams show the logical webs and large diagrams their physical representatives. For clarity, only one $Z$-type and one $X$-type layer is drawn per phase (plus an additional $Z$-type layer in the merge region). In practice, each phase contains $d$ layers of each type with an additional $Z$-type layer in the merge region.}
    \label{fig:zx_surgery}
\end{figure*}

\begin{figure*}[tb]
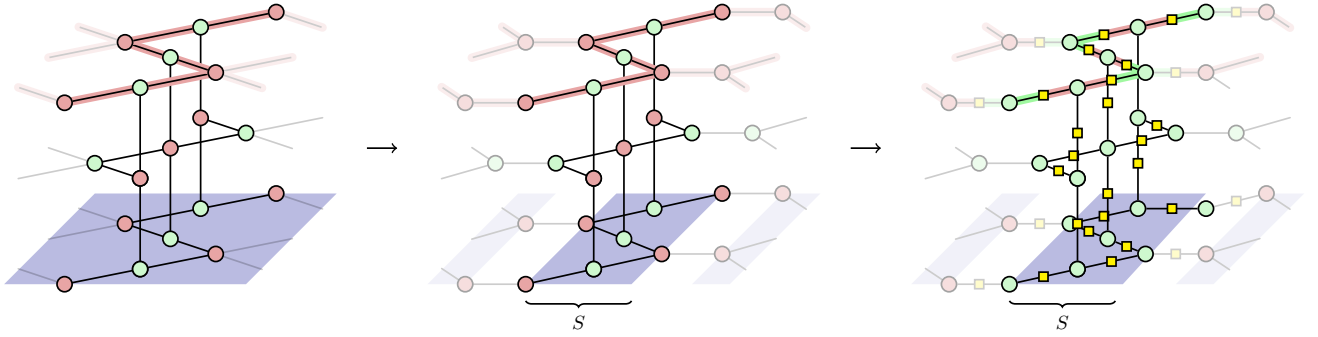

    \centering
    \scalebox{0.8}{\tikzfig{4_hybrid_dqec/06_surgery_circuit_split_2d_pre_noscale_ZZ}[scale=1.0]}
    \caption{\textbf{Graphical transformation of the merge region.} (left) The seam and neighbouring generator spiders, carrying the joint logical web. (centre) Split each boundary generator into an inside and outside half, joined by an edge of $\partial E(S)$. (right) Dualise the red spiders inside $S$, including their boundary legs, to obtain graph-like form. All three panels precede outcome and world-line annotations. The highlighted web is carried through the splits and dualities with its external Pauli labels retained. Faded spiders and edges belong to the complementary region.}
    \label{fig:surgery_circuit_split}
    \label{fig:zx_clusterise_surgery}
\end{figure*}

The construction of \cref{sec:hybrid_construction} can be applied to SEDs implementing logical operations using techniques such as lattice surgery. Lattice surgery couples surface-code patches by modifying  stabilisers near their shared boundary using a line of data qubits (the seam), confining the interaction to a merge region. By clusterising this region, the logical patches can remain on separate matter-based processors while resource states mediate their interaction.

The primitive lattice surgery operations are the \textit{merge} and \textit{split}~\cite{horsmanSurfaceCodeQuantum2012}. A merge joins two patches by measuring stabilisers across their shared boundary, whose outcomes determine a joint logical Pauli. A split separates them again by measuring out a line of data qubits. We consider a merge-and-split measuring the logical $Z\!Z$ operator, though exchanging $X$ and $Z$ gives the corresponding $X\!X$ operation.

The ZX calculus describes these operations at the logical level~\cite{beaudrapZXCalculusLanguage2020}. Each logical Pauli web is represented at the syndrome level by a correlation surface. These surfaces describe both the joint measurement and the logical correlations retained through it, and can be composed to describe Bell preparation, fan-out, $C\!X$ gates and teleportation. The hybrid construction must retain these logical webs as well as the detecting regions.

\begin{figure*}[tb]
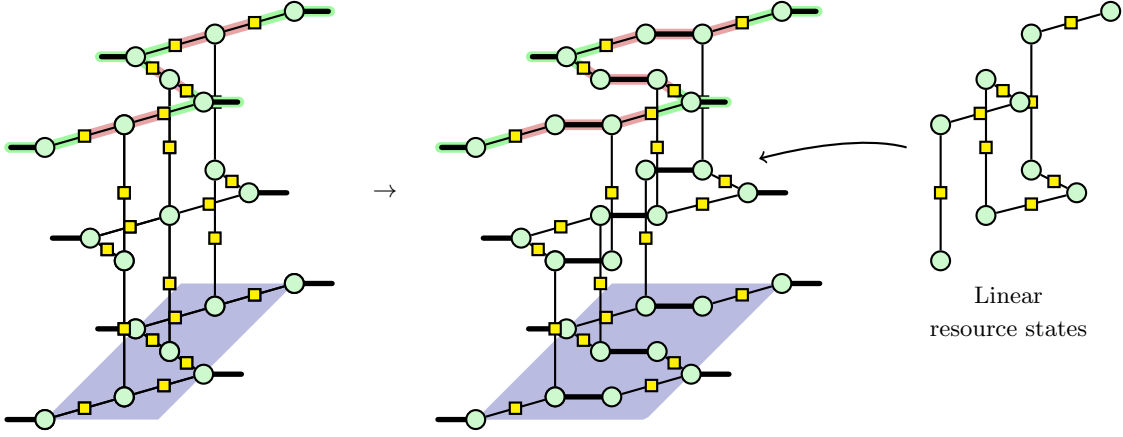
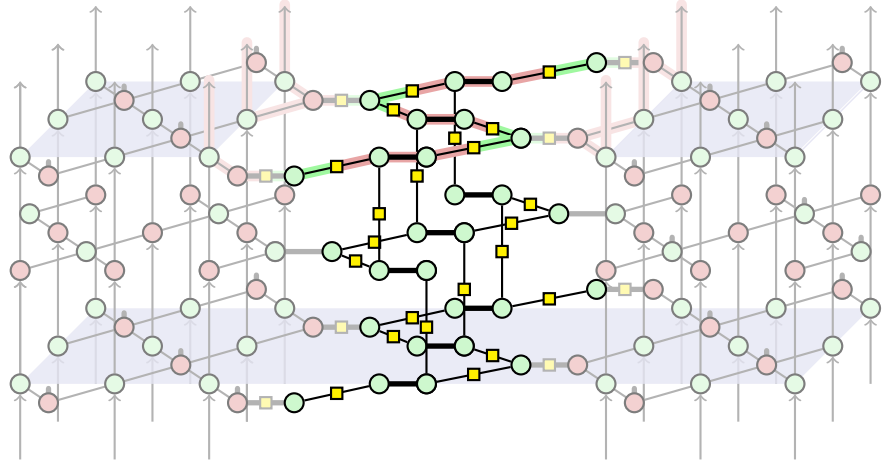

    \centering
    \begin{subfigure}{\textwidth}
        \centering
        \scalebox{1.0}{\tikzfig{4_hybrid_dqec/07_hybrid_protocol_example_cluster_to_fn}[scale=1.2]}
        \caption{Splitting seam spiders gives a network of linear resource states. Internal bold edges denote fusions between resources; open bold legs include couplings to the omitted circuit region. Local readout legs not needed for the highlighted logical web are suppressed.}
        \label{fig:hybrid_protocol_example_cluster_to_fn}
    \end{subfigure}
    \par\medskip
    \begin{subfigure}{\textwidth}
        \centering
        \scalebox{1.0}{\tikzfig{4_hybrid_dqec/08_phys_to_log_surgery_linear_long_web_zoom}[scale=1.0]}
        \caption{The linear-resource fusion network placed between the two circuit-based patches. The logical web extends through the resource fusions and the boundary couplings.}
        \label{fig:zx_fusion_linear_surgery}
    \end{subfigure}
    \caption{\textbf{A resource-state realisation of the merge region.} The choice of splits determines the resource topology; fusion and local-readout annotations determine the outcomes reporting each web. The surrounding circuit is faded to emphasise the resource network.}
    \label{fig:hybrid_resource_decomposition}
\end{figure*}

\subsubsection{Building surgery SEDs}

Begin with two distance-$d$ rotated surface-code memory SEDs, each containing $d$ layers of $X$-type and $Z$-type generators. Between the patches, introduce a seam of $d$ strands with $\ket{+}$ preparation spiders. Compose these with the SED for a single rectangular patch of distance $(2d+1)\times d$, covering both patches and the seam. After the merge layers, terminate the seam strands with green spider projectors and continue and compose with two further memory SEDs. For $Z\!Z$ surgery, we choose the first and last merge layers to be $Z$-type -- this allows initialisations and projections in the merge region to be fused into data spiders, leaving only generator and data spiders. The merge therefore contains $d$ layers of each generator type, plus one additional $Z$-type layer, giving $2d+1$ alternating extraction layers in the merge region.

To obtain the circuit realisation, introduce singleton world-lines and outputs for generators and the seam $X$-readout outcomes, and direct the data strands through the extraction layers. The seam strands now describe qubits prepared in $\ket{+}$ at the merge and measured in $X$ at the split. The resulting instrument measures logical $Z\!Z$, with the outcome reported by the final layer of ancilla readouts in the merge region. \Cref{fig:zx_surgery} shows the corresponding logical webs: the two individual $Z$ correlations and the joint $Z\!Z$ relation, together with the surviving joint $X\!X$ correlation. The latter includes a seam readout, so its sign also depends on the split record.

\subsubsection{Clusterising the merge region}

Returning to the base surgery SED, we choose a subset of nodes $S$ that contains all seam data spiders and their neighbouring generator spiders in the merge. We call this the \textit{merge region}. It is connected and colour-bipartite, and its boundary spiders $\partial V(S)$ are generator spiders.

Applying the construction of \cref{sec:hybrid_construction}: we split the boundary generators, retain their inside halves in $S$, and dualise every red spider in the selected region. \Cref{fig:surgery_circuit_split} shows these graphical steps. We then supply resource state assignments within $S$, circuit world-lines outside it, and the outcomes of the boundary fusions and remaining local readouts. This completes the clusterisation and gives a hybrid surgery diagram. The seam is realised by a two-dimensional sheet of resource state qubits instead of persistent circuit data qubits, while the local data paths order their interactions with the boundary ancillas.

\subsubsection{Decomposing the cluster region}

The merge region can be further decomposed into smaller resource states by applying split operations to seam nodes before clusterisation. Temporarily removing all split edges leaves disjoint sets of nodes in the merge region -- these can be individually clusterised to produce a fusion network. The exact resource state decomposition depends on how the incident legs are partitioned. Regular small resources, such as the six-ring states of Ref.~\cite{bartolucciFusionbasedQuantumComputation2023}, provide one approach. Here we use a partition of the seam spiders' legs that produces linear resource states, as shown in \cref{fig:hybrid_protocol_example_cluster_to_fn}. These chains span the merge boundary, each containing $3d$ qubits, though each chain's boundary edges are restricted to one of the two logical qubit boundaries. The resulting fusion network replaces the seam while leaving the two circuit-based patches in place, as shown in \cref{fig:zx_fusion_linear_surgery}. A chain can be shortened by replacing one of its Hadamard edges with a chain of three Hadamard edges and two new green degree-two spiders. We demonstrate this construction in \cref{sec:protocols} and present a family of truncated linear resource states.

%% file: sections/5_protocols.tex
\section{Resource states and interface detector error models}\label{sec:protocols}

So far the protocols we have described have been abstract -- while we have drawn a distinction between circuit-based, measurement-based and hybrid realisations of syndrome extraction diagrams, we have not yet described the physical implementation or the corresponding error models. Part of the motivation behind the hybrid approach is that different physical implementations can be used for circuit-based and measurement-based components, allowing us to tailor the protocols to heterogeneous noise models. In a distributed quantum computer with both matter-based and photonic qubits, the strongest noise effects typically occur at the \textit{interface} between different hardware modules. Fusion failures and loss of photons in the network cause \textit{erasures}, typically at much higher rates than local computational errors. In remote entanglement generation, heralds from erasures are usually used to post-select on successful outcomes. However, since fusion-based quantum computing is designed to handle loss and fusion failure in real time, network erasures can be treated as errors to be decoded. To analyse tolerance to network erasures, we can derive \textit{detector error models} for the reduced case where only network erasures are considered. This allows us to predict erasure thresholds and effective distances for different interfaces. Before doing so, we must first clarify which components of the system are considered part of the interface and how they are represented in the corresponding error model.

\subsection{Resource state generation}

\begin{figure*}[htb]
    \centering
    \begin{subfigure}[c]{0.53\textwidth}
        \centering
        \resizebox{0.98\linewidth}{!}{\input{figures/5_protocols/01_emitter_bell_fusion.tikz}}
        \caption{Remote Bell-pair generation.}
        \label{fig:emitter_bell_fusion}
    \end{subfigure}
    \hfill
    \begin{subfigure}[c]{0.43\textwidth}
        \centering
        \tikzexternaldisable
        \resizebox{0.98\linewidth}{!}{\input{figures/5_protocols/02_emitter_linear_cluster.tikz}}
        \tikzexternalenable
        \caption{Sequential linear-cluster generation.}
        \label{fig:emitter_linear_cluster}
    \end{subfigure}
    \caption{\textbf{Emitter-based resource-state generation.} (a) Each retained emitter $e_A,e_B$ emits a photon $p_A,p_B$ entangled with it in a Bell pair (dashed arcs). The photons propagate to a Bell-state analyser and are consumed in a linear-optical fusion measurement. On successful fusion, its outcomes $m_{XX},m_{ZZ}$ herald a Bell pair between the emitters and determine its Pauli frame. (b) A single emitter initialised in $\ket{+}$ generates a linear cluster by repeatedly emitting a photon and undergoing a Hadamard rotation. After $N$ cycles, the emitter remains at one end of a chain of $N$ photons.}
    \label{fig:emitter_resource_generation}
\end{figure*}
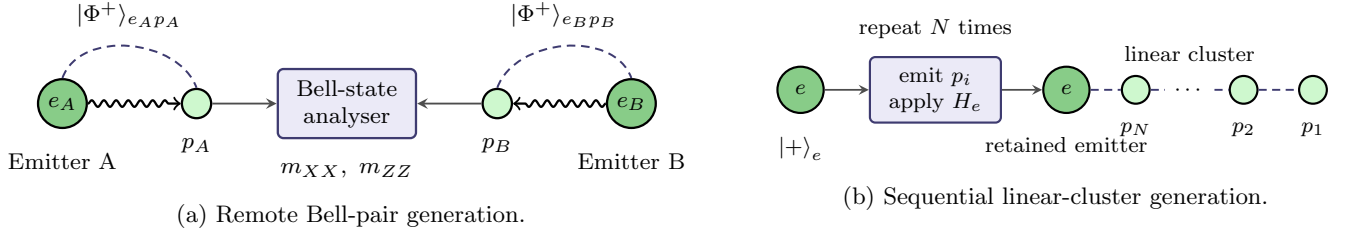

Interacting matter-based qubits with photons is typically done via a matter-based emitter -- a qubit that can emit photons entangled with it. To generate all-photonic resource states, emitters can be measured out after emission, leaving only photons in the state~\cite{buterakosDeterministicGenerationAllPhotonic2017}. However, since matter-based qubits can interact deterministically with each other, it is desirable to retain matter-based qubits in resource states in heterogeneous systems. The result of this is that local two-qubit gates and \textit{boundary fusions} can be performed deterministically, leaving only the linear-optical fusions to be performed probabilistically. The simplest example of this is the generation of a matter-based Bell pair using distributed emitters. Each remote emitter produces an emitter–photon Bell pair. Successfully fusing the two photons then leaves the two emitters in a Bell state up to corrections from the fusion outcomes. The two emitters can then be interacted locally with other matter-based qubits. Failed cases can be treated as erasures of fusion outcomes, leading to heralded Pauli errors on the emitter Bell states. This can be mapped directly to the ZX diagram in \cref{diag:bell_prep_fusion}, where local ancillas correspond to emitters and the central fusion is performed between the emitted photons. The process is illustrated in \cref{fig:emitter_bell_fusion}.

Similar emission-based processes can generate larger resource states. For simple resource state topologies, such as a linear chain of photons entangled with a source emitter, resource states can be generated deterministically by repeating an emission and emitter-rotation cycle~\cite{lindnerProposalPulsedOnDemand2009}, as illustrated in \cref{fig:emitter_linear_cluster}. This process has been experimentally demonstrated using quantum dots~\cite{schwartzDeterministicGenerationCluster2016}. Generating more complex resource state topologies requires either post-selection on probabilistic photon-photon interactions, or interactions between matter-based emitters. The latter is desirable, as state generation remains deterministic. For example, using pre-entangled emitters permits deterministic generation of 2D ladder resource states~\cite{economouOpticallyGenerated2Dimensional2010}, and allowing interactions between emitters after emission allows arbitrary graph states to be generated, though in both cases this is restricted by the connectivity between emitters~\cite{russoGenerationArbitraryAllphotonic2019}.

\begin{figure*}
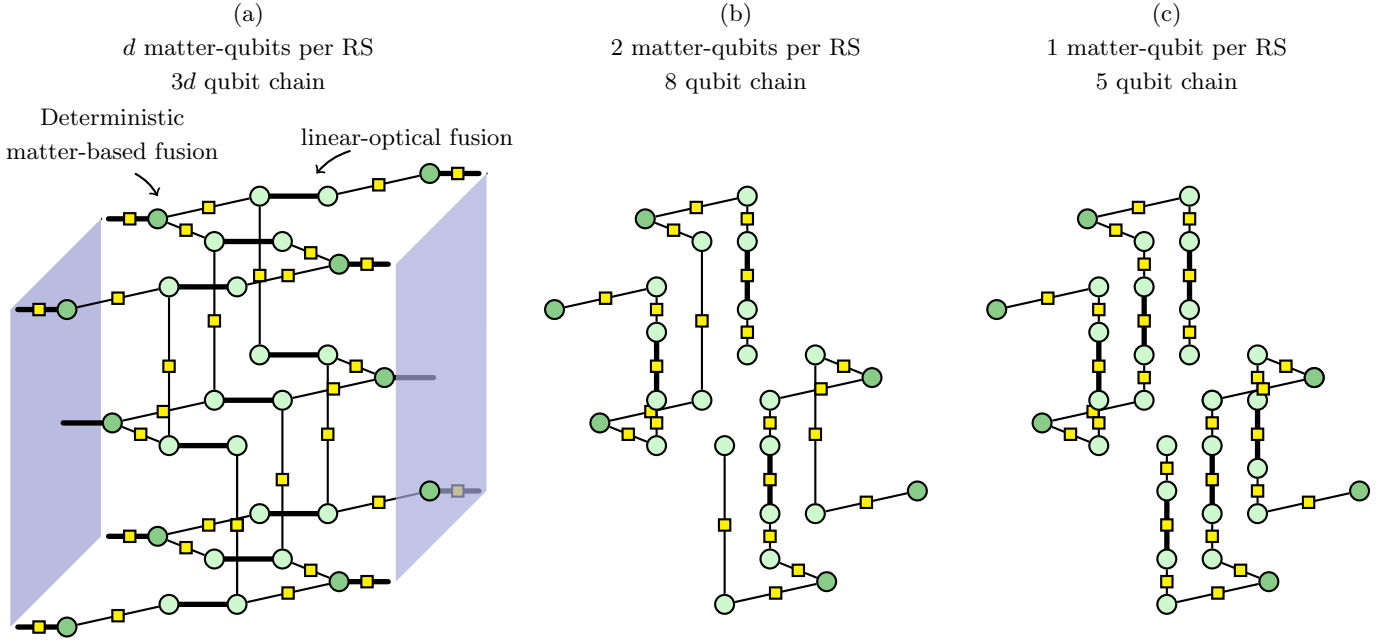

    \tikzfig{5_protocols/03_linear_chain_length}[scale=1.2]
    \caption{\textbf{Decomposing the merge region into linear chain resource states}.
   The fusion network is composed of layers of linear chains, each with a maximum of $d$ matter-based qubits. The darker shade of green indicates matter-based qubits, while the lighter shade of green indicates photonic qubits. In the bulk, each resource state connects to two other resource states via $d$ linear-optical fusions, with an additional $d$ matter-based fusions at the interface. Resource states can be truncated by subdividing edges with Hadamard fusions: (a) $d$ matter qubits per chain, (b) $2$ matter qubits per chain, and (c) $1$ matter qubit per chain.}
    \label{fig:linear_chain_decomp}
\end{figure*}

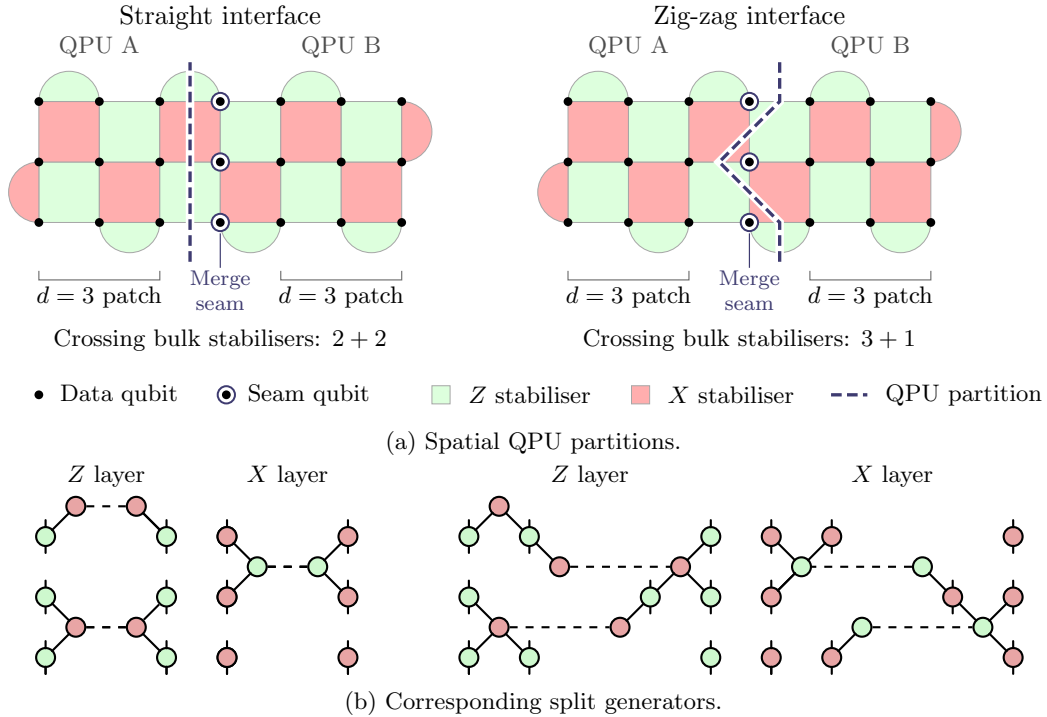
\begin{figure*}[tb]
    \centering
    \begin{subfigure}[c]{\textwidth}
        \centering
        \input{figures/5_protocols/05_zigzag_interface.tikz}
        \caption{Spatial QPU partitions.}
        \label{fig:zigzag_spatial_partition}
    \end{subfigure}
    \hfill
    \begin{subfigure}[c]{\textwidth}
        \centering
        \tikzfig{figures/5_protocols/06_split_check_zigzag}[scale=0.8]
        \caption{Corresponding split generators.}
        \label{fig:split_check_zigzag}
    \end{subfigure}
    \caption{\textbf{Straight and zig-zag stabiliser partitions.} (a) Two distance-three rotated surface-code patches are joined through a central column of three seam qubits, giving the $7\times3$ merged patch of \cref{sec:hybrid_logical_circs}. Brackets identify the original patches, and circled dots identify the seam qubits. The left and right diagrams show the straight and zig-zag QPU assignments of the same merged code; the zig-zag assignment places the middle seam qubit on QPU B and the other two on QPU A. Black dots are data qubits; green and red faces are $Z$- and $X$-type stabiliser supports, with semicircular faces denoting weight-two boundary stabilisers. The straight interface partitions each crossing bulk stabiliser as $2+2$, whereas the zig-zag interface gives $3+1$ or $1+3$. The dashed line is a hardware partition, not a code boundary. (b) The corresponding local split-generator geometries. Dashed edges are the non-local split edges realised using the distributed Bell-pair gadget.}
    \label{fig:zigzag_interface}
\end{figure*}

While the resource state decomposition we have described in \cref{sec:hybrid_logical_circs} uses only linear resource states, they may be challenging to generate in practice for two reasons. First, the number of qubits in the resource state scales with the width of the boundary, thus the distance of the code. For large code distances, these may become prohibitively large. Second, if we wish to retain matter-based qubits in the resource states for deterministic interactions at the boundaries, then this necessitates multiple matter-based qubits per resource state, which introduces the additional challenge of interacting matter-based emitters before and after emission.

In our case, since we are interested in distributed operations, we will assume that linear resource states including both matter-based and photonic qubits can be generated deterministically, provided the matter-based qubits are local to each other (i.e. nearby on the same processor). Since the placement of our matter-based qubits corresponds to ancilla placement in the surface code, this requires only additional nearest-neighbour interactions between emitters along the boundary. However, since coupling of emitters introduces new experimental challenges, we will also consider more restricted cases. A lower-level treatment of resource state preparation is left to future work.

The linear resource state decomposition of the merge region is shown in \cref{fig:linear_chain_decomp}. All qubits that are involved in boundary fusions are matter-based qubits and are coloured darker to distinguish them. The basic decomposition uses resource states with $3d$ qubits, of which $d$ are matter-based qubits that have been split from generator spiders corresponding to local ancillas. Generating such a resource state requires interacting $d$ matter-based qubits before emission. We can reduce the number of matter-based qubits per resource state by \textit{subdividing} edges in the resource state with Hadamard fusions using the following transformation:

\ctikzfig{figures/5_protocols/04_subdivision_resource_states}\label{diag:subdivision_rs}

Subdivision replaces a Hadamard edge with two outer edges connected by a central Hadamard fusion, leaving Pauli webs intact on the outer boundaries. This allows us to create a family of resource states which are truncated according to the number of matter qubits in the state, naming each decomposition by the resulting resource state size. The untruncated resource state, in which chains span the merge region and carry $d$ matter qubits each, is $3d$-\textsc{Chain}; truncating to $k$ matter qubits per chain gives the $(3k+2)$-\textsc{Chain} family, of which we consider $8$-\textsc{Chain} ($k=2$) and $5$-\textsc{Chain} ($k=1$). These are compared against two circuit-based protocols employing shared Bell pairs generated using the fusion-based construction in \cref{diag:bell_prep_fusion}. This allows us to treat the two protocols using the same erasure-based noise model. For comparison, we use two interface geometries: straight and zig-zag, as investigated in Refs.~\cite{shalbyOptimizedNoiseresilientSurface2025,jacintoNetworkRequirementsDistributed2025}. We refer to these as \textsc{Bell} and \textsc{BellZigZag}.

\begin{figure*}
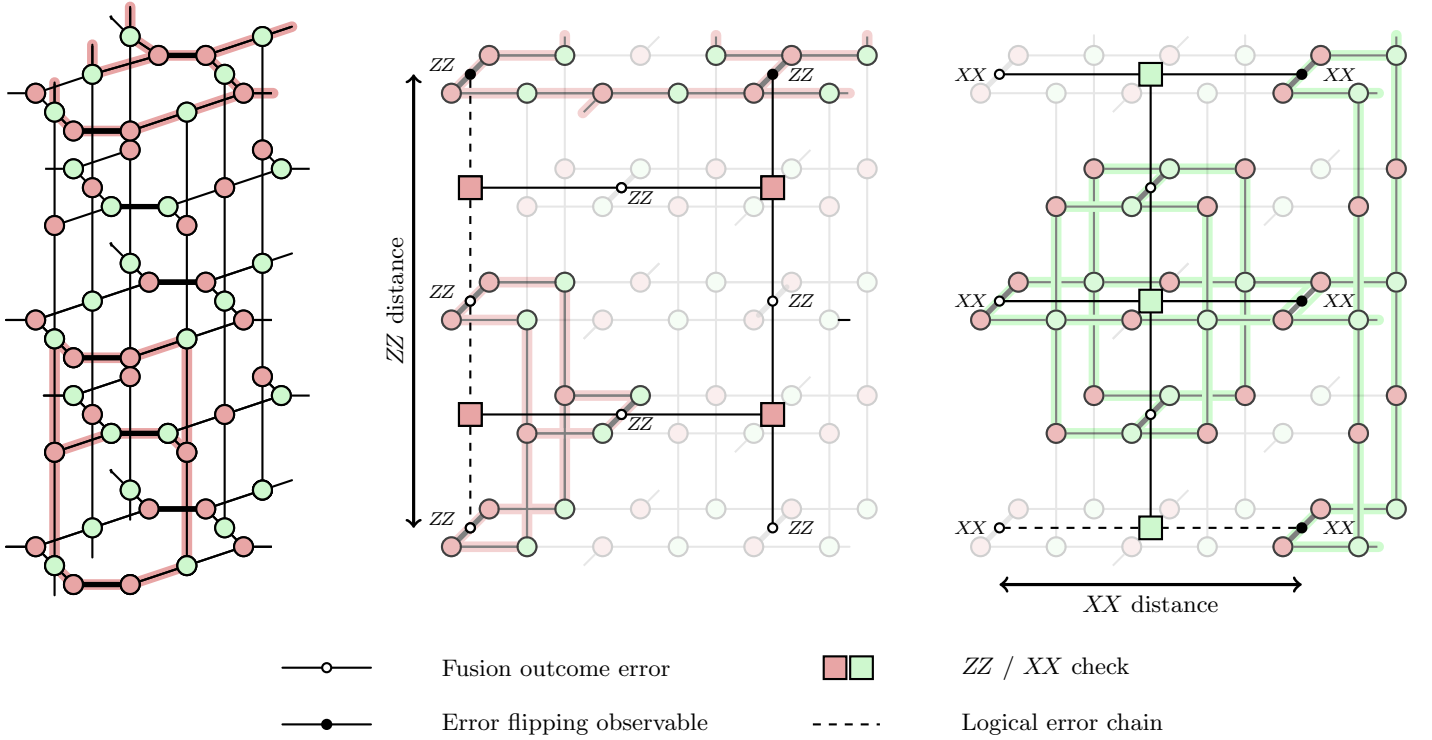

  \tikzfig{5_protocols/07_dem_bell_fusion_small}[scale=1.0]
  \caption{\textbf{Interface detector error model for \textsc{Bell}.} Restricting to the subgraph containing the fusions gives the diagram shown on the left. Each detecting region corresponds to a check containing $XX$/$ZZ$ measurement outcomes, which we represent with green/red square nodes. We use the bold fusion edge as shorthand for the fusion-based Bell preparation gadget from \cref{diag:bell_prep_fusion}. Each highlighted fusion edge corresponds to an error on one of the two measurement outcomes. Connecting the checks gives the primal and dual lattices of an unrotated surface code. The observable being measured in the merge is carried by the primal lattice in the centre, while the perpendicular joint observable is carried by the dual lattice on the right. Errors which flip the logical observable are shown as black dots, and the shortest undetectable error chain flipping the observable is shown with dashed edges. The interface DEM for \textsc{BellZigZag} is structurally identical, with double the $X\!X$ distance, though more challenging to visualise compactly.}
  \label{fig:dem_bell}
\end{figure*}

\subsection{Interface detector error models}\label{sec:interface_codes}

We construct the interface detector error model (DEM) by allowing erasures only on elementary linear-optical fusion outcomes, keeping all other operations ideal. We perform this analysis using a \textit{hardware-agnostic fusion erasure} error model, in which each elementary fusion outcome is erased independently with probability $p_e$~\cite{bartolucciFusionbasedQuantumComputation2023}. In \cref{sec:linear_optical_model}, we consider the effects of loss and fusion failure more directly. In the hardware-agnostic model, an erased outcome is flipped with $50\%$ probability, effectively randomising the outcome, while also providing a herald to the decoder. A flip of an outcome toggles each detecting parity $\mu(w)$ whose support $\mathcal C(w)$ contains it, together with any logical-record parity containing that outcome. An outcome shared by two detecting regions gives an edge between them and one appearing in a single detector gives a boundary edge.

For the chosen $Z\!Z$ surgery, we track the merge observable $\bar{Z}^{(1)}\bar{Z}^{(2)}$ and the surviving perpendicular correlation $\bar{X}^{(1)}\bar{X}^{(2)}$, whose record includes the seam readout. For either logical record $L$, we define the interface distance $d_{\mathrm{int}}(L)$ as the minimum number of fusion-outcome flips that implement an undetectable flip of $L$. Equivalently, it is the minimum number of erased fusion outcomes whose support can contain such a logical error. Thus $d_{\mathrm{int}}(L)$ is the usual distance restricted to the interface outcomes permitted by the erasure model; the erasure heralds change the decoding problem rather than this distance. An erasure can be decoded on a matchable detector error model by merging any detecting regions together which share the edge, such that they form a larger detecting region (``super-plaquette'')~\cite{staceThresholdsTopologicalCodes2009} and decoding error syndromes on the modified detector error model. The extended detecting region corresponds to the sum of the two affected Pauli webs. If the erased outcome is also part of the frame of the logical observable, the observable can be deformed around it by summing the observable web with a detecting region also containing the outcome. If a chain of erasures percolates to the boundaries of the lattice, the logical information cannot reliably be recovered. The probability of this happening can be mapped to the probability of \textit{bond percolation} on a lattice~\cite{kestenPercolationTheoryMathematicians1982,staceThresholdsTopologicalCodes2009}. Below a critical probability percolation is unlikely, and above it percolation occurs with high probability. This critical probability thus corresponds to the erasure threshold.

\subsubsection{Bell interfaces}

We first consider the \textsc{Bell} protocol and the \textsc{BellZigZag} variant. In \textsc{Bell}, each generator spider along the interface is split equally, such that split generators in the bulk retain two local edges each, and those on the boundary retain one each. The split edge is realised as a Bell pair using the fusion gadget of~\cref{diag:bell_prep_fusion}, which fuses two auxiliary Bell pairs. With only erasures, errors occur only on the outcomes of each linear-optical fusion. When constructing base surgery SEDs, we fixed the merge region to contain $d$ $X$-type and $Z$-type generator layers, with an additional $Z$-type layer at end of the merge. As a result, the merge region consists of $2d+1$ generator \textit{layers}. In the bulk, each fusion outcome appears in two checks and each check comprises four measurement results of the same type ($X\!X$ or $Z\!Z$). On the boundaries, each check contains three measurement outcomes. Bulk checks share fusion outcomes with those centred one round before and one round after, as well as to either side. The resulting DEM splits into a primal and dual square lattice, with the same structure as a single sheet of an \textit{unrotated} surface code on a $\lfloor(d+1)/2\rfloor \times \lceil n_{\mathrm{layers}}/2 \rceil$ lattice. The lattice advances one row per \textit{pair} of layers -- the observable being measured in the merge runs horizontally across the primal lattice and has interface distance $\lceil n_{\mathrm{layers}}/2 \rceil = d+1$. The perpendicular joint observable runs vertically across the dual lattice and has interface distance $\lfloor(d+1)/2\rfloor$. A similar fact has been noted for such a straight interface in previous work using depolarising noise models~\cite{shalbyOptimizedNoiseresilientSurface2025,jacintoNetworkRequirementsDistributed2025,haugLatticeSurgeryBell2025}. Since the DEM is an isotropic square lattice in the bulk, independent bond percolation predicts an asymptotic erasure threshold of $50\%$ \cite{kestenCriticalProbabilityBond1980,kestenPercolationTheoryMathematicians1982,staceThresholdsTopologicalCodes2009,barrettFaultTolerantQuantum2010a}. We illustrate the erasure DEM for the straight \textsc{Bell} interface in \cref{fig:dem_bell}.

\begin{figure*}
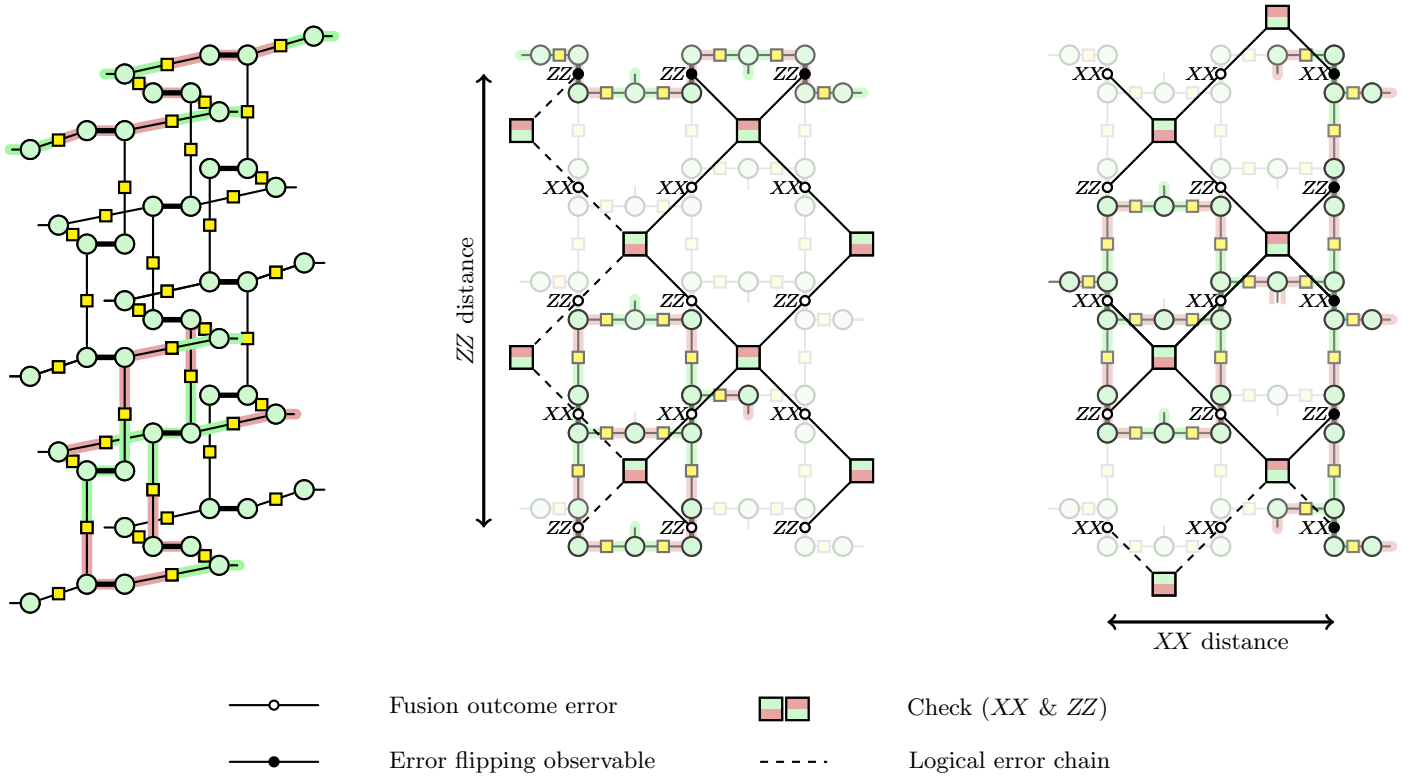

  \tikzfig{5_protocols/08_dem_hybrid_small}[scale=1.0]
  \caption{\textbf{Interface detector error model for the hybrid $3d$-chain protocol.} The hybrid interface is geometrically different, since linear-optical fusions occur at the seam, rather than along generator spiders. This means that fusions are supported on detecting regions to either side of the seam. The interface can be laid on a plane to visualise its DEM. Each detecting region corresponds to a check containing a combination of $XX$ and $ZZ$ measurement outcomes, which we represent with square nodes coloured by the measurement results they contain. Each highlighted fusion edge corresponds to an error on one of the two measurement outcomes. Connecting the checks gives the primal and dual lattices of a rotated surface code with an augmented distance in the primal lattice. The observable being measured in the merge is carried by the primal lattice in the centre, while the perpendicular joint observable is carried by the dual lattice on the right. Errors which flip the logical observable are shown as black dots, and the shortest undetectable error chain flipping the observable is shown with dashed edges.}
  \label{fig:dem_hybrid}

\end{figure*}

The \textsc{BellZigZag} protocol uses the same Bell-pair gadget with a zig-zag QPU partition, splitting the crossing bulk stabilisers $3+1$ across two columns of the interface. Notably, since the split positions are identical, the connectivity of the interface DEM is unchanged. However, since stabilisers are split over two columns of the interface region, fusions appear in an additional column of detecting regions, doubling the width of the lattice. This restores the interface distance on the perpendicular observable from $\lfloor(d+1)/2\rfloor$ to $d$, while leaving the merge-observable distance at $d+1$. This distance doubling has also been observed using circuit-level noise models~\cite{shalbyOptimizedNoiseresilientSurface2025,jacintoNetworkRequirementsDistributed2025}.

\subsubsection{Hybrid interfaces}

Consider now the $3d$-\textsc{Chain} decomposition. Fusions occur along the seam, rather than between generator spiders. This changes the interface geometry, since fusion outcomes are now shared between checks on either side of the seam. Each elementary fusion outcome is contained in two checks in the bulk and each check contains four measurement results. On the boundaries, each check contains two measurement results. Although the degree of checks in the bulk is the same as the previous case, checks share edges with those in each diagonal direction, rather than vertically and horizontally adjacent. As a result, the detector error model again has a primal and dual lattice corresponding to a single sheet of a \textit{rotated} surface code on a $d \times n_{\mathrm{layers}}$ lattice, with checks drawing from both $X\!X$ and $Z\!Z$ outcomes. Relative to the straight \textsc{Bell} interface, the rotated interface augments the interface distance of the merge observable from $d+1$ to $n_{\mathrm{layers}} = 2d+1$ and restores the perpendicular-observable distance from $\lfloor(d+1)/2\rfloor$ to $d$. Relative to \textsc{BellZigZag}, it retains the same perpendicular distance $d$ and augments the merge observable distance. Despite the rotation, the DEM remains an isotropic square lattice in the bulk, so the erasure threshold matches both Bell-pair interfaces at $50\%$ under independent bond erasure~\cite{kestenCriticalProbabilityBond1980,kestenPercolationTheoryMathematicians1982,staceThresholdsTopologicalCodes2009}.

\begin{table*}[tb]
    \centering
    \caption{\textbf{Interface detector error models.} Geometry, interface distances, and asymptotic erasure thresholds for the five protocols considered. The distances refer to the merge record $\bar{Z}^{(1)}\bar{Z}^{(2)}$ and the surviving perpendicular correlation $\bar{X}^{(1)}\bar{X}^{(2)}$. Thresholds assume independent erasure of elementary fusion outcomes and ideal operations away from the interface. These distances apply only to faults on the interface outcomes and do not establish the distance under general circuit-level noise.
    }
    \label{tab:interface_dem_summary}
    \begin{tabular}{lcccc}
        \toprule
        Protocol & Interface DEM & $d_{\mathrm{int}}(\bar{Z}^{(1)}\bar{Z}^{(2)})$ & $d_{\mathrm{int}}(\bar{X}^{(1)}\bar{X}^{(2)})$ & $p_e^*$ \\
        \midrule
        \textsc{Bell}       & unrotated square & $d+1$  & $\lfloor(d+1)/2\rfloor$ & $0.50$ \\
        \textsc{BellZigZag} & unrotated square & $d+1$  & $d$                     & $0.50$ \\
        $3d$-\textsc{Chain} & rotated square   & $2d+1$ & $d$                     & $0.50$ \\
        $8$-\textsc{Chain}  & rotated square   & $2d+1$ & $d$                     & $\approx0.382$ \\
        $5$-\textsc{Chain}  & rotated square   & $2d+1$ & $d$                     & $\approx0.318$ \\
        \bottomrule
    \end{tabular}
\end{table*}

\subsubsection{Truncated resource states}

\begin{figure}[t]
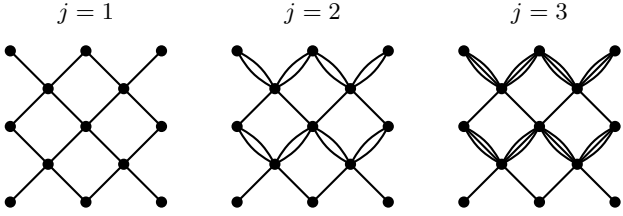

    \centering
    \tikzfig{5_protocols/09_truncated_chain_percolation}[scale=1.0]
    \caption{\textbf{Staggered bulk bond patterns of the interface DEMs.}
    Rotated DEM embeddings for $j=1$ ($3d$-\textsc{Chain}), $j=2$ ($8$-\textsc{Chain}) and
    $j=3$ ($5$-\textsc{Chain}). For $j>1$, the bonds with $j$ elementary erasure contributions
    join into horizontal zig-zag paths across an axis-aligned finite boundary. Together with the
    complementary $j=1$ bonds, they form the self-dual staggered pattern used in the exact
    threshold calculation.}
    \label{fig:truncated_chain_percolation}
\end{figure}

The additional fusions used to truncate the resource states do not change the square-lattice structure of the detector error model. To see this, associate to each elementary outcome $m$ its full label $\lambda(m)=(\mathcal D(m),\mathcal L(m))$, where $\mathcal D(m)$ is the set of detecting-region parities containing $m$ and $\mathcal L(m)$ is the set of logical-record parities containing it. The structure of the detecting regions makes the interface DEM graph-like, so $|\mathcal D(m)|\leq 2$ for every elementary outcome. Each subdivision is made on an unbranched part of a linear resource chain. The web rules at the two inserted degree-two spiders force every pure-$X$ or pure-$Z$ web to continue uniquely across the subdivision. Consequently, each added outcome $m_r$ has the same full label as the corresponding contribution $m$ to the existing effective bond, $\lambda(m_r)=\lambda(m)$. The new outcomes therefore add parallel contributions to an existing DEM edge rather than new connectivity, as shown in \cref{fig:truncated_chain_percolation}. Because this is an equality of the full labels, this also includes boundary edges with $|\mathcal D(m)|=1$ and detector-free logical mechanisms with $|\mathcal D(m)|=0$. Subdivision thus leaves both interface distances unchanged, although the increased edge multiplicities make the bulk erasure model \textit{anisotropic} and lower its threshold.

If the probability of erasure of a single fusion measurement result is $p_e$, then the probability of an edge with $j$ independent elementary-outcome erasure contributions being erased is $P_j=1-(1-p_e)^j$. For $8$-\textsc{Chain}, the doubled edges form horizontal zig-zag paths across the rotated DEM; for $5$-\textsc{Chain}, the same pattern is formed with $j=3$. This can be mapped onto a checkerboard lattice with alternating bond probabilities $P_j$ and $P_1$, for which the exact critical manifold reduces to $P_j+P_1=1$~\cite{wuCriticalPointPlanar1979}. Substitution gives
\begin{equation}
    p_e^\ast=(1-p_e^\ast)^j.
\end{equation}
This yields $p_e^\ast=(3-\sqrt{5})/2\approx0.382$ for $8$-\textsc{Chain} and $p_e^\ast\approx0.318$ for $5$-\textsc{Chain}.

The distance on the observable measured in the merge grows with the number of layers and governs the fidelity of the merge outcome itself -- augmenting it to $2d+1$ buys accuracy in the merge when erasure is the dominant noise source without adding any additional rounds. The distance on the perpendicular joint observable is set by the length of the seam and governs the protection of the existing correlation between the two logical qubits during the merge; restoring it to $d$ removes the distance loss along a straight interface. \textsc{BellZigZag} also restores this perpendicular distance~\cite{shalbyOptimizedNoiseresilientSurface2025,jacintoNetworkRequirementsDistributed2025}, but leaves the merge-observable distance at $d+1$. The augmented distances do not immediately translate to lower logical error rates for hybrid protocols. Notably, unrotated surface codes typically exhibit lower logical error rates than rotated surface codes at the same code distance due to fewer low-weight logical error mechanisms. Additionally, these distances include only faults on elementary linear-optical fusion outcomes. In the complete circuit, faults in the deterministic boundary fusions coupling the circuit and resource-state regions introduce local distance-reducing mechanisms on the perpendicular observable -- we discuss potential remedies for this in \cref{sec:discussion}. In \cref{sec:numerical_simulations}, we will probe the subthreshold regimes in which the augmented interface distance buys better logical error suppression and those where local noise effects and path-dependent mechanisms dominate.

%% file: figures/5_protocols/01_emitter_bell_fusion.tikz
\begin{tikzpicture}[x=1cm,y=1cm,font=\footnotesize]
    \definecolor{emitterfill}{RGB}{136,204,136}
    \definecolor{photonfill}{RGB}{206,248,206}
    \definecolor{analyserfill}{RGB}{235,234,247}
    \definecolor{linkcolour}{RGB}{62,59,112}
    \tikzset{emitternode/.style={circle,draw=black,fill=emitterfill,minimum size=6mm,inner sep=0pt},
        photonnode/.style={circle,draw=black,fill=photonfill,minimum size=3.6mm,inner sep=0pt},
        belllink/.style={draw=linkcolour,densely dashed,line width=0.7pt},
        photonpath/.style={->,>=stealth,draw=black!70,line width=0.65pt}}
    \node[emitternode] (ea) at (0,0) {$e_A$};
    \node[emitternode] (eb) at (7,0) {$e_B$};
    \node[photonnode] (pa) at (1.65,0) {};
    \node[photonnode] (pb) at (5.35,0) {};
    \node[draw=linkcolour,fill=analyserfill,rounded corners=2pt,align=center,
        minimum width=1.7cm,minimum height=0.85cm,inner sep=3pt] (bsa) at (3.5,0) {Bell-state\\analyser};
    \draw[snakeArrowStyle] (ea) -- (pa);
    \draw[snakeArrowStyle] (eb) -- (pb);
    \draw[photonpath] (pa) -- (bsa.west);
    \draw[photonpath] (pb) -- (bsa.east);
    \node[below=3pt] at (pa.south) {$p_A$};
    \node[below=3pt] at (pb.south) {$p_B$};
    \node[below=5pt] at (ea.south) {Emitter A};
    \node[below=5pt] at (eb.south) {Emitter B};
    \draw[belllink] (ea.north) to[out=70,in=110] node[above=3pt,pos=0.5] {$|\Phi^+\rangle_{e_Ap_A}$} (pa.north);
    \draw[belllink] (pb.north) to[out=70,in=110] node[above=3pt,pos=0.5] {$|\Phi^+\rangle_{e_Bp_B}$} (eb.north);
    \node[below=5pt] at (bsa.south) {$m_{XX},\ m_{ZZ}$};
\end{tikzpicture}

%% file: figures/5_protocols/02_emitter_linear_cluster.tikz
\begin{tikzpicture}[x=1cm,y=1cm,font=\footnotesize]
    \definecolor{emitterfill}{RGB}{136,204,136}
    \definecolor{photonfill}{RGB}{206,248,206}
    \definecolor{cyclefill}{RGB}{235,234,247}
    \definecolor{linkcolour}{RGB}{62,59,112}
    \tikzset{
        emitternode/.style={circle,draw=black,fill=emitterfill,minimum size=6mm,inner sep=0pt},
        photonnode/.style={circle,draw=black,fill=photonfill,minimum size=3.6mm,inner sep=0pt},
        clusterlink/.style={draw=linkcolour,densely dashed,line width=0.7pt},
        processarrow/.style={->,>=stealth,draw=black!70,line width=0.65pt}
    }

    \node[emitternode] (ein) at (0,0) {$e$};
    \node[below=5pt] at (ein.south) {$\ket{+}_e$};

    \node[draw=linkcolour,fill=cyclefill,rounded corners=2pt,align=center,
        minimum width=1.7cm,minimum height=0.85cm,inner sep=3pt] (cycle) at (1.75,0)
        {emit $p_i$\\apply $H_e$};
    \node[above=3pt] at (cycle.north) {repeat $N$ times};
    \draw[processarrow] (ein) -- (cycle.west);

    \node[emitternode] (eout) at (3.45,0) {$e$};
    \node[photonnode] (pn) at (4.35,0) {};
    \node (dots) at (5.05,0) {$\cdots$};
    \node[photonnode] (p2) at (5.75,0) {};
    \node[photonnode] (p1) at (6.65,0) {};
    \draw[processarrow] (cycle.east) -- (eout);
    \draw[clusterlink] (eout) -- (pn);
    \draw[clusterlink] (pn) -- (dots);
    \draw[clusterlink] (dots) -- (p2);
    \draw[clusterlink] (p2) -- (p1);

    \node[below=5pt] at (eout.south) {retained emitter};
    \node[below=3pt] at (pn.south) {$p_N$};
    \node[below=3pt] at (p2.south) {$p_2$};
    \node[below=3pt] at (p1.south) {$p_1$};
    \node[above=7pt] at (5.05,0) {linear cluster};
\end{tikzpicture}

%% file: figures/5_protocols/05_zigzag_interface.tikz
\begin{tikzpicture}[x=0.8cm,y=0.8cm,font=\small]
    \definecolor{partitionline}{RGB}{62,59,112}
    \definecolor{patchZ}{RGB}{221,255,221}
    \definecolor{patchX}{RGB}{255,174,174}
    \tikzset{patchface/.style={draw=black!35,line width=0.35pt},
        partition/.style={draw=partitionline,line width=1.25pt,dash pattern=on 4pt off 2pt,
            preaction={draw=white,line width=3.2pt,solid}},
        datadot/.style={circle,fill=black,inner sep=0pt,minimum size=3.3pt},
        seamdot/.style={circle,draw=partitionline,fill=white,line width=0.75pt,inner sep=0pt,minimum size=6.5pt}}
    \foreach \panel/\shift/\heading/\detail in {
        0/0/{Straight interface}/{Crossing bulk stabilisers: $2+2$},
        1/7.0/{Zig-zag interface}/{Crossing bulk stabilisers: $3+1$}}
    {
        \begin{scope}[xshift=\shift cm]
            \node[font=\normalsize] at (3,3.4) {\heading};
            \node[text=black!70] at (1,2.9) {QPU A};
            \node[text=black!70] at (5,2.9) {QPU B};
            \foreach \i in {0,...,5} {
                \foreach \j in {0,1} {
                    \pgfmathtruncatemacro{\parity}{mod(\i+\j,2)}
                    \ifnum\parity=0
                        \path[patchface,fill=patchZ] (\i,\j) rectangle ++(1,1);
                    \else
                        \path[patchface,fill=patchX] (\i,\j) rectangle ++(1,1);
                    \fi
                }
            }
            \foreach \i in {0,2,4} {
                \path[patchface,fill=patchZ] (\i,2) arc[start angle=180,end angle=0,radius=0.4cm] -- cycle;
            }
            \foreach \i in {1,3,5} {
                \path[patchface,fill=patchZ] (\i,0) arc[start angle=180,end angle=360,radius=0.4cm] -- cycle;
            }
            \path[patchface,fill=patchX] (0,0) arc[start angle=270,end angle=90,radius=0.4cm] -- cycle;
            \path[patchface,fill=patchX] (6,1) arc[start angle=270,end angle=450,radius=0.4cm] -- cycle;
            \ifnum\panel=0
                \draw[partition] (2.5,-0.65) -- (2.5,2.65);
            \else
                \draw[partition] (3.5,-0.65) -- (3.5,0) -- (2.5,1) -- (3.5,2) -- (3.5,2.65);
            \fi
            \foreach \i in {0,...,6} {
                \foreach \j in {0,...,2} {
                    \ifnum\i=3 \node[seamdot] at (\i,\j) {}; \fi
                    \node[datadot] at (\i,\j) {};
                }
            }
            \draw[black!55,line width=0.45pt] (0,-0.8) -- (0,-0.95) -- (2,-0.95) -- (2,-0.8);
            \draw[black!55,line width=0.45pt] (4,-0.8) -- (4,-0.95) -- (6,-0.95) -- (6,-0.8);
            \node at (1,-1.24) {$d=3$ patch};
            \node at (5,-1.24) {$d=3$ patch};
            \node[text=partitionline,align=center,font=\footnotesize] at (3,-1.12) {Merge\\seam};
            \draw[partitionline,line width=0.45pt] (3,-0.75) -- (3,-0.18);
            \node at (3,-1.94) {\detail};
        \end{scope}
    }
    \node[datadot] at (0,-2.85) {};
    \node[anchor=west] at (0.2,-2.85) {Data qubit};
    \node[seamdot] at (3.1,-2.85) {};
    \node[datadot] at (3.1,-2.85) {};
    \node[anchor=west] at (3.3,-2.85) {Seam qubit};
    \path[patchface,fill=patchZ] (6.5,-3.00) rectangle (6.8,-2.70);
    \node[anchor=west] at (6.95,-2.85) {$Z$ stabiliser};
    \path[patchface,fill=patchX] (9.8,-3.00) rectangle (10.1,-2.70);
    \node[anchor=west] at (10.25,-2.85) {$X$ stabiliser};
    \draw[partition] (13.1,-2.85) -- (13.7,-2.85);
    \node[anchor=west] at (13.85,-2.85) {QPU partition};
\end{tikzpicture}

%% file: sections/6_simulations.tex
\section{Numerical simulations}\label{sec:numerical_simulations}

The previous section demonstrated that a family of hybrid surgery protocols could gain an advantage in the interface distance. While this is interesting, it is idealised. A true heterogeneous architecture has a number of different noise sources which interact to create more complex detector error models. We therefore perform four numerical studies. First, we simulate the protocols under pure fusion-outcome erasure to verify the threshold predictions of the previous section. Second, we add computational errors to resource states and determine the correctable region in $(p_r,p_e)$. Third, we combine local circuit noise with fusion-outcome erasure and determine the correctable region in $(p_c,p_e)$. Finally, we use the loss-based linear-optical model of Ref.~\cite{bartolucciFusionbasedQuantumComputation2023} to compare the subthreshold suppression factor over a range of effective erasure rates and boosting levels, at three strengths of circuit and resource state noise. The results are presented in \cref{sec:results}.

\subsection{Simulation setup}
\label{sec:sim_setup}

We evaluate the performance of the proposed hybrid lattice surgery protocol via Monte Carlo simulations. We build circuits and create the corresponding \textit{detector error models} in \textsc{Stim}~\cite{gidneyStimFastStabilizer2021} and the resulting detection events are decoded with the minimum-weight perfect matching (MWPM) decoder~\cite{edmondsPathsTreesFlowers1965}, as implemented in PyMatching 2~\cite{higgottSparseBlossomCorrecting2025}. Since depolarising noise can introduce correlated two-qubit errors, we configure PyMatching to decode correlated errors. Additionally, following the strategy of Stace et al.~\cite{staceThresholdsTopologicalCodes2009}, we account for heralded erasures by collecting erasure syndromes after each shot and dynamically re-weighting edge probabilities in the matching graph accordingly. We simulate a lattice surgery logical $Z\!Z$ merge with both qubits initially prepared in the logical $\ket{+_L}$ state, thus producing a logical Bell state. This is done using $d$ rounds of error correction before and after the merge -- during the merge $d$ rounds are performed followed by an additional half round in the merge basis. We calculate logical error rates and thresholds on both the $\bar{Z} \otimes \bar{Z}$ (merge) and $\bar{X} \otimes \bar{X}$ (perpendicular) observables separately.

\subsection{Resource state error model}\label{sec:prep_error_model}

To avoid making extensive hardware assumptions, we do not explicitly model the resource state generation process. Each resource state is prepared by initialising all qubits in $\ket{+}$ and applying $CZ$ gates along the chain. We model resource state errors using independent fusion-outcome flip probability $p_r$ in addition to the erasure probability. Unlike erasures, flips are not heralded. Measurement flips account partially for errors in preparation, since any error occurring during the preparation of the resource state propagates only to fusion measurements, as well as errors in photon detection. This does not capture preparation errors on emitters, however, so we introduce a reset flip error on emitters when initialised. This means that, with probability $p_r$, each emitter is initialised to $\ket{-}$ instead of $\ket{+}$. In our simplified preparation circuits, resource errors do not propagate to other qubits in the resource state since phase flips commute with $CZ$ gates. Outside of preparation errors, emitters are treated under the circuit-level error model described in \cref{sec:circuit_error_model}. This is a simplified noise model and does not account for specific errors arising in the resource state generation process or idling of emitters after preparation. A lower-level treatment of resource state preparation is left to future work.

\subsection{Circuit-level error model}\label{sec:circuit_error_model}

For matter-based qubits, local circuit noise acts on two-qubit gates, resets and measurements. Following the \textsc{Stim} convention~\cite{gidneyStimFastStabilizer2021}, \textsc{DEP2} denotes \texttt{DEPOLARIZE2}, whose action on a two-qubit density operator is

\begin{equation}
    \operatorname{DEP2}(p)(\rho)
        = (1-p)\rho + \frac{p}{15}
        \sum_{P\in\{I,X,Y,Z\}^{\otimes 2}\setminus\{II\}} P\rho P. \label{eq:dep2}
\end{equation}
Thus $p$ is the total probability of applying a non-identity two-qubit Pauli, with each of the 15 non-identity Pauli products occurring with probability $p/15$. The maximally mixed channel occurs at $p=15/16$, rather than at $p=1$. We apply \textsc{DEP2}$(p_c)$ after every two-qubit gate. Reset and measurement errors are basis-specific flips: with probability $p_c$, a reset prepares the opposite eigenstate of the specified Pauli basis and a measurement reports the opposite outcome. Basis changes for preparation and readout are therefore incorporated into the specified-basis resets and measurements, and no separate single-qubit depolarising channel is applied. These channels also apply to operations on matter-based emitters, except for emitter resets, which are accounted for in the resource-state preparation error model. Idle locations are noiseless. We use the single parameter $p_c$ for all of these local circuit-noise channels.

The surface-code stabilisers are extracted in alternating $Z$- and $X$-type layers using four $CX$ substeps. The numbered corners below give the temporal order of the $CX$ gates around each ancilla:
\begin{equation}
\begin{aligned}
Z\text{-type}:&\quad
\begin{array}{c@{\;}c@{\;}c}
2&\longleftarrow&1\\[-0.35em]
&\searrow&\\[-0.35em]
4&\longleftarrow&3
\end{array}
&& \text{data}\rightarrow\text{ancilla},\\[0.4em]
X\text{-type}:&\quad
\begin{array}{c@{\;}c@{\;}c}
3&&1\\[-0.0em]
\downarrow&\nwarrow&\downarrow\\[-0.0em]
4&&2
\end{array}
&& \text{ancilla}\rightarrow\text{data}.
\end{aligned}
\label{eq:stabiliser_gate_schedule}
\end{equation}

Split interface stabilisers require an additional choice. For \textsc{BellZigZag}, the three-qubit half interacts first with the data qubit that is isolated on its line and then with the collinear pair perpendicular to the corresponding logical direction; the one-qubit half has a single interaction. This is the hook-avoiding $3+1$ schedule of Refs.~\cite{shalbyOptimizedNoiseresilientSurface2025,jacintoNetworkRequirementsDistributed2025}. The straight \textsc{Bell} interface instead has a $2+2$ split for which reordering alone cannot remove the perpendicular distance reduction. Deterministic boundary fusions are performed after the four local gate substeps: a regular Bell ($X\!X$--$Z\!Z$) fusion is implemented by a $CX$ followed by $X$- and $Z$-basis measurements, while a Hadamard ($X\!Z$--$Z\!X$) fusion is implemented by a controlled-$Z$ followed by two $X$-basis measurements. Using this noise model, as mentioned in the previous section, we find local distance halving effects on the perpendicular ($\bar{X}\!\bar{X}$) observable arising from errors in boundary Hadamard fusions. Though these effects are small when interface noise dominates, they become more relevant as circuit noise increases. In \cref{sec:discussion}, we discuss potential remedies for mitigating these distance-reducing effects, though we do not implement these in simulation.

\subsection{Linear-optical fusion error model}
\label{sec:linear_optical_model}

In order to analyse the effects of loss more directly, we follow Bartolucci et al.~\cite{bartolucciFusionbasedQuantumComputation2023} in considering a linear-optical error model which explicitly accounts for the effects of loss and fusion failure. This model was introduced to investigate the fusion-boosting protocol proposed by Grice~\cite{griceArbitrarilyCompleteBellstate2011a}, which allows the success probability of a fusion to be increased using ancillary photons. Boosting reduces the intrinsic fusion-failure contribution to erasure, but increases exposure to photon loss because more photons must survive. The preferred level of boosting therefore depends on the erasure threshold available after other noise sources are included. We use the effective elementary-outcome marginal erasure probability derived in Ref.~\cite{bartolucciFusionbasedQuantumComputation2023}:

\begin{equation}\label{eq:boosted_loss}
    p_e = 1 - (1-p_{\mathrm{fail}}/2)(1-p_l)^{1/p_{\mathrm{fail}}},
\end{equation}

where $p_{\mathrm{fail}} = 1/2^{n}$ is the failure probability of a fusion boosted with $2^n - 2$ ancillary photons. This allows us to convert an erasure threshold into a \textit{photon-loss threshold} for a specified boosting level. Inverting \cref{eq:boosted_loss} gives

\begin{equation}\label{eq:loss_threshold_conversion}
    p_l^* = 1-\left(\frac{1-p_e^*}{1-2^{-(n+1)}}\right)^{2^{-n}}.
\end{equation}

When $n=1$, no ancillary photons are used and the effective elementary-outcome erasure probability is $25\%$ even at $p_l=0$. One fusion outcome can be reconstructed after a failed Bell measurement by a separable measurement, so only one of the two outcomes is erased. Consequently, boosting can become necessary once local noise lowers $p_e^*$ below $25\%$, even though all five protocols exceed this value in the pure-erasure limit.

We summarise the different contributions to the error model in \cref{tab:noise-models}. In \cref{sec:results}, we use \cref{eq:loss_threshold_conversion} to translate the circuit-and-erasure boundary into photon-loss thresholds and use the full loss-based model for the subthreshold simulations.

\begin{table}[htpb!]
  \centering
  \caption{Noise channels applied under the hybrid fusion error model. The parameters $p_c$, $p_r$ and $p_e$ are varied in the simulations to investigate the thresholds under different noise strengths. Under the linear-optical fusion error model, the erasure probability $p_e$ is related to the physical photon-loss rate $p_l$ according to \cref{eq:boosted_loss}. Emitters are treated as matter qubits except during resource state preparation where they flip with $p_r$ after reset.}
  \label{tab:noise-models}
  \begin{tabular}{l *{4}{c}}
    \toprule
        \textbf{Qubit type}
      & \textbf{Operation}
      & \textbf{Channel}
      & \textbf{Parameter} \\
    \midrule
    Matter  & Reset                             & FLIP  &  $p_c$ \\
    Matter  & Two-qubit gate                    & DEP2  &  $p_c$ \\
    Matter  & Measurement                       & FLIP  &  $p_c$ \\
    Emitter & Reset                             & FLIP  &  $p_r$ \\
    Photon  & Fusion measurement                & FLIP  &  $p_r$ \\
    Photon  & Fusion measurement                & ERASE &  $p_e$ \\
    \bottomrule
  \end{tabular}
\end{table}

%% file: sections/7_results.tex
\section{Results}\label{sec:results}

\begin{figure*}[ht]
    \input{results/master_threshold_pm_corr_mergeZZ_obsZZ-XX.tex}
    \caption{\textbf{Thresholds under pure erasure noise.} Logical error rates are shown while sweeping the elementary fusion-outcome erasure probability $p_e$, with $p_c=p_r=0$, for the straight \textsc{Bell} interface, \textsc{BellZigZag}, and the three hybrid linear-chain decompositions. Solid and dashed curves give the merge observable $\bar{Z}\!\bar{Z}$ and perpendicular correlation $\bar{X}\!\bar{X}$, respectively. Colours denote code distances $d=11,13,15,17,19$; vertical error bars show statistical uncertainty, and dotted vertical lines show the scaling-fit estimates for the two observables.}
    \label{fig:erasure_thresholds}
\end{figure*}
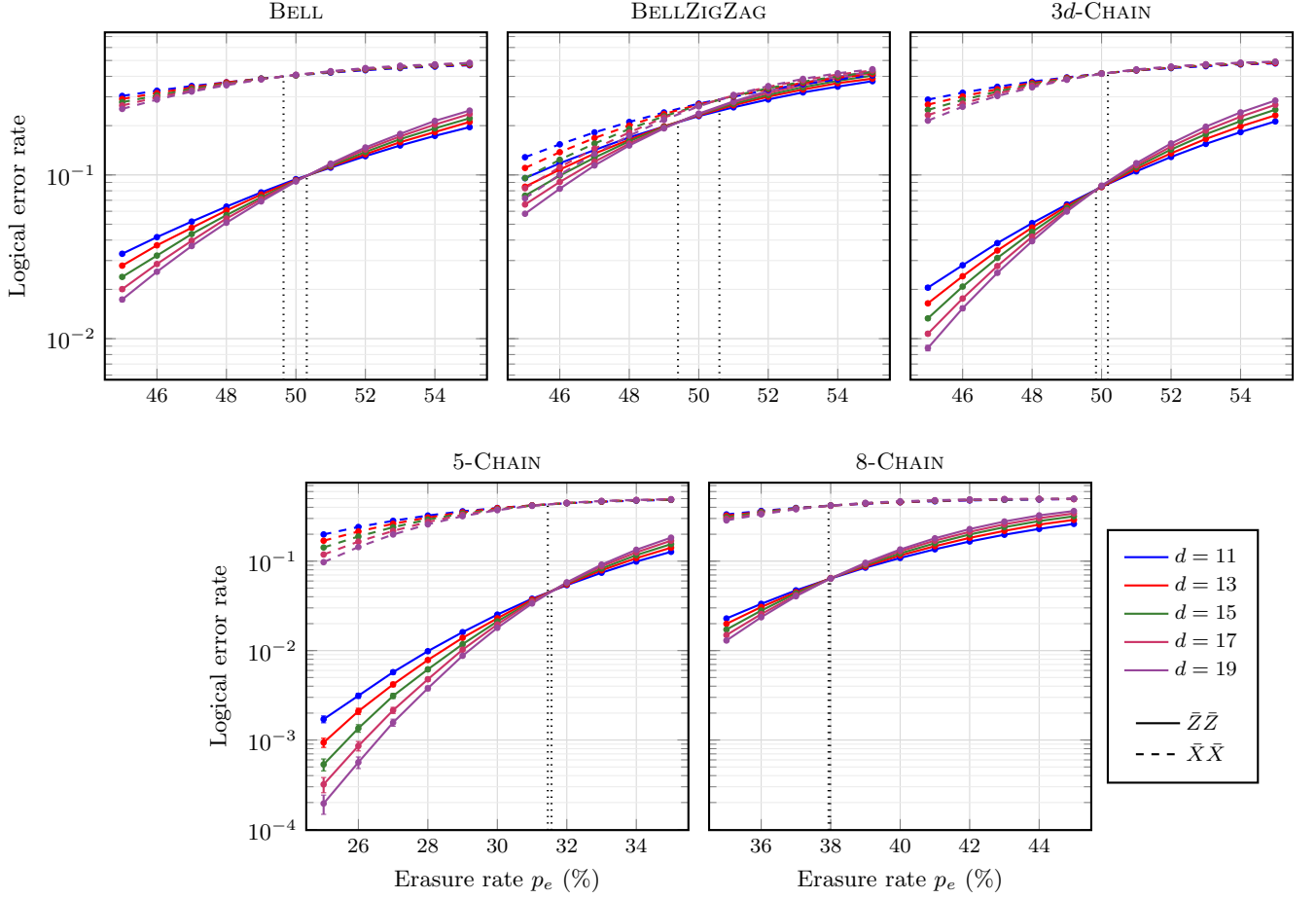

When estimating thresholds, we assume the finite-size scaling form near the predicted threshold~\cite{wangConfinementHiggsTransition2003}:
\begin{align}
    x &= (p_e-p_e^*)d^{1/\nu}, \label{eq:crg_scaling_variable}\\
    P_L(p_e,d) &= A+Bx+Cx^2. \label{eq:crg_scaling_fit}
\end{align}
The parameters $p_e^*$, $\nu$, $A$, $B$, and $C$ are fitted jointly across all distances by weighted least squares. A Monte Carlo point with $N_{\mathrm{err}}$ logical failures in $N_{\mathrm{shots}}$ shots is assigned the binomial standard error:
\begin{equation}
    \sigma_{P_L}=\frac{\sqrt{N_{\mathrm{err}}(1-P_L)}}{N_{\mathrm{shots}}},
    \label{eq:crg_binomial_error}
\end{equation}
with $N_{\mathrm{err}}$ floored at one when constructing the weight.

\subsection{Erasure}

The exact bulk-percolation predictions from \cref{sec:interface_codes} are $p_e^* \approx 0.318$ for $5$-\textsc{Chain}, $p_e^* \approx 0.382$ for $8$-\textsc{Chain}, and $p_e^*=0.50$ for $3d$-\textsc{Chain}. Both Bell-pair geometries also have the square-lattice bulk threshold $p_e^*=0.50$. The plots in \cref{fig:erasure_thresholds} show logical error rates for each protocol at distances $d=11,13,15,17,19$. Restricting to data points near the crossing, the finite-size scaling fits at zero local noise give \ResultThresholdWithError{\CircuitZZData}{5-Chain}{0.0}{4} and \ResultThresholdWithError{\CircuitXXData}{5-Chain}{0.0}{4} for the two $5$-\textsc{Chain} observables, respectively, and \ResultThresholdWithError{\CircuitZZData}{8-Chain}{0.0}{4} and \ResultThresholdWithError{\CircuitXXData}{8-Chain}{0.0}{4} for $8$-\textsc{Chain}. For the three protocols with a predicted $50\%$ threshold, the fitted estimates lie between \ResultThresholdWithError{\CircuitZZData}{BellZigZag}{0.0}{4} and \ResultThresholdWithError{\CircuitXXData}{BellZigZag}{0.0}{4}, depending on the protocol and observable. Parentheses give the quoted uncertainty in the final displayed digits. The agreement of the two observables is consistent with the common bulk threshold even though their finite-size interface distances differ. Notably, the asymmetry in the interface distances also leads to large differences in the logical error rates for the two observables, despite the thresholds being unchanged. \textsc{BellZigZag} closes this gap by restoring the perpendicular interface distance, though this increases the logical error rate on the merge observable relative to \textsc{Bell}.

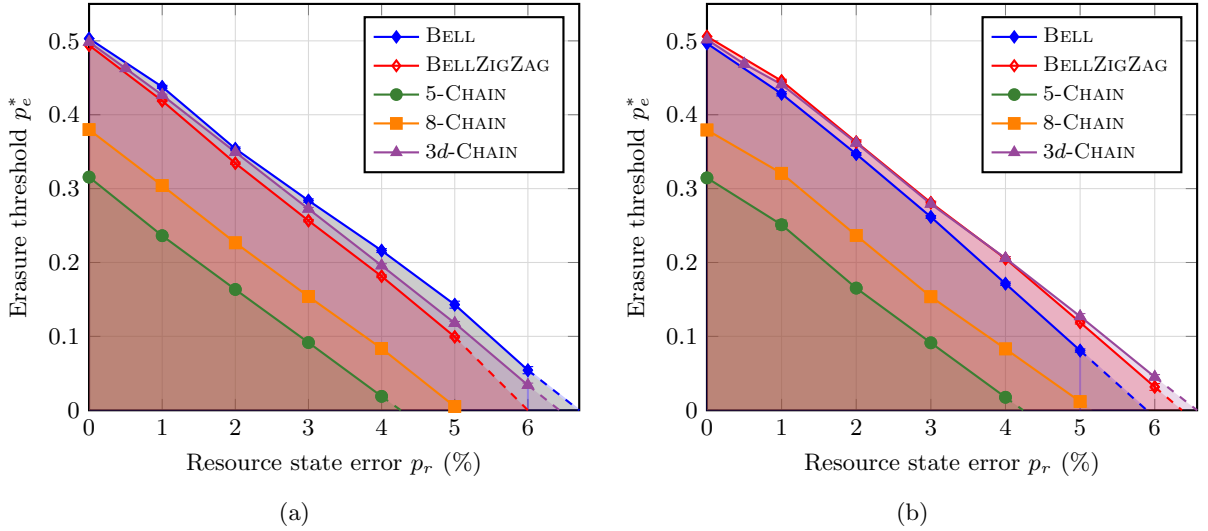
\begin{figure*}
  \begin{subfigure}[b]{0.45\linewidth}
    \centering
    \input{results/prep_and_erasure_correctable_regime_pm_corr_mergeZZ_obsZZ.tex}
    \caption{}
    \label{fig:correctable_regime_resource_merge}
  \end{subfigure}
  \begin{subfigure}[b]{0.45\linewidth}
    \centering
    \input{results/prep_and_erasure_correctable_regime_pm_corr_mergeZZ_obsXX.tex}
    \caption{}
    \label{fig:correctable_regime_resource_obs}
  \end{subfigure}
  \caption{\textbf{Correctable regime under resource-state preparation noise.} Estimated erasure threshold $p_e^*$ for (a) the merge observable $\bar{Z}\!\bar{Z}$ and (b) the perpendicular correlation $\bar{X}\!\bar{X}$ as a function of the phenomenological resource-state error rate $p_r$, with local circuit noise set to $p_c=0$.}
  \label{fig:correctable_regime_resource}
\end{figure*}

\subsection{Correctable-region threshold estimation}

Each point in the correctable-region plots is obtained by fixing the local noise strength, sweeping the elementary fusion-outcome erasure probability $p_e$, and estimating the logical error rate $P_L(p_e,d)$ at several code distances. We first estimate a crossing point using distances $d=11$ and $d=13$, then refine the estimate by taking further shots near the estimated crossing with $d=11,13,15$. When both circuit noise and erasure are high, decoding at higher distances becomes prohibitively expensive, so finite-size drift effects may still be present. Each $(d,p_e)$ value is estimated from at most $10^6$ Monte Carlo shots, with sampling stopped early after $10^5$ logical failures, and decoded using correlated matching as described in \cref{sec:sim_setup}. The zero-local-noise point is obtained from the pure-erasure sweep using the same procedure and the same distances as the corresponding correctable-region plot.

\subsection{Resource-state errors and erasure}

\Cref{fig:correctable_regime_resource} examines the thresholds when we introduce resource-state errors. Here we assume that local noise is negligible ($p_c=0$) and vary the resource-state error rate $p_r$, calculating the erasure threshold $p_e^*$ as a function of $p_r$.

In this case, all thresholds decrease approximately linearly as the resource-state preparation error rate $p_r$ increases. The sampled range differs slightly between protocols and observables, as shown in \cref{fig:correctable_regime_resource}.

\subsection{Circuit-level noise and erasure}

\begin{figure*}
  \begin{subfigure}[b]{0.45\linewidth}
    \input{results/circuit_and_erasure_correctable_regime_pm_corr_mergeZZ_obsZZ.tex}
    \caption{}
    \label{fig:circuit_and_erasure_merge}
  \end{subfigure}
  \begin{subfigure}[b]{0.45\linewidth}
    \centering
    \input{results/circuit_and_erasure_correctable_regime_pm_corr_mergeZZ_obsXX.tex}
    \caption{}
    \label{fig:circuit_and_erasure_obs}
  \end{subfigure}
  \caption{\textbf{Correctable regime under local circuit noise and interface erasure.} Estimated erasure threshold $p_e^*$ for (a) the merge observable $\bar{Z}\!\bar{Z}$ and (b) the perpendicular correlation $\bar{X}\!\bar{X}$ as a function of local circuit noise $p_c$, with resource-state error set to $p_r=0$.}
  \label{fig:circuit_and_erasure}
\end{figure*}
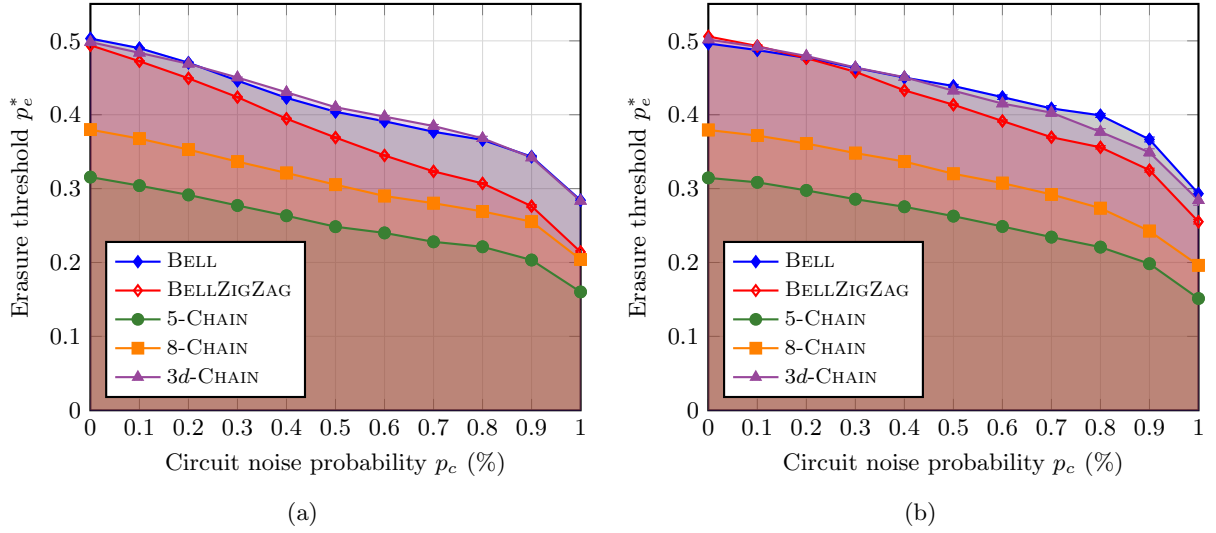

In \cref{fig:circuit_and_erasure}, we vary the local circuit error rate $p_c$ and estimate the largest elementary fusion-outcome erasure probability $p_e^*$ that remains correctable, with the resource-state error rate set to $p_r=0$. This comparison includes both Bell-pair geometries and the three hybrid chain decompositions.

\begin{table*}[tb]
    \centering
    \caption{\textbf{Conversion of circuit-and-erasure thresholds to photon-loss thresholds.} Elementary-outcome erasure thresholds $p_e^*$ for the merge observable $\bar{Z}\!\bar{Z}$ and perpendicular correlation $\bar{X}\!\bar{X}$ are read from the data underlying \cref{fig:circuit_and_erasure} at $p_c=0$ and $p_c=0.01$, with $p_r=0$. Parentheses give the uncertainty defined in the threshold-estimation procedure in the final displayed digits; uncertainties in $p_l^*$ are propagated from those in $p_e^*$ through \cref{eq:loss_threshold_conversion}. The photon-loss thresholds are shown for boosting levels $n=1,2,3$. Bold entries specify the largest physical loss threshold for each protocol. A blank entry indicates that the intrinsic fusion-failure erasure at zero photon loss already exceeds $p_e^*$.}
    \label{tab:circuit_erasure_loss_conversion}
    \setlength{\tabcolsep}{3.5pt}
    \let\ResultThresholdPercent\ResultThresholdWithErrorPercent
    \let\LossThresholdPercent\LossThresholdWithErrorPercent
    \let\BestLossThresholdPercent\BestLossThresholdWithErrorPercent
    \begin{tabular}{lcccc@{\hspace{1.2em}}cccc}
        \toprule
        & \multicolumn{4}{c}{$p_c=0$} & \multicolumn{4}{c}{$p_c=0.01$} \\
        \cmidrule(lr){2-5}\cmidrule(lr){6-9}
        Protocol & $p_e^*$ & \multicolumn{3}{c}{$p_l^*$} & $p_e^*$ & \multicolumn{3}{c}{$p_l^*$} \\
        \cmidrule(lr){3-5}\cmidrule(lr){7-9}
        & & $n=1$ & $n=2$ & $n=3$ & & $n=1$ & $n=2$ & $n=3$ \\
        \midrule
        \multicolumn{9}{l}{$\bar{Z}\!\bar{Z}$ merge observable} \\
        \addlinespace[0.25em]
        \textsc{Bell}
          & \ResultThresholdPercent{\CircuitZZData}{Bell}{0.0}{2}
          & \BestLossThresholdPercent{\CircuitZZData}{Bell}{0.0}{1}{2}
          & \LossThresholdPercent{\CircuitZZData}{Bell}{0.0}{2}{2}
          & \LossThresholdPercent{\CircuitZZData}{Bell}{0.0}{3}{2}
          & \ResultThresholdPercent{\CircuitZZData}{Bell}{0.01}{2}
          & \LossThresholdPercent{\CircuitZZData}{Bell}{0.01}{1}{2}
          & \BestLossThresholdPercent{\CircuitZZData}{Bell}{0.01}{2}{2}
          & \LossThresholdPercent{\CircuitZZData}{Bell}{0.01}{3}{2} \\
        \textsc{BellZigZag}
          & \ResultThresholdPercent{\CircuitZZData}{BellZigZag}{0.0}{2}
          & \BestLossThresholdPercent{\CircuitZZData}{BellZigZag}{0.0}{1}{2}
          & \LossThresholdPercent{\CircuitZZData}{BellZigZag}{0.0}{2}{2}
          & \LossThresholdPercent{\CircuitZZData}{BellZigZag}{0.0}{3}{2}
          & \ResultThresholdPercent{\CircuitZZData}{BellZigZag}{0.01}{2}
          & \LossThresholdPercent{\CircuitZZData}{BellZigZag}{0.01}{1}{2}
          & \BestLossThresholdPercent{\CircuitZZData}{BellZigZag}{0.01}{2}{2}
          & \LossThresholdPercent{\CircuitZZData}{BellZigZag}{0.01}{3}{2} \\
        $5$-\textsc{Chain}
          & \ResultThresholdPercent{\CircuitZZData}{5-Chain}{0.0}{2}
          & \LossThresholdPercent{\CircuitZZData}{5-Chain}{0.0}{1}{2}
          & \BestLossThresholdPercent{\CircuitZZData}{5-Chain}{0.0}{2}{2}
          & \LossThresholdPercent{\CircuitZZData}{5-Chain}{0.0}{3}{2}
          & \ResultThresholdPercent{\CircuitZZData}{5-Chain}{0.01}{2}
          & \LossThresholdPercent{\CircuitZZData}{5-Chain}{0.01}{1}{2}
          & \LossThresholdPercent{\CircuitZZData}{5-Chain}{0.01}{2}{2}
          & \BestLossThresholdPercent{\CircuitZZData}{5-Chain}{0.01}{3}{2} \\
        $8$-\textsc{Chain}
          & \ResultThresholdPercent{\CircuitZZData}{8-Chain}{0.0}{2}
          & \BestLossThresholdPercent{\CircuitZZData}{8-Chain}{0.0}{1}{2}
          & \LossThresholdPercent{\CircuitZZData}{8-Chain}{0.0}{2}{2}
          & \LossThresholdPercent{\CircuitZZData}{8-Chain}{0.0}{3}{2}
          & \ResultThresholdPercent{\CircuitZZData}{8-Chain}{0.01}{2}
          & \LossThresholdPercent{\CircuitZZData}{8-Chain}{0.01}{1}{2}
          & \LossThresholdPercent{\CircuitZZData}{8-Chain}{0.01}{2}{2}
          & \LossThresholdPercent{\CircuitZZData}{8-Chain}{0.01}{3}{2} \\
        $3d$-\textsc{Chain}
          & \ResultThresholdPercent{\CircuitZZData}{3d-Chain}{0.0}{2}
          & \BestLossThresholdPercent{\CircuitZZData}{3d-Chain}{0.0}{1}{2}
          & \LossThresholdPercent{\CircuitZZData}{3d-Chain}{0.0}{2}{2}
          & \LossThresholdPercent{\CircuitZZData}{3d-Chain}{0.0}{3}{2}
          & \ResultThresholdPercent{\CircuitZZData}{3d-Chain}{0.01}{2}
          & \LossThresholdPercent{\CircuitZZData}{3d-Chain}{0.01}{1}{2}
          & \BestLossThresholdPercent{\CircuitZZData}{3d-Chain}{0.01}{2}{2}
          & \LossThresholdPercent{\CircuitZZData}{3d-Chain}{0.01}{3}{2} \\
        \addlinespace[0.45em]
        \multicolumn{9}{l}{$\bar{X}\!\bar{X}$ perpendicular correlation} \\
        \addlinespace[0.25em]
        \textsc{Bell}
          & \ResultThresholdPercent{\CircuitXXData}{Bell}{0.0}{2}
          & \BestLossThresholdPercent{\CircuitXXData}{Bell}{0.0}{1}{2}
          & \LossThresholdPercent{\CircuitXXData}{Bell}{0.0}{2}{2}
          & \LossThresholdPercent{\CircuitXXData}{Bell}{0.0}{3}{2}
          & \ResultThresholdPercent{\CircuitXXData}{Bell}{0.01}{2}
          & \LossThresholdPercent{\CircuitXXData}{Bell}{0.01}{1}{2}
          & \BestLossThresholdPercent{\CircuitXXData}{Bell}{0.01}{2}{2}
          & \LossThresholdPercent{\CircuitXXData}{Bell}{0.01}{3}{2} \\
        \textsc{BellZigZag}
          & \ResultThresholdPercent{\CircuitXXData}{BellZigZag}{0.0}{2}
          & \BestLossThresholdPercent{\CircuitXXData}{BellZigZag}{0.0}{1}{2}
          & \LossThresholdPercent{\CircuitXXData}{BellZigZag}{0.0}{2}{2}
          & \LossThresholdPercent{\CircuitXXData}{BellZigZag}{0.0}{3}{2}
          & \ResultThresholdPercent{\CircuitXXData}{BellZigZag}{0.01}{2}
          & \LossThresholdPercent{\CircuitXXData}{BellZigZag}{0.01}{1}{2}
          & \LossThresholdPercent{\CircuitXXData}{BellZigZag}{0.01}{2}{2}
          & \LossThresholdPercent{\CircuitXXData}{BellZigZag}{0.01}{3}{2} \\
        $5$-\textsc{Chain}
          & \ResultThresholdPercent{\CircuitXXData}{5-Chain}{0.0}{2}
          & \LossThresholdPercent{\CircuitXXData}{5-Chain}{0.0}{1}{2}
          & \BestLossThresholdPercent{\CircuitXXData}{5-Chain}{0.0}{2}{2}
          & \LossThresholdPercent{\CircuitXXData}{5-Chain}{0.0}{3}{2}
          & \ResultThresholdPercent{\CircuitXXData}{5-Chain}{0.01}{2}
          & \LossThresholdPercent{\CircuitXXData}{5-Chain}{0.01}{1}{2}
          & \LossThresholdPercent{\CircuitXXData}{5-Chain}{0.01}{2}{2}
          & \BestLossThresholdPercent{\CircuitXXData}{5-Chain}{0.01}{3}{2} \\
        $8$-\textsc{Chain}
          & \ResultThresholdPercent{\CircuitXXData}{8-Chain}{0.0}{2}
          & \BestLossThresholdPercent{\CircuitXXData}{8-Chain}{0.0}{1}{2}
          & \LossThresholdPercent{\CircuitXXData}{8-Chain}{0.0}{2}{2}
          & \LossThresholdPercent{\CircuitXXData}{8-Chain}{0.0}{3}{2}
          & \ResultThresholdPercent{\CircuitXXData}{8-Chain}{0.01}{2}
          & \LossThresholdPercent{\CircuitXXData}{8-Chain}{0.01}{1}{2}
          & \BestLossThresholdPercent{\CircuitXXData}{8-Chain}{0.01}{2}{2}
          & \LossThresholdPercent{\CircuitXXData}{8-Chain}{0.01}{3}{2} \\
        $3d$-\textsc{Chain}
          & \ResultThresholdPercent{\CircuitXXData}{3d-Chain}{0.0}{2}
          & \BestLossThresholdPercent{\CircuitXXData}{3d-Chain}{0.0}{1}{2}
          & \LossThresholdPercent{\CircuitXXData}{3d-Chain}{0.0}{2}{2}
          & \LossThresholdPercent{\CircuitXXData}{3d-Chain}{0.0}{3}{2}
          & \ResultThresholdPercent{\CircuitXXData}{3d-Chain}{0.01}{2}
          & \LossThresholdPercent{\CircuitXXData}{3d-Chain}{0.01}{1}{2}
          & \LossThresholdPercent{\CircuitXXData}{3d-Chain}{0.01}{2}{2}
          & \LossThresholdPercent{\CircuitXXData}{3d-Chain}{0.01}{3}{2} \\
        \bottomrule
    \end{tabular}
\end{table*}

All five erasure thresholds decrease from the pure-erasure case as local circuit noise increases. In this case, the threshold degradation is not linear in $p_c$, and the slope steepens as $p_c$ approaches the local circuit-noise threshold. To express these boundaries in terms of a physical photon-loss probability, we substitute the threshold estimates from \cref{fig:circuit_and_erasure} into \cref{eq:loss_threshold_conversion}. For each observable, \cref{tab:circuit_erasure_loss_conversion} reports the converted loss threshold for boosting levels $n=1,2,3$, with the largest physical threshold highlighted in bold. The $50\%$ erasure threshold converts to \ErasureToLossPercent{0.5}{1}{2} photon-loss probability at $n=1$, which is the optimal boosting level, while the $\bar{Z}\!\bar{Z}$ threshold of $8$-\textsc{Chain} gives an approximately \LossThresholdPercent{\CircuitZZData}{8-Chain}{0.0}{1}{1} photon-loss threshold at the same boosting level. At $n=1$, the marginal fusion-failure probability is near the $\bar{Z}\!\bar{Z}$ erasure threshold for $5$-\textsc{Chain}, so boosting to $n=2$ improves the photon-loss threshold from \LossThresholdPercent{\CircuitZZData}{5-Chain}{0.0}{1}{2} to \BestLossThresholdPercent{\CircuitZZData}{5-Chain}{0.0}{2}{2}. At $1\%$ local circuit noise, the available $\bar{Z}\!\bar{Z}$ estimates favour $n=2$ for \textsc{Bell}, \textsc{BellZigZag}, and $3d$-\textsc{Chain}, while $5$-\textsc{Chain} is optimal at $n=3$. The available $\bar{X}\!\bar{X}$ estimates likewise favour $n=2$ for \textsc{Bell} and $8$-\textsc{Chain}, and $n=3$ for $5$-\textsc{Chain}.

\subsection{Subthreshold suppression}

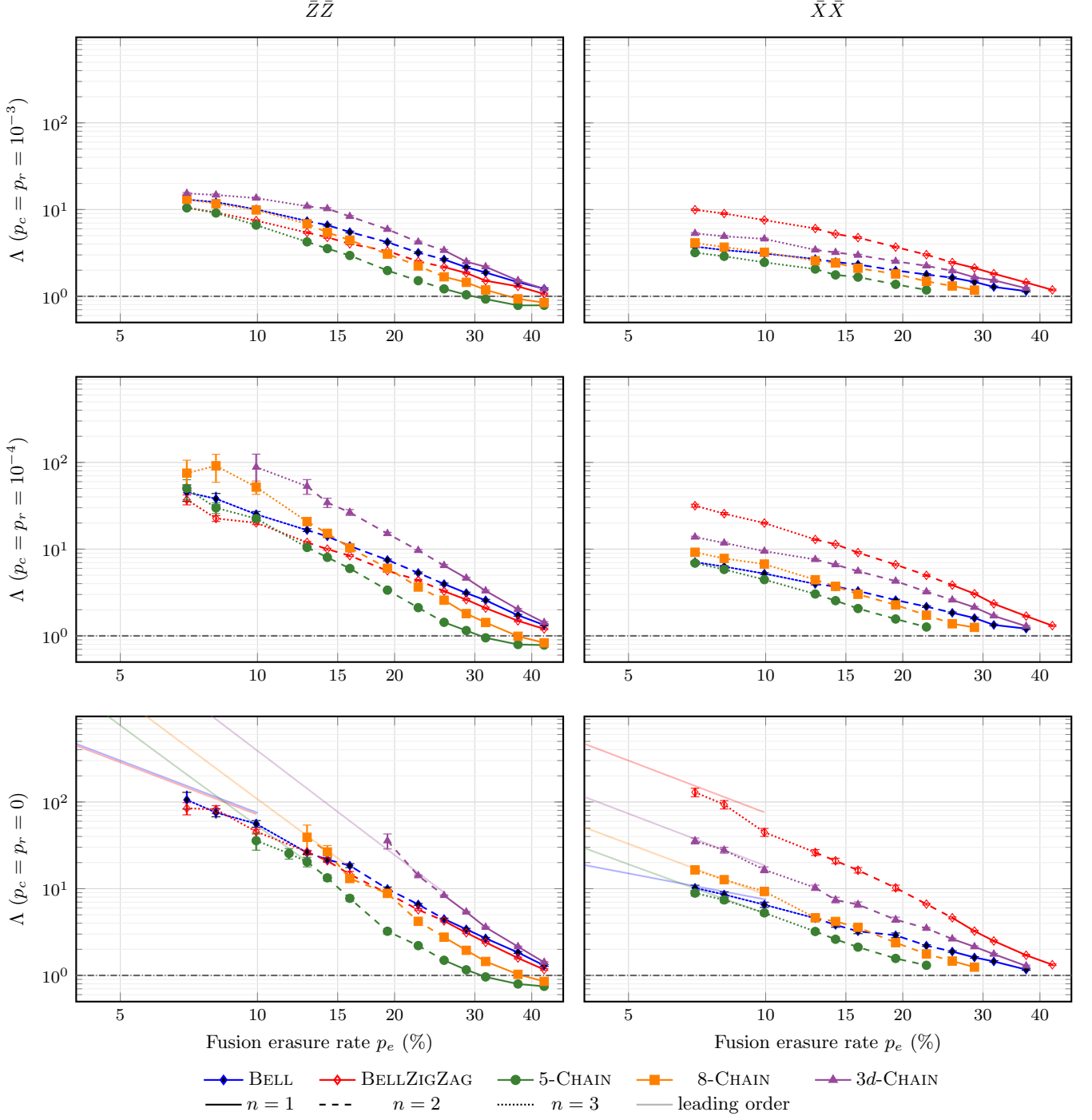
\begin{figure*}
  \input{results/subthreshold_lambda_erasure_grid_pm_corr_mergeZZ_obsZZ-XX.tex}
  \caption{\textbf{Subthreshold suppression under the loss-based linear-optical model.} The finite-distance suppression factor $\Lambda_{5\rightarrow7}=p_L(5)/p_L(7)$ is shown against the effective elementary-outcome erasure probability $p_e$ for the merge observable $\bar{Z}\!\bar{Z}$ (left) and perpendicular observable $\bar{X}\!\bar{X}$ (right), at $p_c=p_r=10^{-3}$, $10^{-4}$ and $0$ (top to bottom). Colours and markers identify the interface protocol. Line styles indicate the lowest boosting level that can attain the corresponding $p_e$ under \cref{eq:boosted_loss}. Faint curves in the zero-local-noise row show the leading-order erasure predictions. Error bars show the estimated statistical uncertainty in $\Lambda_{5\rightarrow7}$, and the horizontal line marks $\Lambda_{5\rightarrow7}=1$, below which the logical error rate increases from $d=5$ to $d=7$.}
  \label{fig:subthreshold_performance}
\end{figure*}

Thresholds locate the boundary of the correctable region but do not determine how rapidly the logical error rate decreases with code distance. Since subthreshold errors become rare at high distances, we calculate the finite-distance suppression from $d$ to $d+2$ using
\begin{equation}
    \Lambda_{d\rightarrow d+2}:=\frac{p_L(d)}{p_L(d+2)}.
    \label{eq:lambda_effective}
\end{equation}

This is the factor by which $p_L$ decreases over this distance increment at fixed $(p_c,p_r,p_e)$. Thus $\Lambda_{5\rightarrow7}>1$ indicates suppression with increasing distance, and a larger value indicates stronger finite-distance suppression. We report $\Lambda_{5\rightarrow7}$ uniformly for all protocols. At points where the logical error rate could also be resolved at $d=9$, we additionally calculated $\Lambda_{7\rightarrow9}=p_L(7)/p_L(9)$. The ratio $\Lambda_{7\rightarrow9}/\Lambda_{5\rightarrow7}$ was close to unity in most cases, so we do not expect the inclusion of larger distances to significantly change the comparison.

We write $\Lambda_{5\rightarrow7}$ as $\Lambda$ below. Since we consider multiple noise sources with different distances under reduced noise models, we expect $\Lambda$ to behave differently in different noise regimes. To probe this behaviour, we fix three local noise regimes, corresponding to moderate local noise $p_c=10^{-3}$, low local noise $p_c=10^{-4}$ and zero local noise $p_c=0$, and sweep the erasure rate $p_e$ within each regime. To focus on the effect of erasures, we fix $p_c=p_r$ in all cases. Since the effective erasure rate is determined by the boosting level $n$ and the photon-loss rate, we use only erasure rates corresponding to specific loss/boosting combinations.

At the moderate local-noise level $p_c=p_r=10^{-3}$, circuit and resource-state faults set the dominant scale of $\Lambda$. Consequently, reducing $p_e$ improves the suppression factors of the different protocols at broadly similar rates, and the interface-distance differences are largely masked. The $3d$-\textsc{Chain} nevertheless gives the strongest $\bar{Z}\!\bar{Z}$ suppression, whereas \textsc{BellZigZag} gives the strongest $\bar{X}\!\bar{X}$ suppression. The latter is the only construction considered here that avoids a distance-reducing local fault mechanism while retaining the full perpendicular interface distance. The effect of the reduced interface distance for \textsc{Bell} becomes apparent below $p_e=0.15$. In both cases, the truncated linear chains start with lower suppression factors due to their reduced thresholds, though they catch up as $p_e$ decreases and local noise dominates.

At $p_c=p_r=10^{-4}$, the interface contribution becomes visible. As $p_e$ decreases, the $\Lambda$ curves of the linear-chain protocols rise more steeply than those of the Bell-pair interfaces. For $\bar{Z}\!\bar{Z}$, all three linear decompositions eventually overtake both Bell-pair protocols, exposing the augmented interface distance. The same crossover mechanism is visible for $\bar{X}\!\bar{X}$ relative to the straight \textsc{Bell} interface, whose reduced perpendicular distance causes it to fall below the other constructions at sufficiently small $p_e$. \textsc{BellZigZag} remains the strongest $\bar{X}\!\bar{X}$ protocol.

With $p_c=p_r=0$, sampling at low erasure rates becomes challenging as logical errors become exceedingly rare, so we resort to using a leading-order erasure approximation to predict the trends in $\Lambda$. Under pure erasure, the decoder fails only when the erased fusions support a logical operator, in which case it fails with probability $50\%$~\cite{staceThresholdsTopologicalCodes2009}. At low erasure rates this is dominated by the minimum-weight logicals, giving $p_L(d) \approx \frac{1}{2} \mathcal{N}_d p_e^{w_d}$, where the interface distance $w_d$ and the number of minimum-weight logicals $\mathcal{N}_d$ are counted directly from each protocol's detector error model. Hence $\Lambda \approx (\mathcal{N}_5/\mathcal{N}_7) p_e^{-(w_7-w_5)}$. The chains have $w_d = 2d+1$, against $w_d = d+1$ for \textsc{Bell}, so $\Lambda$ grows as $p_e^{-4}$ rather than $p_e^{-2}$. But they also have far more minimal logicals (e.g. $\mathcal{N}_7 = 4.6\times 10^4$ for the $3d$-\textsc{Chain} vs $4$ for \textsc{Bell} in the primal interface DEM), which delays the crossover to lower $p_e$. For the $\bar{X}\!\bar{X}$ observable, all protocols have $w_d = d$ except for \textsc{Bell}, which has $w_d = \lfloor (d+1)/2\rfloor$. The leading-order estimates are generally consistent with the slopes of the collected data, though show small vertical offsets in some cases.

Below $15\%$ erasure, $8$-\textsc{Chain} overtakes \textsc{Bell} and \textsc{BellZigZag} for $\bar{Z}\!\bar{Z}$, as well as overtaking \textsc{Bell} for $\bar{X}\!\bar{X}$ at the same erasure rate. The leading-order estimates predict $5$-\textsc{Chain} to overtake \textsc{Bell} and \textsc{BellZigZag} around $p_e=10\%$ for $\bar{Z}\!\bar{Z}$, and to overtake \textsc{Bell} for $\bar{X}\!\bar{X}$ closer to $7\%$. Low logical error rates at these levels make numerical estimation challenging, but the trends are consistent with the expected distance scaling. For the $\bar{X}\!\bar{X}$ observable, despite the hybrid protocols matching the interface distance of \textsc{BellZigZag}, the actual $\Lambda$ values for \textsc{BellZigZag} are much higher. This can be attributed to the fact that the number of minimum-weight logical operators in the rotated lattice increases exponentially with $d$, while increasing only linearly on the unrotated lattice, leading to lower logical error rates for unrotated surface codes at equivalent distances~\cite{crigerMultipathSummationDecoding2018,orourkeComparePairRotated2025}. These effects can be observed in the prefactors $\mathcal{N}_d$ appearing in the leading-order estimates for $p_L(d)$.

\FloatBarrier

%% file: results/master_threshold_pm_corr_mergeZZ_obsZZ-XX.tex
\begin{tikzpicture}
\begin{groupplot}[
  group style={group name=groupmasterthresholdpmcorrmergeZZobsZZXX, group size=3 by 2, yticklabels at=edge left, ylabels at=edge left, xlabels at=edge bottom, horizontal sep=8pt, vertical sep=1.4cm},
  scale only axis,
  width=0.290\linewidth,
  height=0.2\textheight,
  xmode=linear,
  ymode=log,
  xlabel={Erasure rate $p_e$ (\%)},
  ylabel={Logical error rate},
  grid=both,
  minor grid style={gray!15},
  major grid style={gray!30},
  title style={font=\small, yshift=-4pt},
  legend cell align=left,
  legend style={font=\footnotesize, /tikz/every even column/.append style={column sep=8pt}},
  every axis plot/.append style={line width=0.5pt, mark size=1.5pt},
  error bars/y dir=both,
  error bars/y explicit,
  scaled x ticks=false,
  x tick label style={/pgf/number format/fixed, font=\footnotesize},
  xticklabel={\pgfmathparse{100*\tick}\pgfmathprintnumber{\pgfmathresult}},
]

\nextgroupplot[title={\textsc{Bell}}, xmin=0.445, xmax=0.555, ymin=0.00561411, ymax=0.742799, legend to name=dlegend_masterthresholdpmcorrmergeZZobsZZXX, legend columns=1, legend style={draw=none}]
  \addplot[color=black, dotted, mark=none, forget plot] coordinates {(0.503124,0.00561411) (0.503124,0.0987452)};
  \addplot[color=black, dotted, mark=none, forget plot] coordinates {(0.496441,0.00561411) (0.496441,0.399357)};
  \addplot[color=blue, mark=*, mark size=1pt, solid, mark options={solid}, forget plot] coordinates {
        (0.45,0.032991) +- (0.00066858,0.00066008)
        (0.46,0.041691) +- (0.00074749,0.00073914)
        (0.47,0.051885) +- (0.00082934,0.00082056)
        (0.48,0.064056) +- (0.00091544,0.00090727)
        (0.49,0.078209) +- (0.00100268,0.00099537)
        (0.5,0.094224) +- (0.00108999,0.00108262)
        (0.51,0.110781) +- (0.00119523,0.0011875)
        (0.52,0.130192) +- (0.00131202,0.00130517)
        (0.53,0.151251) +- (0.0014863,0.00147881)
        (0.54,0.173219) +- (0.00174094,0.00173265)
        (0.55,0.195818) +- (0.00187759,0.00186746)
  };
  \addplot[color=red, mark=*, mark size=1pt, solid, mark options={solid}, forget plot] coordinates {
        (0.45,0.027925) +- (0.00061679,0.00060806)
        (0.46,0.037164) +- (0.00070771,0.00069909)
        (0.47,0.047531) +- (0.00079479,0.00078653)
        (0.48,0.06065) +- (0.00089246,0.00088413)
        (0.49,0.075905) +- (0.00098913,0.00098161)
        (0.5,0.092983) +- (0.00108346,0.00107654)
        (0.51,0.112499) +- (0.00118557,0.00117845)
        (0.52,0.133733) +- (0.00140491,0.00139653)
        (0.53,0.158469) +- (0.00157141,0.00156192)
        (0.54,0.182775) +- (0.0018719,0.00186308)
        (0.55,0.210688) +- (0.0020354,0.00202457)
  };
  \addplot[color=OliveGreen, mark=*, mark size=1pt, solid, mark options={solid}, forget plot] coordinates {
        (0.45,0.02383) +- (0.00057135,0.00056239)
        (0.46,0.032174) +- (0.00066103,0.00065221)
        (0.47,0.043553) +- (0.00076261,0.00075455)
        (0.48,0.056812) +- (0.00086473,0.00085637)
        (0.49,0.072965) +- (0.00097167,0.00096351)
        (0.5,0.092224) +- (0.00108063,0.00107317)
        (0.51,0.113964) +- (0.00120815,0.00120076)
        (0.52,0.138041) +- (0.00135032,0.00134353)
        (0.53,0.165358) +- (0.00170552,0.00169706)
        (0.54,0.192476) +- (0.00192294,0.00191442)
        (0.55,0.221828) +- (0.00219969,0.00218997)
  };
  \addplot[color=purple!80, mark=*, mark size=1pt, solid, mark options={solid}, forget plot] coordinates {
        (0.45,0.020065) +- (0.00052557,0.00051663)
        (0.46,0.028572) +- (0.00062373,0.00061513)
        (0.47,0.039627) +- (0.00072996,0.00072141)
        (0.48,0.054161) +- (0.00084605,0.00083775)
        (0.49,0.071088) +- (0.00095973,0.00095201)
        (0.5,0.091863) +- (0.00107828,0.00107062)
        (0.51,0.11589) +- (0.00120257,0.00119534)
        (0.52,0.142885) +- (0.00147315,0.00146478)
        (0.53,0.171832) +- (0.00164379,0.0016347)
        (0.54,0.202922) +- (0.0019714,0.00196325)
        (0.55,0.235504) +- (0.00233586,0.00232708)
  };
  \addplot[color=Purple, mark=*, mark size=1pt, solid, mark options={solid}, forget plot] coordinates {
        (0.45,0.017347) +- (0.00048985,0.00048093)
        (0.46,0.025631) +- (0.00059175,0.00058306)
        (0.47,0.036794) +- (0.00070456,0.00069602)
        (0.48,0.05108) +- (0.0008235,0.00081498)
        (0.49,0.068917) +- (0.00094522,0.00093738)
        (0.5,0.091183) +- (0.00107514,0.0010674)
        (0.51,0.117201) +- (0.00121853,0.00121138)
        (0.52,0.147118) +- (0.00142808,0.00142029)
        (0.53,0.178137) +- (0.00176504,0.00175607)
        (0.54,0.213707) +- (0.00213069,0.00211952)
        (0.55,0.247147) +- (0.00236084,0.00235054)
  };
  \addplot[color=blue, mark=*, mark size=1pt, dashed, mark options={solid}, forget plot] coordinates {
        (0.45,0.303731) +- (0.00278147,0.00277322)
        (0.46,0.326763) +- (0.00287921,0.0028704)
        (0.47,0.349044) +- (0.00300486,0.0029978)
        (0.48,0.367648) +- (0.0030945,0.0030884)
        (0.49,0.387733) +- (0.00329896,0.00329281)
        (0.5,0.405774) +- (0.00341934,0.00341343)
        (0.51,0.421382) +- (0.00354567,0.0035409)
        (0.52,0.435819) +- (0.00352222,0.00351669)
        (0.53,0.449328) +- (0.00357657,0.00357313)
        (0.54,0.458461) +- (0.00375003,0.00374757)
        (0.55,0.467129) +- (0.00371099,0.00370871)
  };
  \addplot[color=red, mark=*, mark size=1pt, dashed, mark options={solid}, forget plot] coordinates {
        (0.45,0.290976) +- (0.00273574,0.00272585)
        (0.46,0.316191) +- (0.0029233,0.00291324)
        (0.47,0.341678) +- (0.00305455,0.00304605)
        (0.48,0.367284) +- (0.00323011,0.00322234)
        (0.49,0.386883) +- (0.00329701,0.00329007)
        (0.5,0.40593) +- (0.00334978,0.0033442)
        (0.51,0.424898) +- (0.00347221,0.00346656)
        (0.52,0.44001) +- (0.00349767,0.00349369)
        (0.53,0.453391) +- (0.00367242,0.0036688)
        (0.54,0.463916) +- (0.00370787,0.00370555)
        (0.55,0.470768) +- (0.00379701,0.0037948)
  };
  \addplot[color=OliveGreen, mark=*, mark size=1pt, dashed, mark options={solid}, forget plot] coordinates {
        (0.45,0.278216) +- (0.00255078,0.00254208)
        (0.46,0.305584) +- (0.00275227,0.00274184)
        (0.47,0.334626) +- (0.00308276,0.00307335)
        (0.48,0.36188) +- (0.00315878,0.00314993)
        (0.49,0.387042) +- (0.00331571,0.0033084)
        (0.5,0.405893) +- (0.00343327,0.00342762)
        (0.51,0.425134) +- (0.00347734,0.00347213)
        (0.52,0.442305) +- (0.00356962,0.00356637)
        (0.53,0.455025) +- (0.00371079,0.00370713)
        (0.54,0.467274) +- (0.0037799,0.00377761)
        (0.55,0.476718) +- (0.00380461,0.00380385)
  };
  \addplot[color=purple!80, mark=*, mark size=1pt, dashed, mark options={solid}, forget plot] coordinates {
        (0.45,0.265497) +- (0.00246928,0.00245908)
        (0.46,0.297971) +- (0.00265144,0.00264215)
        (0.47,0.327648) +- (0.002986,0.00297662)
        (0.48,0.35867) +- (0.00316628,0.00315754)
        (0.49,0.384897) +- (0.00329523,0.00328819)
        (0.5,0.408585) +- (0.0033596,0.00335427)
        (0.51,0.427348) +- (0.00346397,0.00345922)
        (0.52,0.445634) +- (0.00362477,0.00362066)
        (0.53,0.4599) +- (0.00371388,0.00371051)
        (0.54,0.471659) +- (0.00379136,0.00378881)
        (0.55,0.48269) +- (0.00387997,0.00387893)
  };
  \addplot[color=Purple, mark=*, mark size=1pt, dashed, mark options={solid}, forget plot] coordinates {
        (0.45,0.253021) +- (0.00236058,0.0023504)
        (0.46,0.288392) +- (0.00271417,0.00270576)
        (0.47,0.322834) +- (0.00289335,0.00288241)
        (0.48,0.352756) +- (0.00310592,0.00309807)
        (0.49,0.382967) +- (0.00337113,0.00336415)
        (0.5,0.408793) +- (0.0033759,0.00336956)
        (0.51,0.430193) +- (0.00359112,0.0035863)
        (0.52,0.449334) +- (0.00367443,0.00367073)
        (0.53,0.464494) +- (0.00378499,0.00378125)
        (0.54,0.473234) +- (0.00382066,0.00381952)
        (0.55,0.482901) +- (0.0038312,0.0038301)
  };
  \addlegendimage{color=blue, line legend, mark=none}
  \addlegendentry{$d=11$}
  \addlegendimage{color=red, line legend, mark=none}
  \addlegendentry{$d=13$}
  \addlegendimage{color=OliveGreen, line legend, mark=none}
  \addlegendentry{$d=15$}
  \addlegendimage{color=purple!80, line legend, mark=none}
  \addlegendentry{$d=17$}
  \addlegendimage{color=Purple, line legend, mark=none}
  \addlegendentry{$d=19$}

\nextgroupplot[title={\textsc{BellZigZag}}, xmin=0.445, xmax=0.555, ymin=0.00561411, ymax=0.742799]
  \addplot[color=black, dotted, mark=none, forget plot] coordinates {(0.494023,0.00561411) (0.494023,0.209866)};
  \addplot[color=black, dotted, mark=none, forget plot] coordinates {(0.50595,0.00561411) (0.50595,0.288297)};
  \addplot[color=blue, mark=*, mark size=1pt, solid, mark options={solid}, forget plot] coordinates {
        (0.45,0.095521) +- (0.00109632,0.00108878)
        (0.46,0.117812) +- (0.001259,0.00125142)
        (0.47,0.141924) +- (0.00140252,0.00139445)
        (0.48,0.169437) +- (0.00175502,0.00174532)
        (0.49,0.197518) +- (0.00191998,0.00191007)
        (0.5,0.228714) +- (0.00208948,0.00208175)
        (0.51,0.258739) +- (0.00251708,0.00250604)
        (0.52,0.288941) +- (0.00263112,0.00262015)
        (0.53,0.319218) +- (0.00279519,0.00278649)
        (0.54,0.346274) +- (0.00316945,0.0031597)
        (0.55,0.373773) +- (0.00329216,0.00328515)
  };
  \addplot[color=red, mark=*, mark size=1pt, solid, mark options={solid}, forget plot] coordinates {
        (0.45,0.084675) +- (0.00103956,0.00103252)
        (0.46,0.107807) +- (0.00116083,0.00115413)
        (0.47,0.134967) +- (0.00144767,0.00143941)
        (0.48,0.164114) +- (0.00161261,0.00160367)
        (0.49,0.197948) +- (0.00203488,0.00202437)
        (0.5,0.231124) +- (0.00219458,0.00218579)
        (0.51,0.266391) +- (0.00260462,0.00259474)
        (0.52,0.30072) +- (0.00277554,0.00276526)
        (0.53,0.332232) +- (0.00292224,0.0029133)
        (0.54,0.363503) +- (0.00308896,0.00308195)
        (0.55,0.387466) +- (0.00320038,0.00319409)
  };
  \addplot[color=OliveGreen, mark=*, mark size=1pt, solid, mark options={solid}, forget plot] coordinates {
        (0.45,0.074676) +- (0.00098238,0.00097394)
        (0.46,0.099026) +- (0.00111537,0.00110835)
        (0.47,0.127515) +- (0.00125588,0.00124893)
        (0.48,0.16053) +- (0.00162856,0.00161913)
        (0.49,0.196121) +- (0.00204211,0.00203237)
        (0.5,0.233604) +- (0.00223305,0.00222269)
        (0.51,0.272288) +- (0.00250472,0.00249363)
        (0.52,0.308058) +- (0.00290938,0.00289957)
        (0.53,0.343261) +- (0.00310828,0.00310022)
        (0.54,0.375661) +- (0.00323111,0.00322292)
        (0.55,0.40397) +- (0.0035309,0.00352462)
  };
  \addplot[color=purple!80, mark=*, mark size=1pt, solid, mark options={solid}, forget plot] coordinates {
        (0.45,0.065982) +- (0.00092717,0.00091935)
        (0.46,0.090555) +- (0.00107305,0.00106549)
        (0.47,0.120473) +- (0.00128311,0.00127503)
        (0.48,0.155286) +- (0.00150653,0.00149804)
        (0.49,0.194054) +- (0.00198012,0.00197079)
        (0.5,0.234748) +- (0.00216564,0.00215637)
        (0.51,0.278411) +- (0.00261321,0.00260468)
        (0.52,0.316912) +- (0.00285838,0.00285056)
        (0.53,0.354843) +- (0.00300875,0.00300114)
        (0.54,0.388849) +- (0.00351736,0.00350951)
        (0.55,0.418128) +- (0.00361127,0.00360608)
  };
  \addplot[color=Purple, mark=*, mark size=1pt, solid, mark options={solid}, forget plot] coordinates {
        (0.45,0.057938) +- (0.00087313,0.00086471)
        (0.46,0.082359) +- (0.00102749,0.00101944)
        (0.47,0.114273) +- (0.00121297,0.00120549)
        (0.48,0.151218) +- (0.00144243,0.00143581)
        (0.49,0.192504) +- (0.00188225,0.00187258)
        (0.5,0.236784) +- (0.00239607,0.00238373)
        (0.51,0.281174) +- (0.00265395,0.00264477)
        (0.52,0.325775) +- (0.00285771,0.00284996)
        (0.53,0.365647) +- (0.00309375,0.00308504)
        (0.54,0.398689) +- (0.00326279,0.00325682)
        (0.55,0.427643) +- (0.00338089,0.00337637)
  };
  \addplot[color=blue, mark=*, mark size=1pt, dashed, mark options={solid}, forget plot] coordinates {
        (0.45,0.128278) +- (0.00136278,0.00135426)
        (0.46,0.153767) +- (0.00148888,0.00148124)
        (0.47,0.181897) +- (0.00179755,0.00178832)
        (0.48,0.210578) +- (0.00197204,0.00196255)
        (0.49,0.240723) +- (0.00234263,0.00233235)
        (0.5,0.272608) +- (0.00256309,0.002553)
        (0.51,0.300645) +- (0.00284913,0.0028392)
        (0.52,0.330448) +- (0.00282288,0.00281578)
        (0.53,0.357902) +- (0.00321785,0.00320885)
        (0.54,0.382919) +- (0.00333707,0.00333028)
        (0.55,0.405357) +- (0.00339809,0.00339131)
  };
  \addplot[color=red, mark=*, mark size=1pt, dashed, mark options={solid}, forget plot] coordinates {
        (0.45,0.110272) +- (0.00117165,0.0011648)
        (0.46,0.137704) +- (0.00143426,0.00142554)
        (0.47,0.168023) +- (0.0016038,0.00159649)
        (0.48,0.20019) +- (0.00199379,0.00198469)
        (0.49,0.235477) +- (0.00219887,0.00218888)
        (0.5,0.269281) +- (0.0026003,0.00259092)
        (0.51,0.30125) +- (0.00277208,0.00276451)
        (0.52,0.334796) +- (0.00286308,0.00285428)
        (0.53,0.364722) +- (0.00330842,0.0033002)
        (0.54,0.392262) +- (0.00341699,0.0034102)
        (0.55,0.41304) +- (0.0035744,0.00356865)
  };
  \addplot[color=OliveGreen, mark=*, mark size=1pt, dashed, mark options={solid}, forget plot] coordinates {
        (0.45,0.09563) +- (0.00109964,0.00109213)
        (0.46,0.123796) +- (0.00123481,0.0012275)
        (0.47,0.155459) +- (0.00160259,0.00159376)
        (0.48,0.190812) +- (0.00178577,0.00177754)
        (0.49,0.228171) +- (0.00218677,0.0021767)
        (0.5,0.265645) +- (0.00238377,0.00237433)
        (0.51,0.303765) +- (0.0028451,0.00283438)
        (0.52,0.338425) +- (0.00301322,0.00300509)
        (0.53,0.371104) +- (0.00314752,0.00314138)
        (0.54,0.399949) +- (0.00340366,0.00339697)
        (0.55,0.424571) +- (0.00359406,0.00358841)
  };
  \addplot[color=purple!80, mark=*, mark size=1pt, dashed, mark options={solid}, forget plot] coordinates {
        (0.45,0.082917) +- (0.00103036,0.00102277)
        (0.46,0.111563) +- (0.00117755,0.00117084)
        (0.47,0.14471) +- (0.0015273,0.00151937)
        (0.48,0.181609) +- (0.00179431,0.00178583)
        (0.49,0.224152) +- (0.00217677,0.00216707)
        (0.5,0.264773) +- (0.00252952,0.00251903)
        (0.51,0.305781) +- (0.00279585,0.00278493)
        (0.52,0.345403) +- (0.00290703,0.00289983)
        (0.53,0.378792) +- (0.00309012,0.00308409)
        (0.54,0.410463) +- (0.0035558,0.00354886)
        (0.55,0.434321) +- (0.00366328,0.00365899)
  };
  \addplot[color=Purple, mark=*, mark size=1pt, dashed, mark options={solid}, forget plot] coordinates {
        (0.45,0.071948) +- (0.00096401,0.00095593)
        (0.46,0.100402) +- (0.00112271,0.00111557)
        (0.47,0.13437) +- (0.00130829,0.00130086)
        (0.48,0.173751) +- (0.00176815,0.00175921)
        (0.49,0.216953) +- (0.00197593,0.00196726)
        (0.5,0.262167) +- (0.00248391,0.00247223)
        (0.51,0.307904) +- (0.00272324,0.0027153)
        (0.52,0.348708) +- (0.00291106,0.0029034)
        (0.53,0.38599) +- (0.00316222,0.00315603)
        (0.54,0.417787) +- (0.00330462,0.00329937)
        (0.55,0.441909) +- (0.00339973,0.00339594)
  };

\nextgroupplot[title={$3d$-\textsc{Chain}}, xmin=0.445, xmax=0.555, ymin=0.00561411, ymax=0.742799]
  \addplot[color=black, dotted, mark=none, forget plot] coordinates {(0.498389,0.00561411) (0.498389,0.0810364)};
  \addplot[color=black, dotted, mark=none, forget plot] coordinates {(0.501818,0.00561411) (0.501818,0.419775)};
  \addplot[color=blue, mark=*, mark size=1pt, solid, mark options={solid}, forget plot] coordinates {
        (0.45,0.02049) +- (0.00053118,0.00052234)
        (0.46,0.028084) +- (0.00061832,0.0006096)
        (0.47,0.03835) +- (0.00071777,0.00070931)
        (0.48,0.050763) +- (0.00082051,0.00081237)
        (0.49,0.066005) +- (0.00092687,0.00091867)
        (0.5,0.0845015) +- (0.0007323,0.00072771)
        (0.51,0.10516) +- (0.000806869,0.000802865)
        (0.52,0.12876) +- (0.000956208,0.000951428)
        (0.53,0.154701) +- (0.00109196,0.00108787)
        (0.54,0.182657) +- (0.00132354,0.00131833)
        (0.55,0.212278) +- (0.00144911,0.00144303)
  };
  \addplot[color=red, mark=*, mark size=1pt, solid, mark options={solid}, forget plot] coordinates {
        (0.45,0.0164385) +- (0.00033659,0.00033211)
        (0.46,0.024053) +- (0.00040561,0.00040113)
        (0.47,0.034554) +- (0.000482505,0.000478015)
        (0.48,0.0477225) +- (0.00056238,0.00055877)
        (0.49,0.064497) +- (0.00064821,0.000644785)
        (0.5,0.084347) +- (0.000730985,0.000727025)
        (0.51,0.108725) +- (0.000820815,0.000816868)
        (0.52,0.135506) +- (0.000996049,0.000991457)
        (0.53,0.165892) +- (0.00110415,0.00109923)
        (0.54,0.197981) +- (0.00141891,0.00141232)
        (0.55,0.231134) +- (0.00147969,0.00147531)
  };
  \addplot[color=OliveGreen, mark=*, mark size=1pt, solid, mark options={solid}, forget plot] coordinates {
        (0.45,0.0133045) +- (0.00030345,0.000298975)
        (0.46,0.0207995) +- (0.00037723,0.00037293)
        (0.47,0.031111) +- (0.00045928,0.000454665)
        (0.48,0.044829) +- (0.000545625,0.00054183)
        (0.49,0.0627415) +- (0.000639945,0.00063621)
        (0.5,0.085253) +- (0.00073545,0.000732135)
        (0.51,0.111801) +- (0.000830208,0.000827232)
        (0.52,0.142872) +- (0.00104956,0.00104607)
        (0.53,0.177238) +- (0.00127523,0.00126967)
        (0.54,0.213064) +- (0.00137631,0.00137198)
        (0.55,0.250275) +- (0.0016151,0.00161098)
  };
  \addplot[color=purple!80, mark=*, mark size=1pt, solid, mark options={solid}, forget plot] coordinates {
        (0.45,0.01072) +- (0.00027304,0.000268515)
        (0.46,0.0175915) +- (0.000347765,0.000343405)
        (0.47,0.0277701) +- (0.000615088,0.000606275)
        (0.48,0.041937) +- (0.00074959,0.00074093)
        (0.49,0.060892) +- (0.00089388,0.00088513)
        (0.5,0.085528) +- (0.00104421,0.00103614)
        (0.51,0.114646) +- (0.0011905,0.00118279)
        (0.52,0.149033) +- (0.001563,0.0015546)
        (0.53,0.186671) +- (0.00180024,0.00179275)
        (0.54,0.22674) +- (0.0022483,0.00223827)
        (0.55,0.267985) +- (0.00245015,0.00244082)
  };
  \addplot[color=Purple, mark=*, mark size=1pt, solid, mark options={solid}, forget plot] coordinates {
        (0.45,0.008763) +- (0.0003509,0.00034184)
        (0.46,0.015327) +- (0.00046108,0.00045212)
        (0.47,0.025229) +- (0.00058746,0.0005787)
        (0.48,0.039383) +- (0.00072794,0.00071929)
        (0.49,0.059763) +- (0.0008854,0.00087681)
        (0.5,0.085977) +- (0.00104694,0.00103931)
        (0.51,0.117958) +- (0.00121138,0.0012042)
        (0.52,0.155678) +- (0.00160542,0.00159572)
        (0.53,0.197035) +- (0.00184469,0.0018355)
        (0.54,0.240903) +- (0.00236232,0.00235064)
        (0.55,0.284538) +- (0.00259924,0.00258889)
  };
  \addplot[color=blue, mark=*, mark size=1pt, dashed, mark options={solid}, forget plot] coordinates {
        (0.45,0.28837) +- (0.00268692,0.00267818)
        (0.46,0.317714) +- (0.0028092,0.00279933)
        (0.47,0.345359) +- (0.00296719,0.00295935)
        (0.48,0.371426) +- (0.00332494,0.00331721)
        (0.49,0.394153) +- (0.00341358,0.00340565)
        (0.5,0.417051) +- (0.00358137,0.00357665)
        (0.51,0.433979) +- (0.00362842,0.0036235)
        (0.52,0.448965) +- (0.00366122,0.00365703)
        (0.53,0.461235) +- (0.00370515,0.0037024)
        (0.54,0.472737) +- (0.00386753,0.0038664)
        (0.55,0.479552) +- (0.00384238,0.0038414)
  };
  \addplot[color=red, mark=*, mark size=1pt, dashed, mark options={solid}, forget plot] coordinates {
        (0.45,0.268787) +- (0.00262364,0.00261285)
        (0.46,0.301653) +- (0.00278844,0.00277999)
        (0.47,0.333627) +- (0.0029741,0.00296443)
        (0.48,0.364255) +- (0.00313445,0.00312647)
        (0.49,0.391445) +- (0.00327565,0.00326889)
        (0.5,0.416818) +- (0.00337457,0.00336945)
        (0.51,0.435673) +- (0.0036682,0.00366318)
        (0.52,0.452326) +- (0.00375278,0.00374949)
        (0.53,0.466402) +- (0.00380415,0.0038031)
        (0.54,0.475601) +- (0.00383969,0.00383862)
        (0.55,0.483174) +- (0.00391295,0.0039114)
  };
  \addplot[color=OliveGreen, mark=*, mark size=1pt, dashed, mark options={solid}, forget plot] coordinates {
        (0.45,0.24943) +- (0.00231341,0.00230419)
        (0.46,0.286994) +- (0.00260678,0.00259703)
        (0.47,0.322986) +- (0.00279624,0.00278811)
        (0.48,0.357849) +- (0.00317182,0.00316355)
        (0.49,0.387755) +- (0.00325722,0.00325127)
        (0.5,0.416076) +- (0.00346727,0.00346173)
        (0.51,0.437439) +- (0.00372126,0.00371655)
        (0.52,0.453101) +- (0.0036053,0.00360238)
        (0.53,0.469266) +- (0.00365128,0.00364823)
        (0.54,0.479379) +- (0.00372042,0.00371873)
        (0.55,0.489217) +- (0.00373918,0.00373761)
  };
  \addplot[color=purple!80, mark=*, mark size=1pt, dashed, mark options={solid}, forget plot] coordinates {
        (0.45,0.231849) +- (0.00223723,0.00222612)
        (0.46,0.272439) +- (0.00248057,0.002471)
        (0.47,0.31384) +- (0.00293416,0.00292404)
        (0.48,0.351487) +- (0.00315358,0.00314459)
        (0.49,0.383868) +- (0.00313346,0.0031273)
        (0.5,0.414868) +- (0.00356838,0.00356168)
        (0.51,0.438735) +- (0.00373253,0.00372807)
        (0.52,0.454736) +- (0.00353673,0.00353301)
        (0.53,0.472418) +- (0.00389418,0.00389199)
        (0.54,0.482471) +- (0.00366805,0.00366642)
        (0.55,0.489949) +- (0.00374052,0.0037387)
  };
  \addplot[color=Purple, mark=*, mark size=1pt, dashed, mark options={solid}, forget plot] coordinates {
        (0.45,0.21458) +- (0.0019877,0.00197899)
        (0.46,0.260713) +- (0.00248466,0.00247442)
        (0.47,0.303407) +- (0.0027313,0.00272304)
        (0.48,0.342766) +- (0.00299908,0.00299136)
        (0.49,0.381246) +- (0.00312698,0.00312127)
        (0.5,0.414991) +- (0.00334014,0.00333503)
        (0.51,0.441154) +- (0.00350067,0.00349619)
        (0.52,0.459597) +- (0.00358813,0.00358479)
        (0.53,0.47398) +- (0.00368245,0.0036808)
        (0.54,0.484871) +- (0.0038593,0.00385831)
        (0.55,0.491303) +- (0.00389613,0.00389455)
  };

\nextgroupplot[title={\textsc{$5$-Chain}}, xshift={0.5*(0.290\linewidth+8pt)}, xmin=0.245, xmax=0.355, ymin=9.84e-05, ymax=0.753892]
  \addplot[color=black, dotted, mark=none, forget plot] coordinates {(0.315593,9.84e-05) (0.315593,0.046059)};
  \addplot[color=black, dotted, mark=none, forget plot] coordinates {(0.314526,9.84e-05) (0.314526,0.43067)};
  \addplot[color=blue, mark=*, mark size=1pt, solid, mark options={solid}, forget plot] coordinates {
        (0.25,0.001708) +- (0.00015812,0.00014893)
        (0.26,0.003127) +- (0.00021212,0.00020297)
        (0.27,0.005738) +- (0.00028532,0.00027621)
        (0.28,0.009844) +- (0.00037141,0.00036239)
        (0.29,0.01607) +- (0.00047206,0.00046313)
        (0.3,0.025252) +- (0.00058779,0.00057909)
        (0.31,0.03798) +- (0.000715,0.00070648)
        (0.32,0.0539) +- (0.00084334,0.00083459)
        (0.33,0.074258) +- (0.00097886,0.00097116)
        (0.34,0.099011) +- (0.00111528,0.00110746)
        (0.35,0.127076) +- (0.00132671,0.00131945)
  };
  \addplot[color=red, mark=*, mark size=1pt, solid, mark options={solid}, forget plot] coordinates {
        (0.25,0.000937) +- (0.00011837,0.00010918)
        (0.26,0.002099) +- (0.00017474,0.00016557)
        (0.27,0.004184) +- (0.00024449,0.00023536)
        (0.28,0.007842) +- (0.00033245,0.00032339)
        (0.29,0.01389) +- (0.00043963,0.00043066)
        (0.3,0.022939) +- (0.00056085,0.00055213)
        (0.31,0.037113) +- (0.00070724,0.00069877)
        (0.32,0.054905) +- (0.00085167,0.00084324)
        (0.33,0.07901) +- (0.00100907,0.00100105)
        (0.34,0.107948) +- (0.00116102,0.00115404)
        (0.35,0.141364) +- (0.0014723,0.00146396)
  };
  \addplot[color=OliveGreen, mark=*, mark size=1pt, solid, mark options={solid}, forget plot] coordinates {
        (0.25,0.000533) +- (9.045e-05,8.125e-05)
        (0.26,0.001352) +- (0.00014121,0.00013203)
        (0.27,0.003113) +- (0.00021166,0.00020251)
        (0.28,0.006165) +- (0.00029553,0.00028645)
        (0.29,0.011784) +- (0.00040549,0.00039656)
        (0.3,0.0212) +- (0.00053978,0.00053102)
        (0.31,0.035722) +- (0.00069475,0.00068602)
        (0.32,0.056541) +- (0.00086276,0.00085482)
        (0.33,0.082822) +- (0.00103038,0.00102208)
        (0.34,0.116771) +- (0.00120217,0.00119561)
        (0.35,0.154904) +- (0.00150386,0.00149668)
  };
  \addplot[color=purple!80, mark=*, mark size=1pt, solid, mark options={solid}, forget plot] coordinates {
        (0.25,0.000319) +- (7.106e-05,6.186e-05)
        (0.26,0.000859) +- (0.00011354,0.00010434)
        (0.27,0.00215) +- (0.00017678,0.00016762)
        (0.28,0.004773) +- (0.00026079,0.00025166)
        (0.29,0.010227) +- (0.00037843,0.00036944)
        (0.3,0.019584) +- (0.00051956,0.00051075)
        (0.31,0.035243) +- (0.0006898,0.00068124)
        (0.32,0.057002) +- (0.00086669,0.00085863)
        (0.33,0.087675) +- (0.00105717,0.00104944)
        (0.34,0.125425) +- (0.00131541,0.00130715)
        (0.35,0.169351) +- (0.00157629,0.00156823)
  };
  \addplot[color=Purple, mark=*, mark size=1pt, solid, mark options={solid}, forget plot] coordinates {
        (0.25,0.000195) +- (5.66e-05,4.74e-05)
        (0.26,0.000563) +- (9.283e-05,8.363e-05)
        (0.27,0.001569) +- (0.00015175,0.00014256)
        (0.28,0.003785) +- (0.00023284,0.00022371)
        (0.29,0.008755) +- (0.00035083,0.0003418)
        (0.3,0.017955) +- (0.00049797,0.00048909)
        (0.31,0.033618) +- (0.00067443,0.0006658)
        (0.32,0.057983) +- (0.00087308,0.00086441)
        (0.33,0.091479) +- (0.001077,0.00106946)
        (0.34,0.134008) +- (0.00129149,0.00128489)
        (0.35,0.182979) +- (0.00176556,0.00175571)
  };
  \addplot[color=blue, mark=*, mark size=1pt, dashed, mark options={solid}, forget plot] coordinates {
        (0.25,0.198936) +- (0.00196888,0.00195945)
        (0.26,0.240908) +- (0.00220214,0.00219278)
        (0.27,0.282454) +- (0.00264161,0.00263146)
        (0.28,0.323518) +- (0.00291004,0.00290021)
        (0.29,0.360905) +- (0.00305658,0.00304902)
        (0.3,0.393149) +- (0.00341748,0.00341123)
        (0.31,0.419684) +- (0.00357823,0.00357307)
        (0.32,0.443932) +- (0.00371742,0.00371386)
        (0.33,0.461602) +- (0.0037869,0.00378519)
        (0.34,0.474085) +- (0.00382993,0.00382874)
        (0.35,0.484273) +- (0.00373579,0.00373477)
  };
  \addplot[color=red, mark=*, mark size=1pt, dashed, mark options={solid}, forget plot] coordinates {
        (0.25,0.169186) +- (0.00163157,0.0016231)
        (0.26,0.214328) +- (0.00210779,0.00209783)
        (0.27,0.261637) +- (0.00236469,0.00235478)
        (0.28,0.307275) +- (0.00278122,0.00276987)
        (0.29,0.35103) +- (0.00304334,0.00303576)
        (0.3,0.390904) +- (0.00328556,0.00327876)
        (0.31,0.420824) +- (0.00362838,0.00362288)
        (0.32,0.445136) +- (0.00374196,0.00373794)
        (0.33,0.462897) +- (0.00375609,0.00375379)
        (0.34,0.478664) +- (0.00381016,0.00380806)
        (0.35,0.48651) +- (0.00386067,0.00386019)
  };
  \addplot[color=OliveGreen, mark=*, mark size=1pt, dashed, mark options={solid}, forget plot] coordinates {
        (0.25,0.142165) +- (0.0015225,0.00151355)
        (0.26,0.188494) +- (0.00177704,0.00176802)
        (0.27,0.239531) +- (0.00226294,0.00225329)
        (0.28,0.291843) +- (0.0027594,0.00274986)
        (0.29,0.339877) +- (0.00306158,0.00305276)
        (0.3,0.381176) +- (0.00322748,0.00322021)
        (0.31,0.418058) +- (0.00361255,0.00360501)
        (0.32,0.445783) +- (0.00369764,0.00369448)
        (0.33,0.466255) +- (0.00380512,0.00380184)
        (0.34,0.479017) +- (0.00394273,0.00394102)
        (0.35,0.487502) +- (0.00370621,0.00370609)
  };
  \addplot[color=purple!80, mark=*, mark size=1pt, dashed, mark options={solid}, forget plot] coordinates {
        (0.25,0.118572) +- (0.00121548,0.00120825)
        (0.26,0.165206) +- (0.00165252,0.00164486)
        (0.27,0.217808) +- (0.0021442,0.00213431)
        (0.28,0.274205) +- (0.00263188,0.00262027)
        (0.29,0.329489) +- (0.00301739,0.00300888)
        (0.3,0.377483) +- (0.00319525,0.00318829)
        (0.31,0.417079) +- (0.00356812,0.00356262)
        (0.32,0.448425) +- (0.00371842,0.00371523)
        (0.33,0.470289) +- (0.00384734,0.00384476)
        (0.34,0.482185) +- (0.00395989,0.00395835)
        (0.35,0.491518) +- (0.003666,0.00366488)
  };
  \addplot[color=Purple, mark=*, mark size=1pt, dashed, mark options={solid}, forget plot] coordinates {
        (0.25,0.097397) +- (0.00110598,0.00109866)
        (0.26,0.143735) +- (0.00149772,0.00148888)
        (0.27,0.198599) +- (0.00185654,0.00184854)
        (0.28,0.257929) +- (0.00244939,0.00243939)
        (0.29,0.318311) +- (0.00274402,0.00273698)
        (0.3,0.37143) +- (0.00331476,0.00330634)
        (0.31,0.41534) +- (0.00355162,0.00354573)
        (0.32,0.448034) +- (0.00372137,0.00371845)
        (0.33,0.470931) +- (0.00383087,0.00382823)
        (0.34,0.483266) +- (0.00395511,0.00395366)
        (0.35,0.491908) +- (0.00384408,0.00384348)
  };

\nextgroupplot[title={\textsc{$8$-Chain}}, xshift={0.5*(0.290\linewidth+8pt)}, xmin=0.345, xmax=0.455, ymin=9.84e-05, ymax=0.753892]
  \addplot[color=black, dotted, mark=none, forget plot] coordinates {(0.379839,9.84e-05) (0.379839,0.0638119)};
  \addplot[color=black, dotted, mark=none, forget plot] coordinates {(0.379386,9.84e-05) (0.379386,0.416069)};
  \addplot[color=blue, mark=*, mark size=1pt, solid, mark options={solid}, forget plot] coordinates {
        (0.35,0.022847) +- (0.00055994,0.00055118)
        (0.36,0.033294) +- (0.00067097,0.00066237)
        (0.37,0.047102) +- (0.00079214,0.0007834)
        (0.38,0.063897) +- (0.00091319,0.00090578)
        (0.39,0.084644) +- (0.00103885,0.00103118)
        (0.4,0.108579) +- (0.0011636,0.00115625)
        (0.41,0.135717) +- (0.00137739,0.00136938)
        (0.42,0.165905) +- (0.00170263,0.0016939)
        (0.43,0.197457) +- (0.00183185,0.00182404)
        (0.44,0.229388) +- (0.00226882,0.00225829)
        (0.45,0.261017) +- (0.00238782,0.00237721)
  };
  \addplot[color=red, mark=*, mark size=1pt, solid, mark options={solid}, forget plot] coordinates {
        (0.35,0.019921) +- (0.00052389,0.00051499)
        (0.36,0.030697) +- (0.00064559,0.00063685)
        (0.37,0.044954) +- (0.00077456,0.00076613)
        (0.38,0.06466) +- (0.00091869,0.00091069)
        (0.39,0.087172) +- (0.00105247,0.00104508)
        (0.4,0.11586) +- (0.00120236,0.00119533)
        (0.41,0.147044) +- (0.00154517,0.00153576)
        (0.42,0.182136) +- (0.00168705,0.00167936)
        (0.43,0.217966) +- (0.00215983,0.00214927)
        (0.44,0.256271) +- (0.00244703,0.00243608)
        (0.45,0.28951) +- (0.00256389,0.00255428)
  };
  \addplot[color=OliveGreen, mark=*, mark size=1pt, solid, mark options={solid}, forget plot] coordinates {
        (0.35,0.01718) +- (0.00048749,0.00047863)
        (0.36,0.027999) +- (0.00061766,0.00060899)
        (0.37,0.043606) +- (0.00076338,0.00075539)
        (0.38,0.06425) +- (0.00091522,0.00090713)
        (0.39,0.090536) +- (0.00107168,0.00106401)
        (0.4,0.122081) +- (0.00125585,0.00124933)
        (0.41,0.158462) +- (0.00166063,0.00165171)
        (0.42,0.198614) +- (0.00194573,0.00193616)
        (0.43,0.239678) +- (0.00232456,0.00231571)
        (0.44,0.280422) +- (0.00264776,0.00263785)
        (0.45,0.31749) +- (0.00283085,0.00282214)
  };
  \addplot[color=purple!80, mark=*, mark size=1pt, solid, mark options={solid}, forget plot] coordinates {
        (0.35,0.014922) +- (0.00045525,0.00044632)
        (0.36,0.025559) +- (0.00059104,0.00058217)
        (0.37,0.041629) +- (0.0007471,0.00073917)
        (0.38,0.064143) +- (0.00091548,0.00090737)
        (0.39,0.092891) +- (0.00108383,0.00107583)
        (0.4,0.129205) +- (0.00131166,0.00130371)
        (0.41,0.169726) +- (0.00177284,0.00176261)
        (0.42,0.212475) +- (0.00196423,0.00195573)
        (0.43,0.258039) +- (0.00244909,0.00243751)
        (0.44,0.302456) +- (0.00265157,0.0026434)
        (0.45,0.341603) +- (0.00286096,0.00285393)
  };
  \addplot[color=Purple, mark=*, mark size=1pt, solid, mark options={solid}, forget plot] coordinates {
        (0.35,0.012965) +- (0.00042489,0.0004159)
        (0.36,0.02359) +- (0.00056839,0.0005596)
        (0.37,0.040481) +- (0.0007367,0.00072813)
        (0.38,0.064123) +- (0.00091517,0.00090752)
        (0.39,0.095835) +- (0.00109811,0.00109114)
        (0.4,0.135057) +- (0.00129941,0.00129201)
        (0.41,0.180178) +- (0.00177677,0.00176718)
        (0.42,0.228379) +- (0.00206491,0.00205559)
        (0.43,0.277354) +- (0.00261016,0.00260063)
        (0.44,0.324961) +- (0.00285451,0.00284528)
        (0.45,0.363586) +- (0.00305999,0.00305287)
  };
  \addplot[color=blue, mark=*, mark size=1pt, dashed, mark options={solid}, forget plot] coordinates {
        (0.35,0.333196) +- (0.00301225,0.00300222)
        (0.36,0.364571) +- (0.00318704,0.00317888)
        (0.37,0.393429) +- (0.00336181,0.00335535)
        (0.38,0.419012) +- (0.00343609,0.00343164)
        (0.39,0.437423) +- (0.00354453,0.00353948)
        (0.4,0.454845) +- (0.00357637,0.00357331)
        (0.41,0.466763) +- (0.00384344,0.00384065)
        (0.42,0.477338) +- (0.00383493,0.00383301)
        (0.43,0.48476) +- (0.00385706,0.00385521)
        (0.44,0.488458) +- (0.00382467,0.0038226)
        (0.45,0.49526) +- (0.00380634,0.00380672)
  };
  \addplot[color=red, mark=*, mark size=1pt, dashed, mark options={solid}, forget plot] coordinates {
        (0.35,0.32266) +- (0.00282296,0.00281398)
        (0.36,0.356302) +- (0.00320607,0.00319724)
        (0.37,0.389297) +- (0.00332084,0.00331278)
        (0.38,0.418308) +- (0.00345292,0.00344676)
        (0.39,0.441489) +- (0.00354061,0.00353627)
        (0.4,0.456733) +- (0.00359277,0.00359032)
        (0.41,0.471702) +- (0.00374454,0.0037423)
        (0.42,0.481795) +- (0.00374181,0.00374019)
        (0.43,0.489635) +- (0.0037877,0.00378608)
        (0.44,0.492596) +- (0.00379813,0.00379738)
        (0.45,0.493803) +- (0.003797,0.00379633)
  };
  \addplot[color=OliveGreen, mark=*, mark size=1pt, dashed, mark options={solid}, forget plot] coordinates {
        (0.35,0.309354) +- (0.00274814,0.00273934)
        (0.36,0.351102) +- (0.00317908,0.00317027)
        (0.37,0.387763) +- (0.00335589,0.00334893)
        (0.38,0.418358) +- (0.00347195,0.00346636)
        (0.39,0.443929) +- (0.00361676,0.00361362)
        (0.4,0.461276) +- (0.00370769,0.00370445)
        (0.41,0.473073) +- (0.00377843,0.00377669)
        (0.42,0.485499) +- (0.00378961,0.00378786)
        (0.43,0.488029) +- (0.00383253,0.00383193)
        (0.44,0.494588) +- (0.00389127,0.00389039)
        (0.45,0.497985) +- (0.0039635,0.00396418)
  };
  \addplot[color=purple!80, mark=*, mark size=1pt, dashed, mark options={solid}, forget plot] coordinates {
        (0.35,0.297583) +- (0.00280037,0.00279087)
        (0.36,0.340829) +- (0.00305654,0.00304784)
        (0.37,0.384366) +- (0.00313196,0.00312532)
        (0.38,0.416313) +- (0.00328265,0.00327714)
        (0.39,0.444074) +- (0.00367723,0.0036724)
        (0.4,0.462715) +- (0.00378225,0.00377975)
        (0.41,0.477755) +- (0.00365543,0.00365477)
        (0.42,0.486429) +- (0.00372522,0.00372369)
        (0.43,0.493142) +- (0.00394774,0.0039467)
        (0.44,0.497626) +- (0.00397608,0.00397635)
        (0.45,0.497432) +- (0.0036562,0.00365597)
  };
  \addplot[color=Purple, mark=*, mark size=1pt, dashed, mark options={solid}, forget plot] coordinates {
        (0.35,0.285571) +- (0.0026327,0.00262333)
        (0.36,0.333947) +- (0.00288413,0.00287498)
        (0.37,0.380528) +- (0.00336262,0.00335507)
        (0.38,0.418222) +- (0.00338823,0.00338257)
        (0.39,0.4491) +- (0.00356802,0.00356427)
        (0.4,0.466933) +- (0.00370971,0.0037072)
        (0.41,0.479692) +- (0.00386287,0.00386019)
        (0.42,0.489694) +- (0.00390825,0.00390701)
        (0.43,0.494538) +- (0.00399418,0.00399399)
        (0.44,0.496674) +- (0.0036418,0.00364211)
        (0.45,0.498843) +- (0.0037513,0.00375142)
  };

\end{groupplot}

\begin{axis}[
  hide axis, overlay, at={(groupmasterthresholdpmcorrmergeZZobsZZXX c1r1.center)}, anchor=center,
  xmin=0, xmax=1, ymin=0, ymax=1,
  legend to name=stylelegend_masterthresholdpmcorrmergeZZobsZZXX, legend columns=1, transpose legend,
  legend cell align=left,
  legend style={font=\footnotesize, draw=none, /tikz/every even column/.append style={column sep=8pt}},
]
  \addlegendimage{black, solid, line legend, mark=none}
  \addlegendentry{$\bar{Z}\bar{Z}$}
  \addlegendimage{black, dashed, line legend, mark=none}
  \addlegendentry{$\bar{X}\bar{X}$}
\end{axis}

\node[draw, inner sep=3pt, align=center, anchor=west, xshift=6pt] at (groupmasterthresholdpmcorrmergeZZobsZZXX c2r2.east)
  {\pgfplotslegendfromname{dlegend_masterthresholdpmcorrmergeZZobsZZXX}\\[4pt]\pgfplotslegendfromname{stylelegend_masterthresholdpmcorrmergeZZobsZZXX}};
\end{tikzpicture}

%% file: results/prep_and_erasure_correctable_regime_pm_corr_mergeZZ_obsZZ.tex
\begin{tikzpicture}
\begin{axis}[width=\linewidth,
  xlabel={Resource state error $p_{r}$ (\%)},
  ylabel={Erasure threshold $p^{*}_{e}$},
  grid=both,
  minor grid style={gray!15},
  major grid style={gray!30},
  legend pos=north east,
  legend cell align=left,
  legend style={font=\footnotesize},
  every axis plot/.append style={thick},
  ymin=0,
  ymax=0.55,
  restrict y to domain=0:0.55,
  xmin=0,
  xmax=6.70184,
  xtick distance=1,
  scaled x ticks=false,
  scaled y ticks=false,
  y tick label style={/pgf/number format/fixed},
  x tick label style={/pgf/number format/fixed, /pgf/number format/precision=2}]
  \addplot[color=blue, mark=diamond*, name path=upper] coordinates {
        (0,0.503124)
        (1,0.437736)
        (2,0.353972)
        (3,0.283841)
        (4,0.216111)
        (5,0.142889)
        (6,0.054234)
  };
  \addlegendentry{\textsc{Bell}}
  \path[name path=lower] (axis cs:0,0) -- (axis cs:6,0);
  \addplot[opacity=0.2, blue, forget plot] fill between [of=upper and lower];
  \addplot[color=blue, mark=diamond*, dashed, mark=none, name path=upper, forget plot] coordinates {
        (6,0.054234)
        (6.70184,0)
  };
  \path[name path=lower] (axis cs:6,0) -- (axis cs:6.70184,0);
  \addplot[opacity=0.2, blue, forget plot] fill between [of=upper and lower];
  \addplot[only marks, mark=diamond*, color=blue, forget plot, error bars/.cd, y dir=both, y explicit] coordinates {
        (0,0.503124) +- (0,0.000222365)
        (1,0.437736) +- (0,0.00207896)
        (2,0.353972) +- (0,0.0023652)
        (3,0.283841) +- (0,0.00231471)
        (4,0.216111) +- (0,0.00263792)
        (5,0.142889) +- (0,0.00472009)
        (6,0.054234) +- (0,0.00482193)
  };
  \addplot[color=red, mark=diamond, name path=upper] coordinates {
        (0,0.494023)
        (1,0.418874)
        (2,0.334352)
        (3,0.25663)
        (4,0.181332)
        (5,0.0993417)
  };
  \addlegendentry{\textsc{BellZigZag}}
  \path[name path=lower] (axis cs:0,0) -- (axis cs:5,0);
  \addplot[opacity=0.2, red, forget plot] fill between [of=upper and lower];
  \addplot[color=red, mark=diamond, dashed, mark=none, name path=upper, forget plot] coordinates {
        (5,0.0993417)
        (6,0)
  };
  \path[name path=lower] (axis cs:5,0) -- (axis cs:6,0);
  \addplot[opacity=0.2, red, forget plot] fill between [of=upper and lower];
  \addplot[only marks, mark=diamond, color=red, forget plot, error bars/.cd, y dir=both, y explicit] coordinates {
        (0,0.494023) +- (0,0.000350037)
        (1,0.418874) +- (0,0.00108516)
        (2,0.334352) +- (0,0.00159858)
        (3,0.25663) +- (0,0.00162328)
        (4,0.181332) +- (0,0.00183628)
        (5,0.0993417) +- (0,0.00139924)
  };
  \addplot[color=OliveGreen, mark=*, name path=upper] coordinates {
        (0,0.315593)
        (1,0.236262)
        (2,0.163509)
        (3,0.0916307)
        (4,0.0188154)
  };
  \addlegendentry{\textsc{$5$-Chain}}
  \path[name path=lower] (axis cs:0,0) -- (axis cs:4,0);
  \addplot[opacity=0.2, OliveGreen, forget plot] fill between [of=upper and lower];
  \addplot[color=OliveGreen, mark=*, dashed, mark=none, name path=upper, forget plot] coordinates {
        (4,0.0188154)
        (4.26223,0)
  };
  \path[name path=lower] (axis cs:4,0) -- (axis cs:4.26223,0);
  \addplot[opacity=0.2, OliveGreen, forget plot] fill between [of=upper and lower];
  \addplot[only marks, mark=*, color=OliveGreen, forget plot, error bars/.cd, y dir=both, y explicit] coordinates {
        (0,0.315593) +- (0,0.000231542)
        (1,0.236262) +- (0,0.0018924)
        (2,0.163509) +- (0,0.00141826)
        (3,0.0916307) +- (0,0.00158526)
        (4,0.0188154) +- (0,0.00135609)
  };
  \addplot[color=orange, mark=square*, name path=upper] coordinates {
        (0,0.379839)
        (1,0.304136)
        (2,0.2267)
        (3,0.153996)
        (4,0.0836053)
        (5,0.0050372)
  };
  \addlegendentry{\textsc{$8$-Chain}}
  \path[name path=lower] (axis cs:0,0) -- (axis cs:5,0);
  \addplot[opacity=0.2, orange, forget plot] fill between [of=upper and lower];
  \addplot[color=orange, mark=square*, dashed, mark=none, name path=upper, forget plot] coordinates {
        (5,0.0050372)
        (5.08593,0)
  };
  \path[name path=lower] (axis cs:5,0) -- (axis cs:5.08593,0);
  \addplot[opacity=0.2, orange, forget plot] fill between [of=upper and lower];
  \addplot[only marks, mark=square*, color=orange, forget plot, error bars/.cd, y dir=both, y explicit] coordinates {
        (0,0.379839) +- (0,0.000202592)
        (1,0.304136) +- (0,0.00202919)
        (2,0.2267) +- (0,0.00137711)
        (3,0.153996) +- (0,0.00321584)
        (4,0.0836053) +- (0,0.00134186)
        (5,0.0050372) +- (0,0.00290278)
  };
  \addplot[color=Purple, mark=triangle*, name path=upper] coordinates {
        (0,0.498389)
        (0.5,0.462756)
        (1,0.426555)
        (2,0.349401)
        (3,0.272535)
        (4,0.195806)
        (5,0.117911)
        (6,0.0339898)
  };
  \addlegendentry{$3d$-\textsc{Chain}}
  \path[name path=lower] (axis cs:0,0) -- (axis cs:6,0);
  \addplot[opacity=0.2, Purple, forget plot] fill between [of=upper and lower];
  \addplot[color=Purple, mark=triangle*, dashed, mark=none, name path=upper, forget plot] coordinates {
        (6,0.0339898)
        (6.43252,0)
  };
  \path[name path=lower] (axis cs:6,0) -- (axis cs:6.43252,0);
  \addplot[opacity=0.2, Purple, forget plot] fill between [of=upper and lower];
  \addplot[only marks, mark=triangle*, color=Purple, forget plot, error bars/.cd, y dir=both, y explicit] coordinates {
        (0,0.498389) +- (0,0.000192322)
        (0.5,0.462756) +- (0,0.00166477)
        (1,0.426555) +- (0,0.00130576)
        (2,0.349401) +- (0,0.00163681)
        (3,0.272535) +- (0,0.00197919)
        (4,0.195806) +- (0,0.00252378)
        (5,0.117911) +- (0,0.00223746)
        (6,0.0339898) +- (0,0.00274112)
  };
\end{axis}
\end{tikzpicture}

%% file: results/prep_and_erasure_correctable_regime_pm_corr_mergeZZ_obsXX.tex
\begin{tikzpicture}
\begin{axis}[width=\linewidth,
  xlabel={Resource state error $p_{r}$ (\%)},
  ylabel={Erasure threshold $p^{*}_{e}$},
  grid=both,
  minor grid style={gray!15},
  major grid style={gray!30},
  legend pos=north east,
  legend cell align=left,
  legend style={font=\footnotesize},
  every axis plot/.append style={thick},
  ymin=0,
  ymax=0.55,
  restrict y to domain=0:0.55,
  xmin=0,
  xmax=6.56771,
  xtick distance=1,
  scaled x ticks=false,
  scaled y ticks=false,
  y tick label style={/pgf/number format/fixed},
  x tick label style={/pgf/number format/fixed, /pgf/number format/precision=2}]
  \addplot[color=blue, mark=diamond*, name path=upper] coordinates {
        (0,0.496441)
        (1,0.428303)
        (2,0.346789)
        (3,0.261659)
        (4,0.171322)
        (5,0.0807845)
  };
  \addlegendentry{\textsc{Bell}}
  \path[name path=lower] (axis cs:0,0) -- (axis cs:5,0);
  \addplot[opacity=0.2, blue, forget plot] fill between [of=upper and lower];
  \addplot[color=blue, mark=diamond*, dashed, mark=none, name path=upper, forget plot] coordinates {
        (5,0.0807845)
        (5.89363,0)
  };
  \path[name path=lower] (axis cs:5,0) -- (axis cs:5.89363,0);
  \addplot[opacity=0.2, blue, forget plot] fill between [of=upper and lower];
  \addplot[only marks, mark=diamond*, color=blue, forget plot, error bars/.cd, y dir=both, y explicit] coordinates {
        (0,0.496441) +- (0,0.0008216)
        (1,0.428303) +- (0,0.00285138)
        (2,0.346789) +- (0,0.00198683)
        (3,0.261659) +- (0,0.00195985)
        (4,0.171322) +- (0,0.00229792)
        (5,0.0807845) +- (0,0.0022855)
  };
  \addplot[color=red, mark=diamond, name path=upper] coordinates {
        (0,0.50595)
        (1,0.445861)
        (2,0.363055)
        (3,0.280952)
        (4,0.205033)
        (5,0.118911)
        (6,0.0315333)
  };
  \addlegendentry{\textsc{BellZigZag}}
  \path[name path=lower] (axis cs:0,0) -- (axis cs:6,0);
  \addplot[opacity=0.2, red, forget plot] fill between [of=upper and lower];
  \addplot[color=red, mark=diamond, dashed, mark=none, name path=upper, forget plot] coordinates {
        (6,0.0315333)
        (6.36591,0)
  };
  \path[name path=lower] (axis cs:6,0) -- (axis cs:6.36591,0);
  \addplot[opacity=0.2, red, forget plot] fill between [of=upper and lower];
  \addplot[only marks, mark=diamond, color=red, forget plot, error bars/.cd, y dir=both, y explicit] coordinates {
        (0,0.50595) +- (0,0.000364)
        (1,0.445861) +- (0,0.00217485)
        (2,0.363055) +- (0,0.00298458)
        (3,0.280952) +- (0,0.00207426)
        (4,0.205033) +- (0,0.00304659)
        (5,0.118911) +- (0,0.00219994)
        (6,0.0315333) +- (0,0.00459037)
  };
  \addplot[color=OliveGreen, mark=*, name path=upper] coordinates {
        (0,0.314526)
        (1,0.251068)
        (2,0.165328)
        (3,0.0912901)
        (4,0.0175599)
  };
  \addlegendentry{\textsc{$5$-Chain}}
  \path[name path=lower] (axis cs:0,0) -- (axis cs:4,0);
  \addplot[opacity=0.2, OliveGreen, forget plot] fill between [of=upper and lower];
  \addplot[color=OliveGreen, mark=*, dashed, mark=none, name path=upper, forget plot] coordinates {
        (4,0.0175599)
        (4.23698,0)
  };
  \path[name path=lower] (axis cs:4,0) -- (axis cs:4.23698,0);
  \addplot[opacity=0.2, OliveGreen, forget plot] fill between [of=upper and lower];
  \addplot[only marks, mark=*, color=OliveGreen, forget plot, error bars/.cd, y dir=both, y explicit] coordinates {
        (0,0.314526) +- (0,0.000740173)
        (1,0.251068) +- (0,0.00629696)
        (2,0.165328) +- (0,0.00282363)
        (3,0.0912901) +- (0,0.00257494)
        (4,0.0175599) +- (0,0.00236554)
  };
  \addplot[color=orange, mark=square*, name path=upper] coordinates {
        (0,0.379386)
        (1,0.320702)
        (2,0.236668)
        (3,0.153665)
        (4,0.0830921)
        (5,0.0117615)
  };
  \addlegendentry{\textsc{$8$-Chain}}
  \path[name path=lower] (axis cs:0,0) -- (axis cs:5,0);
  \addplot[opacity=0.2, orange, forget plot] fill between [of=upper and lower];
  \addplot[color=orange, mark=square*, dashed, mark=none, name path=upper, forget plot] coordinates {
        (5,0.0117615)
        (5.16755,0)
  };
  \path[name path=lower] (axis cs:5,0) -- (axis cs:5.16755,0);
  \addplot[opacity=0.2, orange, forget plot] fill between [of=upper and lower];
  \addplot[only marks, mark=square*, color=orange, forget plot, error bars/.cd, y dir=both, y explicit] coordinates {
        (0,0.379386) +- (0,0.000760846)
        (1,0.320702) +- (0,0.00268715)
        (2,0.236668) +- (0,0.00292564)
        (3,0.153665) +- (0,0.00221495)
        (4,0.0830921) +- (0,0.00405148)
        (5,0.0117615) +- (0,0.00149215)
  };
  \addplot[color=Purple, mark=triangle*, name path=upper] coordinates {
        (0,0.501818)
        (0.5,0.468771)
        (1,0.440933)
        (2,0.361839)
        (3,0.279223)
        (4,0.205753)
        (5,0.127082)
        (6,0.0450553)
  };
  \addlegendentry{$3d$-\textsc{Chain}}
  \path[name path=lower] (axis cs:0,0) -- (axis cs:6,0);
  \addplot[opacity=0.2, Purple, forget plot] fill between [of=upper and lower];
  \addplot[color=Purple, mark=triangle*, dashed, mark=none, name path=upper, forget plot] coordinates {
        (6,0.0450553)
        (6.56771,0)
  };
  \path[name path=lower] (axis cs:6,0) -- (axis cs:6.56771,0);
  \addplot[opacity=0.2, Purple, forget plot] fill between [of=upper and lower];
  \addplot[only marks, mark=triangle*, color=Purple, forget plot, error bars/.cd, y dir=both, y explicit] coordinates {
        (0,0.501818) +- (0,0.00087187)
        (0.5,0.468771) +- (0,0.00409513)
        (1,0.440933) +- (0,0.00410908)
        (2,0.361839) +- (0,0.00287129)
        (3,0.279223) +- (0,0.0029916)
        (4,0.205753) +- (0,0.00202032)
        (5,0.127082) +- (0,0.00371749)
        (6,0.0450553) +- (0,0.00323354)
  };
\end{axis}
\end{tikzpicture}

%% file: results/circuit_and_erasure_correctable_regime_pm_corr_mergeZZ_obsZZ.tex
\begin{tikzpicture}
\begin{axis}[width=\linewidth,
  xlabel={Circuit noise probability $p_c$ (\%)},
  ylabel={Erasure threshold $p^{*}_{e}$},
  grid=both,
  minor grid style={gray!15},
  major grid style={gray!30},
  legend pos=south west,
  legend cell align=left,
  legend style={font=\footnotesize},
  every axis plot/.append style={thick},
  ymin=0,
  ymax=0.55,
  restrict y to domain=0:0.55,
  xmin=0,
  xmax=1,
  xtick distance=0.1,
  scaled x ticks=false,
  scaled y ticks=false,
  y tick label style={/pgf/number format/fixed},
  x tick label style={/pgf/number format/fixed, /pgf/number format/precision=2}]
  \addplot[color=blue, mark=diamond*, name path=upper] coordinates {
        (0,0.503124)
        (0.1,0.490063)
        (0.2,0.470375)
        (0.3,0.446081)
        (0.4,0.42282)
        (0.5,0.403995)
        (0.6,0.391125)
        (0.7,0.37695)
        (0.8,0.365954)
        (0.9,0.343248)
        (1,0.284221)
  };
  \addlegendentry{\textsc{Bell}}
  \path[name path=lower] (axis cs:0,0) -- (axis cs:1,0);
  \addplot[opacity=0.2, blue, forget plot] fill between [of=upper and lower];
  \addplot[only marks, mark=diamond*, color=blue, forget plot, error bars/.cd, y dir=both, y explicit] coordinates {
        (0,0.503124) +- (0,0.000222365)
        (0.1,0.490063) +- (0,0.000610627)
        (0.2,0.470375) +- (0,0.00106437)
        (0.3,0.446081) +- (0,0.000967547)
        (0.4,0.42282) +- (0,0.00143448)
        (0.5,0.403995) +- (0,0.00167178)
        (0.6,0.391125) +- (0,0.00194061)
        (0.7,0.37695) +- (0,0.00148851)
        (0.8,0.365954) +- (0,0.00125425)
        (0.9,0.343248) +- (0,0.0016983)
        (1,0.284221) +- (0,0.00270388)
  };
  \addplot[color=red, mark=diamond, name path=upper] coordinates {
        (0,0.494023)
        (0.1,0.472429)
        (0.2,0.449285)
        (0.3,0.423783)
        (0.4,0.394633)
        (0.5,0.369127)
        (0.6,0.344882)
        (0.7,0.323333)
        (0.8,0.30708)
        (0.9,0.27624)
        (1,0.214265)
  };
  \addlegendentry{\textsc{BellZigZag}}
  \path[name path=lower] (axis cs:0,0) -- (axis cs:1,0);
  \addplot[opacity=0.2, red, forget plot] fill between [of=upper and lower];
  \addplot[only marks, mark=diamond, color=red, forget plot, error bars/.cd, y dir=both, y explicit] coordinates {
        (0,0.494023) +- (0,0.000350037)
        (0.1,0.472429) +- (0,0.000921487)
        (0.2,0.449285) +- (0,0.000872198)
        (0.3,0.423783) +- (0,0.00103228)
        (0.4,0.394633) +- (0,0.000903579)
        (0.5,0.369127) +- (0,0.0010798)
        (0.6,0.344882) +- (0,0.000958435)
        (0.7,0.323333) +- (0,0.00154337)
        (0.8,0.30708) +- (0,0.00119609)
        (0.9,0.27624) +- (0,0.00195082)
        (1,0.214265) +- (0,0.00208699)
  };
  \addplot[color=OliveGreen, mark=*, name path=upper] coordinates {
        (0,0.315593)
        (0.1,0.304015)
        (0.2,0.29146)
        (0.3,0.277152)
        (0.4,0.263268)
        (0.5,0.248438)
        (0.6,0.240036)
        (0.7,0.227979)
        (0.8,0.22125)
        (0.9,0.203351)
        (1,0.160156)
  };
  \addlegendentry{\textsc{$5$-Chain}}
  \path[name path=lower] (axis cs:0,0) -- (axis cs:1,0);
  \addplot[opacity=0.2, OliveGreen, forget plot] fill between [of=upper and lower];
  \addplot[only marks, mark=*, color=OliveGreen, forget plot, error bars/.cd, y dir=both, y explicit] coordinates {
        (0,0.315593) +- (0,0.000231542)
        (0.1,0.304015) +- (0,0.000665118)
        (0.2,0.29146) +- (0,0.00111841)
        (0.3,0.277152) +- (0,0.000794264)
        (0.4,0.263268) +- (0,0.000683444)
        (0.5,0.248438) +- (0,0.000780356)
        (0.6,0.240036) +- (0,0.0011326)
        (0.7,0.227979) +- (0,0.00130924)
        (0.8,0.22125) +- (0,0.00106099)
        (0.9,0.203351) +- (0,0.00213867)
        (1,0.160156) +- (0,0.00232226)
  };
  \addplot[color=orange, mark=square*, name path=upper] coordinates {
        (0,0.379839)
        (0.1,0.367663)
        (0.2,0.352833)
        (0.3,0.336553)
        (0.4,0.321196)
        (0.5,0.305342)
        (0.6,0.290023)
        (0.7,0.280259)
        (0.8,0.269108)
        (0.9,0.255293)
        (1,0.204095)
  };
  \addlegendentry{\textsc{$8$-Chain}}
  \path[name path=lower] (axis cs:0,0) -- (axis cs:1,0);
  \addplot[opacity=0.2, orange, forget plot] fill between [of=upper and lower];
  \addplot[only marks, mark=square*, color=orange, forget plot, error bars/.cd, y dir=both, y explicit] coordinates {
        (0,0.379839) +- (0,0.000202592)
        (0.1,0.367663) +- (0,0.000788593)
        (0.2,0.352833) +- (0,0.000706912)
        (0.3,0.336553) +- (0,0.000616895)
        (0.4,0.321196) +- (0,0.000657645)
        (0.5,0.305342) +- (0,0.00104588)
        (0.6,0.290023) +- (0,0.000970899)
        (0.7,0.280259) +- (0,0.00134057)
        (0.8,0.269108) +- (0,0.00101677)
        (0.9,0.255293) +- (0,0.00192168)
        (1,0.204095) +- (0,0.00215785)
  };
  \addplot[color=Purple, mark=triangle*, name path=upper] coordinates {
        (0,0.498389)
        (0.1,0.483823)
        (0.2,0.468911)
        (0.3,0.450378)
        (0.4,0.430376)
        (0.5,0.410206)
        (0.6,0.397477)
        (0.7,0.38469)
        (0.8,0.368146)
        (0.9,0.342047)
        (1,0.283278)
  };
  \addlegendentry{$3d$-\textsc{Chain}}
  \path[name path=lower] (axis cs:0,0) -- (axis cs:1,0);
  \addplot[opacity=0.2, Purple, forget plot] fill between [of=upper and lower];
  \addplot[only marks, mark=triangle*, color=Purple, forget plot, error bars/.cd, y dir=both, y explicit] coordinates {
        (0,0.498389) +- (0,0.000192322)
        (0.1,0.483823) +- (0,0.000562109)
        (0.2,0.468911) +- (0,0.000708471)
        (0.3,0.450378) +- (0,0.00072429)
        (0.4,0.430376) +- (0,0.000811934)
        (0.5,0.410206) +- (0,0.00109646)
        (0.6,0.397477) +- (0,0.00135647)
        (0.7,0.38469) +- (0,0.00215129)
        (0.8,0.368146) +- (0,0.00268815)
        (0.9,0.342047) +- (0,0.00238303)
        (1,0.283278) +- (0,0.00306179)
  };
\end{axis}
\end{tikzpicture}

%% file: results/circuit_and_erasure_correctable_regime_pm_corr_mergeZZ_obsXX.tex
\begin{tikzpicture}
\begin{axis}[width=\linewidth,
  xlabel={Circuit noise probability $p_c$ (\%)},
  ylabel={Erasure threshold $p^{*}_{e}$},
  grid=both,
  minor grid style={gray!15},
  major grid style={gray!30},
  legend pos=south west,
  legend cell align=left,
  legend style={font=\footnotesize},
  every axis plot/.append style={thick},
  ymin=0,
  ymax=0.55,
  restrict y to domain=0:0.55,
  xmin=0,
  xmax=1,
  xtick distance=0.1,
  scaled x ticks=false,
  scaled y ticks=false,
  y tick label style={/pgf/number format/fixed},
  x tick label style={/pgf/number format/fixed, /pgf/number format/precision=2}]
  \addplot[color=blue, mark=diamond*, name path=upper] coordinates {
        (0,0.496441)
        (0.1,0.487458)
        (0.2,0.476894)
        (0.3,0.463462)
        (0.4,0.450279)
        (0.5,0.438853)
        (0.6,0.423818)
        (0.7,0.408509)
        (0.8,0.399111)
        (0.9,0.366535)
        (1,0.293118)
  };
  \addlegendentry{\textsc{Bell}}
  \path[name path=lower] (axis cs:0,0) -- (axis cs:1,0);
  \addplot[opacity=0.2, blue, forget plot] fill between [of=upper and lower];
  \addplot[only marks, mark=diamond*, color=blue, forget plot, error bars/.cd, y dir=both, y explicit] coordinates {
        (0,0.496441) +- (0,0.0008216)
        (0.1,0.487458) +- (0,0.00136707)
        (0.2,0.476894) +- (0,0.00252398)
        (0.3,0.463462) +- (0,0.00184321)
        (0.4,0.450279) +- (0,0.00177295)
        (0.5,0.438853) +- (0,0.00223262)
        (0.6,0.423818) +- (0,0.00304127)
        (0.7,0.408509) +- (0,0.00378092)
        (0.8,0.399111) +- (0,0.00272109)
        (0.9,0.366535) +- (0,0.00332023)
        (1,0.293118) +- (0,0.00354789)
  };
  \addplot[color=red, mark=diamond, name path=upper] coordinates {
        (0,0.50595)
        (0.1,0.493015)
        (0.2,0.476549)
        (0.3,0.458055)
        (0.4,0.433045)
        (0.5,0.413565)
        (0.6,0.391553)
        (0.7,0.369571)
        (0.8,0.355902)
        (0.9,0.324988)
        (1,0.255174)
  };
  \addlegendentry{\textsc{BellZigZag}}
  \path[name path=lower] (axis cs:0,0) -- (axis cs:1,0);
  \addplot[opacity=0.2, red, forget plot] fill between [of=upper and lower];
  \addplot[only marks, mark=diamond, color=red, forget plot, error bars/.cd, y dir=both, y explicit] coordinates {
        (0,0.50595) +- (0,0.000364)
        (0.1,0.493015) +- (0,0.00146354)
        (0.2,0.476549) +- (0,0.000914024)
        (0.3,0.458055) +- (0,0.00113233)
        (0.4,0.433045) +- (0,0.00115504)
        (0.5,0.413565) +- (0,0.00111975)
        (0.6,0.391553) +- (0,0.00162346)
        (0.7,0.369571) +- (0,0.00139879)
        (0.8,0.355902) +- (0,0.0023087)
        (0.9,0.324988) +- (0,0.00269529)
        (1,0.255174) +- (0,0.00306725)
  };
  \addplot[color=OliveGreen, mark=*, name path=upper] coordinates {
        (0,0.314526)
        (0.1,0.308498)
        (0.2,0.297599)
        (0.3,0.285682)
        (0.4,0.275424)
        (0.5,0.262737)
        (0.6,0.248736)
        (0.7,0.234344)
        (0.8,0.220817)
        (0.9,0.198309)
        (1,0.151245)
  };
  \addlegendentry{\textsc{$5$-Chain}}
  \path[name path=lower] (axis cs:0,0) -- (axis cs:1,0);
  \addplot[opacity=0.2, OliveGreen, forget plot] fill between [of=upper and lower];
  \addplot[only marks, mark=*, color=OliveGreen, forget plot, error bars/.cd, y dir=both, y explicit] coordinates {
        (0,0.314526) +- (0,0.000740173)
        (0.1,0.308498) +- (0,0.00163046)
        (0.2,0.297599) +- (0,0.00138771)
        (0.3,0.285682) +- (0,0.001556)
        (0.4,0.275424) +- (0,0.00166604)
        (0.5,0.262737) +- (0,0.00262054)
        (0.6,0.248736) +- (0,0.00238686)
        (0.7,0.234344) +- (0,0.00185247)
        (0.8,0.220817) +- (0,0.0020947)
        (0.9,0.198309) +- (0,0.00170503)
        (1,0.151245) +- (0,0.00239663)
  };
  \addplot[color=orange, mark=square*, name path=upper] coordinates {
        (0,0.379386)
        (0.1,0.371774)
        (0.2,0.361048)
        (0.3,0.348097)
        (0.4,0.336655)
        (0.5,0.320115)
        (0.6,0.307465)
        (0.7,0.292128)
        (0.8,0.273512)
        (0.9,0.242558)
        (1,0.196176)
  };
  \addlegendentry{\textsc{$8$-Chain}}
  \path[name path=lower] (axis cs:0,0) -- (axis cs:1,0);
  \addplot[opacity=0.2, orange, forget plot] fill between [of=upper and lower];
  \addplot[only marks, mark=square*, color=orange, forget plot, error bars/.cd, y dir=both, y explicit] coordinates {
        (0,0.379386) +- (0,0.000760846)
        (0.1,0.371774) +- (0,0.00147946)
        (0.2,0.361048) +- (0,0.00235159)
        (0.3,0.348097) +- (0,0.00136941)
        (0.4,0.336655) +- (0,0.00163338)
        (0.5,0.320115) +- (0,0.00148972)
        (0.6,0.307465) +- (0,0.00178344)
        (0.7,0.292128) +- (0,0.00182349)
        (0.8,0.273512) +- (0,0.00228554)
        (0.9,0.242558) +- (0,0.00310752)
        (1,0.196176) +- (0,0.00360116)
  };
  \addplot[color=Purple, mark=triangle*, name path=upper] coordinates {
        (0,0.501818)
        (0.1,0.491992)
        (0.2,0.479504)
        (0.3,0.463949)
        (0.4,0.45084)
        (0.5,0.432544)
        (0.6,0.415132)
        (0.7,0.402812)
        (0.8,0.37683)
        (0.9,0.349051)
        (1,0.284284)
  };
  \addlegendentry{$3d$-\textsc{Chain}}
  \path[name path=lower] (axis cs:0,0) -- (axis cs:1,0);
  \addplot[opacity=0.2, Purple, forget plot] fill between [of=upper and lower];
  \addplot[only marks, mark=triangle*, color=Purple, forget plot, error bars/.cd, y dir=both, y explicit] coordinates {
        (0,0.501818) +- (0,0.00087187)
        (0.1,0.491992) +- (0,0.00116351)
        (0.2,0.479504) +- (0,0.00167341)
        (0.3,0.463949) +- (0,0.00145068)
        (0.4,0.45084) +- (0,0.00158604)
        (0.5,0.432544) +- (0,0.00154397)
        (0.6,0.415132) +- (0,0.00142779)
        (0.7,0.402812) +- (0,0.00495023)
        (0.8,0.37683) +- (0,0.00433589)
        (0.9,0.349051) +- (0,0.00486605)
        (1,0.284284) +- (0,0.0068075)
  };
\end{axis}
\end{tikzpicture}

%% file: results/subthreshold_lambda_erasure_grid_pm_corr_mergeZZ_obsZZ-XX.tex
\begin{tikzpicture}
\begin{groupplot}[
  group style={group size=2 by 3, horizontal sep=0.35cm, vertical sep=0.9cm, xlabels at=edge bottom, ylabels at=edge left, xticklabels at=edge bottom, yticklabels at=edge left},
  width=0.450\linewidth, height=0.2\textheight, scale only axis,
  xlabel={Fusion erasure rate $p_e$ (\%)}, ymode=log, grid=both,
  minor grid style={gray!12}, major grid style={gray!30},
  title style={font=\small}, every tick label/.append style={font=\footnotesize},
  error bars/error bar style={solid},
]
\nextgroupplot[title={$\bar{Z}\bar{Z}$}, ylabel={$\Lambda$ ($p_c=p_r=10^{-3}$)}, legend to name=legend_subthreshold_lambda_erasure_grid_pm_corr_mergeZZ_obsZZ-XX, legend columns=5, legend style={draw=none, /tikz/every even column/.append style={column sep=1em}}, xmin=0.04, xmax=0.469, ymin=0.499, ymax=970, xmode=log, log basis x=10, xtick={0.05,0.1,0.15,0.2,0.3,0.4}, xticklabels={5,10,15,20,30,40}]
  \addlegendimage{blue, mark=diamond*, thick}
  \addlegendentry{\textsc{Bell}}
  \addlegendimage{red, mark=diamond, thick}
  \addlegendentry{\textsc{BellZigZag}}
  \addlegendimage{OliveGreen, mark=*, thick}
  \addlegendentry{\textsc{$5$-Chain}}
  \addlegendimage{orange, mark=square*, thick}
  \addlegendentry{\textsc{$8$-Chain}}
  \addlegendimage{Purple, mark=triangle*, thick}
  \addlegendentry{$3d$-\textsc{Chain}}
  \addlegendimage{black, thick, solid, line legend}
  \addlegendentry{$n=1$}
  \addlegendimage{black, thick, dashed, line legend}
  \addlegendentry{$n=2$}
  \addlegendimage{black, thick, densely dotted, line legend}
  \addlegendentry{$n=3$}
  \addlegendimage{black, thick, line width=2.2pt, opacity=0.3, line legend}
  \addlegendentry{leading order}
  \addplot[blue, mark=diamond*, only marks, mark options={solid}, forget plot,
    error bars/.cd, y dir=both, y explicit]
  coordinates {
    (0.0699738,13.0502) +- (0,0.321705)
    (0.0810868,12.181) +- (0,0.297068)
    (0.0993503,10.0579) +- (0,0.23659)
    (0.128495,7.37706) +- (0,0.169938)
    (0.142369,6.61499) +- (0,0.155511)
    (0.159478,5.50921) +- (0,0.137302)
    (0.192928,4.20702) +- (0,0.100566)
    (0.225369,3.18256) +- (0,0.0778966)
    (0.256822,2.66189) +- (0,0.0656144)
    (0.287307,2.15968) +- (0,0.052597)
    (0.316845,1.88014) +- (0,0.0451239)
    (0.373156,1.45342) +- (0,0.034817)
    (0.425913,1.22298) +- (0,0.0280954)
  };
  \addplot[blue, thick, solid, no marks, forget plot] coordinates {(0.256822,2.66189) (0.287307,2.15968) (0.316845,1.88014) (0.373156,1.45342) (0.425913,1.22298)};
  \addplot[blue, thick, dashed, no marks, forget plot] coordinates {(0.128495,7.37706) (0.142369,6.61499) (0.159478,5.50921) (0.192928,4.20702) (0.225369,3.18256) (0.256822,2.66189)};
  \addplot[blue, thick, densely dotted, no marks, forget plot] coordinates {(0.0699738,13.0502) (0.0810868,12.181) (0.0993503,10.0579) (0.128495,7.37706)};
  \addplot[red, mark=diamond, only marks, mark options={solid}, forget plot,
    error bars/.cd, y dir=both, y explicit]
  coordinates {
    (0.0699738,10.5481) +- (0,0.252023)
    (0.0810868,9.23751) +- (0,0.225958)
    (0.0993503,7.45022) +- (0,0.179052)
    (0.128495,5.4346) +- (0,0.12993)
    (0.142369,4.78701) +- (0,0.115011)
    (0.159478,4.029) +- (0,0.095039)
    (0.192928,3.32768) +- (0,0.0823115)
    (0.225369,2.52778) +- (0,0.0620453)
    (0.256822,2.17726) +- (0,0.0522868)
    (0.287307,1.85469) +- (0,0.0448629)
    (0.316845,1.51374) +- (0,0.0359318)
    (0.373156,1.30137) +- (0,0.0302294)
    (0.425913,1.06396) +- (0,0.0233464)
  };
  \addplot[red, thick, solid, no marks, forget plot] coordinates {(0.256822,2.17726) (0.287307,1.85469) (0.316845,1.51374) (0.373156,1.30137) (0.425913,1.06396)};
  \addplot[red, thick, dashed, no marks, forget plot] coordinates {(0.128495,5.4346) (0.142369,4.78701) (0.159478,4.029) (0.192928,3.32768) (0.225369,2.52778) (0.256822,2.17726)};
  \addplot[red, thick, densely dotted, no marks, forget plot] coordinates {(0.0699738,10.5481) (0.0810868,9.23751) (0.0993503,7.45022) (0.128495,5.4346)};
  \addplot[OliveGreen, mark=*, only marks, mark options={solid}, forget plot,
    error bars/.cd, y dir=both, y explicit]
  coordinates {
    (0.0699738,10.4599) +- (0,0.262972)
    (0.0810868,9.13908) +- (0,0.226384)
    (0.0993503,6.61813) +- (0,0.162175)
    (0.128495,4.24423) +- (0,0.102028)
    (0.142369,3.54761) +- (0,0.0835979)
    (0.159478,2.95553) +- (0,0.0731775)
    (0.192928,1.98153) +- (0,0.0473373)
    (0.225369,1.51341) +- (0,0.0358914)
    (0.256822,1.21738) +- (0,0.0292559)
    (0.287307,1.04145) +- (0,0.0246368)
    (0.316845,0.929345) +- (0,0.0214736)
    (0.373156,0.785655) +- (0,0.0163599)
    (0.425913,0.78747) +- (0,0.0166556)
  };
  \addplot[OliveGreen, thick, solid, no marks, forget plot] coordinates {(0.256822,1.21738) (0.287307,1.04145) (0.316845,0.929345) (0.373156,0.785655) (0.425913,0.78747)};
  \addplot[OliveGreen, thick, dashed, no marks, forget plot] coordinates {(0.128495,4.24423) (0.142369,3.54761) (0.159478,2.95553) (0.192928,1.98153) (0.225369,1.51341) (0.256822,1.21738)};
  \addplot[OliveGreen, thick, densely dotted, no marks, forget plot] coordinates {(0.0699738,10.4599) (0.0810868,9.13908) (0.0993503,6.61813) (0.128495,4.24423)};
  \addplot[orange, mark=square*, only marks, mark options={solid}, forget plot,
    error bars/.cd, y dir=both, y explicit]
  coordinates {
    (0.0699738,13.1634) +- (0,0.311069)
    (0.0810868,11.6745) +- (0,0.28477)
    (0.0993503,9.90765) +- (0,0.239908)
    (0.128495,6.84945) +- (0,0.163888)
    (0.142369,5.45362) +- (0,0.128591)
    (0.159478,4.44316) +- (0,0.108115)
    (0.192928,3.06806) +- (0,0.072648)
    (0.225369,2.23975) +- (0,0.0535222)
    (0.256822,1.67781) +- (0,0.0408485)
    (0.287307,1.44901) +- (0,0.0349281)
    (0.316845,1.18453) +- (0,0.0283098)
    (0.373156,0.934244) +- (0,0.021524)
    (0.425913,0.843844) +- (0,0.017765)
  };
  \addplot[orange, thick, solid, no marks, forget plot] coordinates {(0.256822,1.67781) (0.287307,1.44901) (0.316845,1.18453) (0.373156,0.934244) (0.425913,0.843844)};
  \addplot[orange, thick, dashed, no marks, forget plot] coordinates {(0.128495,6.84945) (0.142369,5.45362) (0.159478,4.44316) (0.192928,3.06806) (0.225369,2.23975) (0.256822,1.67781)};
  \addplot[orange, thick, densely dotted, no marks, forget plot] coordinates {(0.0699738,13.1634) (0.0810868,11.6745) (0.0993503,9.90765) (0.128495,6.84945)};
  \addplot[Purple, mark=triangle*, only marks, mark options={solid}, forget plot,
    error bars/.cd, y dir=both, y explicit]
  coordinates {
    (0.0699738,15.3697) +- (0,0.362317)
    (0.0810868,14.7361) +- (0,0.374943)
    (0.0993503,13.5618) +- (0,0.337992)
    (0.128495,10.9211) +- (0,0.261665)
    (0.142369,10.2463) +- (0,0.262254)
    (0.159478,8.33741) +- (0,0.205212)
    (0.192928,5.91433) +- (0,0.138041)
    (0.225369,4.22122) +- (0,0.0994987)
    (0.256822,3.38505) +- (0,0.0804035)
    (0.287307,2.50753) +- (0,0.0615867)
    (0.316845,2.18987) +- (0,0.0519612)
    (0.373156,1.5192) +- (0,0.0365914)
    (0.425913,1.21987) +- (0,0.0289834)
  };
  \addplot[Purple, thick, solid, no marks, forget plot] coordinates {(0.256822,3.38505) (0.287307,2.50753) (0.316845,2.18987) (0.373156,1.5192) (0.425913,1.21987)};
  \addplot[Purple, thick, dashed, no marks, forget plot] coordinates {(0.128495,10.9211) (0.142369,10.2463) (0.159478,8.33741) (0.192928,5.91433) (0.225369,4.22122) (0.256822,3.38505)};
  \addplot[Purple, thick, densely dotted, no marks, forget plot] coordinates {(0.0699738,15.3697) (0.0810868,14.7361) (0.0993503,13.5618) (0.128495,10.9211)};
  \addplot[black!60, densely dashdotted, forget plot] coordinates {(0.04,1) (0.469,1)};
\nextgroupplot[title={$\bar{X}\bar{X}$}, xmin=0.04, xmax=0.469, ymin=0.499, ymax=970, xmode=log, log basis x=10, xtick={0.05,0.1,0.15,0.2,0.3,0.4}, xticklabels={5,10,15,20,30,40}]
  \addplot[blue, mark=diamond*, only marks, mark options={solid}, forget plot,
    error bars/.cd, y dir=both, y explicit]
  coordinates {
    (0.0699738,3.74452) +- (0,0.0873391)
    (0.0810868,3.41091) +- (0,0.0808463)
    (0.0993503,3.11764) +- (0,0.0736075)
    (0.128495,2.69077) +- (0,0.0629745)
    (0.142369,2.48382) +- (0,0.0591438)
    (0.159478,2.32753) +- (0,0.0547817)
    (0.192928,1.98699) +- (0,0.0459506)
    (0.225369,1.7826) +- (0,0.0413565)
    (0.256822,1.63671) +- (0,0.0365775)
    (0.287307,1.46567) +- (0,0.0307198)
    (0.316845,1.28247) +- (0,0.0273371)
    (0.373156,1.14654) +- (0,0.0247625)
  };
  \addplot[blue, thick, solid, no marks, forget plot] coordinates {(0.256822,1.63671) (0.287307,1.46567) (0.316845,1.28247) (0.373156,1.14654)};
  \addplot[blue, thick, dashed, no marks, forget plot] coordinates {(0.128495,2.69077) (0.142369,2.48382) (0.159478,2.32753) (0.192928,1.98699) (0.225369,1.7826) (0.256822,1.63671)};
  \addplot[blue, thick, densely dotted, no marks, forget plot] coordinates {(0.0699738,3.74452) (0.0810868,3.41091) (0.0993503,3.11764) (0.128495,2.69077)};
  \addplot[red, mark=diamond, only marks, mark options={solid}, forget plot,
    error bars/.cd, y dir=both, y explicit]
  coordinates {
    (0.0699738,9.95129) +- (0,0.221015)
    (0.0810868,9.00247) +- (0,0.217545)
    (0.0993503,7.56271) +- (0,0.181839)
    (0.128495,6.05123) +- (0,0.145016)
    (0.142369,5.22436) +- (0,0.121976)
    (0.159478,4.7512) +- (0,0.110602)
    (0.192928,3.69776) +- (0,0.085928)
    (0.225369,3.0188) +- (0,0.0727734)
    (0.256822,2.45113) +- (0,0.0581385)
    (0.287307,2.12317) +- (0,0.0501279)
    (0.316845,1.82232) +- (0,0.0426964)
    (0.373156,1.44174) +- (0,0.0322302)
    (0.425913,1.18468) +- (0,0.025354)
  };
  \addplot[red, thick, solid, no marks, forget plot] coordinates {(0.256822,2.45113) (0.287307,2.12317) (0.316845,1.82232) (0.373156,1.44174) (0.425913,1.18468)};
  \addplot[red, thick, dashed, no marks, forget plot] coordinates {(0.128495,6.05123) (0.142369,5.22436) (0.159478,4.7512) (0.192928,3.69776) (0.225369,3.0188) (0.256822,2.45113)};
  \addplot[red, thick, densely dotted, no marks, forget plot] coordinates {(0.0699738,9.95129) (0.0810868,9.00247) (0.0993503,7.56271) (0.128495,6.05123)};
  \addplot[OliveGreen, mark=*, only marks, mark options={solid}, forget plot,
    error bars/.cd, y dir=both, y explicit]
  coordinates {
    (0.0699738,3.18891) +- (0,0.0784423)
    (0.0810868,2.89028) +- (0,0.0685225)
    (0.0993503,2.46208) +- (0,0.0574918)
    (0.128495,2.06145) +- (0,0.0484355)
    (0.142369,1.76382) +- (0,0.0402307)
    (0.159478,1.66437) +- (0,0.037712)
    (0.192928,1.37291) +- (0,0.0292287)
    (0.225369,1.1829) +- (0,0.0230989)
  };
  \addplot[OliveGreen, thick, dashed, no marks, forget plot] coordinates {(0.128495,2.06145) (0.142369,1.76382) (0.159478,1.66437) (0.192928,1.37291) (0.225369,1.1829)};
  \addplot[OliveGreen, thick, densely dotted, no marks, forget plot] coordinates {(0.0699738,3.18891) (0.0810868,2.89028) (0.0993503,2.46208) (0.128495,2.06145)};
  \addplot[orange, mark=square*, only marks, mark options={solid}, forget plot,
    error bars/.cd, y dir=both, y explicit]
  coordinates {
    (0.0699738,4.12862) +- (0,0.100201)
    (0.0810868,3.68879) +- (0,0.0866104)
    (0.0993503,3.21959) +- (0,0.0777066)
    (0.128495,2.57333) +- (0,0.0613743)
    (0.142369,2.4258) +- (0,0.0582835)
    (0.159478,2.10813) +- (0,0.0494826)
    (0.192928,1.81066) +- (0,0.0411236)
    (0.225369,1.48792) +- (0,0.0319529)
    (0.256822,1.3149) +- (0,0.027782)
    (0.287307,1.17852) +- (0,0.0254832)
  };
  \addplot[orange, thick, solid, no marks, forget plot] coordinates {(0.256822,1.3149) (0.287307,1.17852)};
  \addplot[orange, thick, dashed, no marks, forget plot] coordinates {(0.128495,2.57333) (0.142369,2.4258) (0.159478,2.10813) (0.192928,1.81066) (0.225369,1.48792) (0.256822,1.3149)};
  \addplot[orange, thick, densely dotted, no marks, forget plot] coordinates {(0.0699738,4.12862) (0.0810868,3.68879) (0.0993503,3.21959) (0.128495,2.57333)};
  \addplot[Purple, mark=triangle*, only marks, mark options={solid}, forget plot,
    error bars/.cd, y dir=both, y explicit]
  coordinates {
    (0.0699738,5.31005) +- (0,0.123414)
    (0.0810868,4.9171) +- (0,0.112314)
    (0.0993503,4.59271) +- (0,0.106997)
    (0.128495,3.42502) +- (0,0.0817295)
    (0.142369,3.20543) +- (0,0.0770326)
    (0.159478,2.97864) +- (0,0.0701174)
    (0.192928,2.53269) +- (0,0.0588778)
    (0.225369,2.24378) +- (0,0.0520233)
    (0.256822,1.96805) +- (0,0.0447604)
    (0.287307,1.65266) +- (0,0.035798)
    (0.316845,1.5315) +- (0,0.0329523)
    (0.373156,1.23955) +- (0,0.0268156)
  };
  \addplot[Purple, thick, solid, no marks, forget plot] coordinates {(0.256822,1.96805) (0.287307,1.65266) (0.316845,1.5315) (0.373156,1.23955)};
  \addplot[Purple, thick, dashed, no marks, forget plot] coordinates {(0.128495,3.42502) (0.142369,3.20543) (0.159478,2.97864) (0.192928,2.53269) (0.225369,2.24378) (0.256822,1.96805)};
  \addplot[Purple, thick, densely dotted, no marks, forget plot] coordinates {(0.0699738,5.31005) (0.0810868,4.9171) (0.0993503,4.59271) (0.128495,3.42502)};
  \addplot[black!60, densely dashdotted, forget plot] coordinates {(0.04,1) (0.469,1)};
\nextgroupplot[ylabel={$\Lambda$ ($p_c=p_r=10^{-4}$)}, xmin=0.04, xmax=0.469, ymin=0.499, ymax=970, xmode=log, log basis x=10, xtick={0.05,0.1,0.15,0.2,0.3,0.4}, xticklabels={5,10,15,20,30,40}]
  \addplot[blue, mark=diamond*, only marks, mark options={solid}, forget plot,
    error bars/.cd, y dir=both, y explicit]
  coordinates {
    (0.0699738,45.2273) +- (0,9.7485)
    (0.0810868,38.1087) +- (0,5.69206)
    (0.0993503,25.2193) +- (0,1.97624)
    (0.128495,16.5576) +- (0,0.622422)
    (0.142369,14.0337) +- (0,0.393053)
    (0.159478,10.8629) +- (0,0.272027)
    (0.192928,7.50771) +- (0,0.179059)
    (0.225369,5.32274) +- (0,0.126208)
    (0.256822,3.96207) +- (0,0.0970792)
    (0.287307,3.11791) +- (0,0.0756038)
    (0.316845,2.57219) +- (0,0.0623025)
    (0.373156,1.72849) +- (0,0.041056)
    (0.425913,1.33403) +- (0,0.0306837)
  };
  \addplot[blue, thick, solid, no marks, forget plot] coordinates {(0.256822,3.96207) (0.287307,3.11791) (0.316845,2.57219) (0.373156,1.72849) (0.425913,1.33403)};
  \addplot[blue, thick, dashed, no marks, forget plot] coordinates {(0.128495,16.5576) (0.142369,14.0337) (0.159478,10.8629) (0.192928,7.50771) (0.225369,5.32274) (0.256822,3.96207)};
  \addplot[blue, thick, densely dotted, no marks, forget plot] coordinates {(0.0699738,45.2273) (0.0810868,38.1087) (0.0993503,25.2193) (0.128495,16.5576)};
  \addplot[red, mark=diamond, only marks, mark options={solid}, forget plot,
    error bars/.cd, y dir=both, y explicit]
  coordinates {
    (0.0699738,37.2667) +- (0,4.87523)
    (0.0810868,22.5794) +- (0,1.73677)
    (0.0993503,20.1077) +- (0,0.946793)
    (0.128495,11.9548) +- (0,0.300196)
    (0.142369,9.99227) +- (0,0.239738)
    (0.159478,8.40276) +- (0,0.20061)
    (0.192928,5.63399) +- (0,0.136493)
    (0.225369,4.38001) +- (0,0.101231)
    (0.256822,3.27267) +- (0,0.0762333)
    (0.287307,2.617) +- (0,0.0623308)
    (0.316845,2.0926) +- (0,0.050123)
    (0.373156,1.49196) +- (0,0.0349518)
    (0.425913,1.20062) +- (0,0.026838)
  };
  \addplot[red, thick, solid, no marks, forget plot] coordinates {(0.256822,3.27267) (0.287307,2.617) (0.316845,2.0926) (0.373156,1.49196) (0.425913,1.20062)};
  \addplot[red, thick, dashed, no marks, forget plot] coordinates {(0.128495,11.9548) (0.142369,9.99227) (0.159478,8.40276) (0.192928,5.63399) (0.225369,4.38001) (0.256822,3.27267)};
  \addplot[red, thick, densely dotted, no marks, forget plot] coordinates {(0.0699738,37.2667) (0.0810868,22.5794) (0.0993503,20.1077) (0.128495,11.9548)};
  \addplot[OliveGreen, mark=*, only marks, mark options={solid}, forget plot,
    error bars/.cd, y dir=both, y explicit]
  coordinates {
    (0.0699738,50.0667) +- (0,13.0556)
    (0.0810868,29.9216) +- (0,4.2593)
    (0.0993503,22.5059) +- (0,1.63827)
    (0.128495,10.4595) +- (0,0.280013)
    (0.142369,8.03438) +- (0,0.196497)
    (0.159478,5.98431) +- (0,0.143453)
    (0.192928,3.36878) +- (0,0.0811284)
    (0.225369,2.11384) +- (0,0.0508629)
    (0.256822,1.43308) +- (0,0.0345294)
    (0.287307,1.15193) +- (0,0.027601)
    (0.316845,0.949325) +- (0,0.0223537)
    (0.373156,0.795316) +- (0,0.0173273)
    (0.425913,0.78267) +- (0,0.0153027)
  };
  \addplot[OliveGreen, thick, solid, no marks, forget plot] coordinates {(0.256822,1.43308) (0.287307,1.15193) (0.316845,0.949325) (0.373156,0.795316) (0.425913,0.78267)};
  \addplot[OliveGreen, thick, dashed, no marks, forget plot] coordinates {(0.128495,10.4595) (0.142369,8.03438) (0.159478,5.98431) (0.192928,3.36878) (0.225369,2.11384) (0.256822,1.43308)};
  \addplot[OliveGreen, thick, densely dotted, no marks, forget plot] coordinates {(0.0699738,50.0667) (0.0810868,29.9216) (0.0993503,22.5059) (0.128495,10.4595)};
  \addplot[orange, mark=square*, only marks, mark options={solid}, forget plot,
    error bars/.cd, y dir=both, y explicit]
  coordinates {
    (0.0699738,75.1667) +- (0,30.8901)
    (0.0810868,91.375) +- (0,32.4822)
    (0.0993503,51.9091) +- (0,9.12284)
    (0.128495,20.7586) +- (0,1.13543)
    (0.142369,15.2245) +- (0,0.544941)
    (0.159478,10.3777) +- (0,0.25107)
    (0.192928,6.01906) +- (0,0.137424)
    (0.225369,3.65768) +- (0,0.0891291)
    (0.256822,2.57742) +- (0,0.0619463)
    (0.287307,1.80565) +- (0,0.0434882)
    (0.316845,1.42539) +- (0,0.0342971)
    (0.373156,0.989706) +- (0,0.0230765)
    (0.425913,0.831346) +- (0,0.0176425)
  };
  \addplot[orange, thick, solid, no marks, forget plot] coordinates {(0.256822,2.57742) (0.287307,1.80565) (0.316845,1.42539) (0.373156,0.989706) (0.425913,0.831346)};
  \addplot[orange, thick, dashed, no marks, forget plot] coordinates {(0.128495,20.7586) (0.142369,15.2245) (0.159478,10.3777) (0.192928,6.01906) (0.225369,3.65768) (0.256822,2.57742)};
  \addplot[orange, thick, densely dotted, no marks, forget plot] coordinates {(0.0699738,75.1667) (0.0810868,91.375) (0.0993503,51.9091) (0.128495,20.7586)};
  \addplot[Purple, mark=triangle*, only marks, mark options={solid}, forget plot,
    error bars/.cd, y dir=both, y explicit]
  coordinates {
    (0.0993503,88.3333) +- (0,36.2655)
    (0.128495,53.1111) +- (0,10.317)
    (0.142369,34.3913) +- (0,4.19999)
    (0.159478,26.5216) +- (0,2.06508)
    (0.192928,15.1202) +- (0,0.508683)
    (0.225369,9.66363) +- (0,0.243189)
    (0.256822,6.47786) +- (0,0.164805)
    (0.287307,4.63469) +- (0,0.109899)
    (0.316845,3.30433) +- (0,0.0756078)
    (0.373156,2.02373) +- (0,0.0495803)
    (0.425913,1.42506) +- (0,0.0335875)
  };
  \addplot[Purple, thick, solid, no marks, forget plot] coordinates {(0.256822,6.47786) (0.287307,4.63469) (0.316845,3.30433) (0.373156,2.02373) (0.425913,1.42506)};
  \addplot[Purple, thick, dashed, no marks, forget plot] coordinates {(0.128495,53.1111) (0.142369,34.3913) (0.159478,26.5216) (0.192928,15.1202) (0.225369,9.66363) (0.256822,6.47786)};
  \addplot[Purple, thick, densely dotted, no marks, forget plot] coordinates {(0.0993503,88.3333) (0.128495,53.1111)};
  \addplot[black!60, densely dashdotted, forget plot] coordinates {(0.04,1) (0.469,1)};
\nextgroupplot[xmin=0.04, xmax=0.469, ymin=0.499, ymax=970, xmode=log, log basis x=10, xtick={0.05,0.1,0.15,0.2,0.3,0.4}, xticklabels={5,10,15,20,30,40}]
  \addplot[blue, mark=diamond*, only marks, mark options={solid}, forget plot,
    error bars/.cd, y dir=both, y explicit]
  coordinates {
    (0.0699738,7.08894) +- (0,0.172389)
    (0.0810868,6.28256) +- (0,0.153699)
    (0.0993503,5.22381) +- (0,0.122362)
    (0.128495,3.96581) +- (0,0.0933596)
    (0.142369,3.70724) +- (0,0.0869894)
    (0.159478,3.29597) +- (0,0.0787753)
    (0.192928,2.58656) +- (0,0.0607015)
    (0.225369,2.17464) +- (0,0.0493372)
    (0.256822,1.846) +- (0,0.0408413)
    (0.287307,1.60679) +- (0,0.0350358)
    (0.316845,1.33687) +- (0,0.0303692)
    (0.373156,1.21105) +- (0,0.0254685)
  };
  \addplot[blue, thick, solid, no marks, forget plot] coordinates {(0.256822,1.846) (0.287307,1.60679) (0.316845,1.33687) (0.373156,1.21105)};
  \addplot[blue, thick, dashed, no marks, forget plot] coordinates {(0.128495,3.96581) (0.142369,3.70724) (0.159478,3.29597) (0.192928,2.58656) (0.225369,2.17464) (0.256822,1.846)};
  \addplot[blue, thick, densely dotted, no marks, forget plot] coordinates {(0.0699738,7.08894) (0.0810868,6.28256) (0.0993503,5.22381) (0.128495,3.96581)};
  \addplot[red, mark=diamond, only marks, mark options={solid}, forget plot,
    error bars/.cd, y dir=both, y explicit]
  coordinates {
    (0.0699738,31.6929) +- (0,1.28497)
    (0.0810868,25.6323) +- (0,0.784635)
    (0.0993503,19.9825) +- (0,0.489684)
    (0.128495,12.9256) +- (0,0.30394)
    (0.142369,11.3743) +- (0,0.273031)
    (0.159478,9.11785) +- (0,0.226293)
    (0.192928,6.62156) +- (0,0.159385)
    (0.225369,4.98685) +- (0,0.122008)
    (0.256822,3.85236) +- (0,0.0902618)
    (0.287307,3.06154) +- (0,0.0720385)
    (0.316845,2.34184) +- (0,0.0555672)
    (0.373156,1.68963) +- (0,0.0380834)
    (0.425913,1.31139) +- (0,0.0280943)
  };
  \addplot[red, thick, solid, no marks, forget plot] coordinates {(0.256822,3.85236) (0.287307,3.06154) (0.316845,2.34184) (0.373156,1.68963) (0.425913,1.31139)};
  \addplot[red, thick, dashed, no marks, forget plot] coordinates {(0.128495,12.9256) (0.142369,11.3743) (0.159478,9.11785) (0.192928,6.62156) (0.225369,4.98685) (0.256822,3.85236)};
  \addplot[red, thick, densely dotted, no marks, forget plot] coordinates {(0.0699738,31.6929) (0.0810868,25.6323) (0.0993503,19.9825) (0.128495,12.9256)};
  \addplot[OliveGreen, mark=*, only marks, mark options={solid}, forget plot,
    error bars/.cd, y dir=both, y explicit]
  coordinates {
    (0.0699738,6.8975) +- (0,0.156122)
    (0.0810868,5.831) +- (0,0.13586)
    (0.0993503,4.45007) +- (0,0.105747)
    (0.128495,3.03916) +- (0,0.0712495)
    (0.142369,2.54644) +- (0,0.0593128)
    (0.159478,2.06632) +- (0,0.0465647)
    (0.192928,1.5637) +- (0,0.0330903)
    (0.225369,1.26509) +- (0,0.0285303)
  };
  \addplot[OliveGreen, thick, dashed, no marks, forget plot] coordinates {(0.128495,3.03916) (0.142369,2.54644) (0.159478,2.06632) (0.192928,1.5637) (0.225369,1.26509)};
  \addplot[OliveGreen, thick, densely dotted, no marks, forget plot] coordinates {(0.0699738,6.8975) (0.0810868,5.831) (0.0993503,4.45007) (0.128495,3.03916)};
  \addplot[orange, mark=square*, only marks, mark options={solid}, forget plot,
    error bars/.cd, y dir=both, y explicit]
  coordinates {
    (0.0699738,9.20199) +- (0,0.206295)
    (0.0810868,7.80669) +- (0,0.190895)
    (0.0993503,6.72321) +- (0,0.15418)
    (0.128495,4.43342) +- (0,0.107611)
    (0.142369,3.72418) +- (0,0.0874108)
    (0.159478,3.01962) +- (0,0.0708307)
    (0.192928,2.2735) +- (0,0.0524568)
    (0.225369,1.73108) +- (0,0.0387516)
    (0.256822,1.37917) +- (0,0.0299429)
    (0.287307,1.25231) +- (0,0.024711)
  };
  \addplot[orange, thick, solid, no marks, forget plot] coordinates {(0.256822,1.37917) (0.287307,1.25231)};
  \addplot[orange, thick, dashed, no marks, forget plot] coordinates {(0.128495,4.43342) (0.142369,3.72418) (0.159478,3.01962) (0.192928,2.2735) (0.225369,1.73108) (0.256822,1.37917)};
  \addplot[orange, thick, densely dotted, no marks, forget plot] coordinates {(0.0699738,9.20199) (0.0810868,7.80669) (0.0993503,6.72321) (0.128495,4.43342)};
  \addplot[Purple, mark=triangle*, only marks, mark options={solid}, forget plot,
    error bars/.cd, y dir=both, y explicit]
  coordinates {
    (0.0699738,13.7746) +- (0,0.314923)
    (0.0810868,11.7811) +- (0,0.276022)
    (0.0993503,9.46994) +- (0,0.215941)
    (0.128495,7.59773) +- (0,0.186703)
    (0.142369,6.61536) +- (0,0.1589)
    (0.159478,5.56313) +- (0,0.137335)
    (0.192928,4.28013) +- (0,0.102494)
    (0.225369,3.22548) +- (0,0.076328)
    (0.256822,2.59326) +- (0,0.0598094)
    (0.287307,2.13413) +- (0,0.0477404)
    (0.316845,1.69815) +- (0,0.0376339)
    (0.373156,1.29149) +- (0,0.0277585)
  };
  \addplot[Purple, thick, solid, no marks, forget plot] coordinates {(0.256822,2.59326) (0.287307,2.13413) (0.316845,1.69815) (0.373156,1.29149)};
  \addplot[Purple, thick, dashed, no marks, forget plot] coordinates {(0.128495,7.59773) (0.142369,6.61536) (0.159478,5.56313) (0.192928,4.28013) (0.225369,3.22548) (0.256822,2.59326)};
  \addplot[Purple, thick, densely dotted, no marks, forget plot] coordinates {(0.0699738,13.7746) (0.0810868,11.7811) (0.0993503,9.46994) (0.128495,7.59773)};
  \addplot[black!60, densely dashdotted, forget plot] coordinates {(0.04,1) (0.469,1)};
\nextgroupplot[ylabel={$\Lambda$ ($p_c=p_r=0$)}, xmin=0.04, xmax=0.469, ymin=0.499, ymax=970, xmode=log, log basis x=10, xtick={0.05,0.1,0.15,0.2,0.3,0.4}, xticklabels={5,10,15,20,30,40}]
  \addplot[blue, mark=diamond*, only marks, mark options={solid}, forget plot,
    error bars/.cd, y dir=both, y explicit]
  coordinates {
    (0.0699738,105.87) +- (0,23.1273)
    (0.0810868,75.5048) +- (0,8.12353)
    (0.0993503,56.109) +- (0,4.90939)
    (0.128495,26.0293) +- (0,1.15112)
    (0.142369,21.685) +- (0,0.948032)
    (0.159478,18.4591) +- (0,0.967648)
    (0.192928,9.94338) +- (0,0.527196)
    (0.225369,6.59111) +- (0,0.163134)
    (0.256822,4.47201) +- (0,0.104598)
    (0.287307,3.40745) +- (0,0.0812596)
    (0.316845,2.67995) +- (0,0.0648944)
    (0.373156,1.83921) +- (0,0.0441765)
    (0.425913,1.30034) +- (0,0.0305658)
  };
  \addplot[blue, thick, solid, no marks, forget plot] coordinates {(0.256822,4.47201) (0.287307,3.40745) (0.316845,2.67995) (0.373156,1.83921) (0.425913,1.30034)};
  \addplot[blue, thick, dashed, no marks, forget plot] coordinates {(0.128495,26.0293) (0.142369,21.685) (0.159478,18.4591) (0.192928,9.94338) (0.225369,6.59111) (0.256822,4.47201)};
  \addplot[blue, thick, densely dotted, no marks, forget plot] coordinates {(0.0699738,105.87) (0.0810868,75.5048) (0.0993503,56.109) (0.128495,26.0293)};
  \addplot[red, mark=diamond, only marks, mark options={solid}, forget plot,
    error bars/.cd, y dir=both, y explicit]
  coordinates {
    (0.0699738,84.6385) +- (0,13.5789)
    (0.0810868,81.8888) +- (0,8.56163)
    (0.0993503,45.5921) +- (0,2.8324)
    (0.128495,26.6136) +- (0,1.15714)
    (0.142369,20.9665) +- (0,0.886334)
    (0.159478,14.6645) +- (0,1.15797)
    (0.192928,8.8097) +- (0,0.669911)
    (0.225369,5.72402) +- (0,0.138358)
    (0.256822,4.24844) +- (0,0.10099)
    (0.287307,3.10273) +- (0,0.072768)
    (0.316845,2.40381) +- (0,0.0578132)
    (0.373156,1.58702) +- (0,0.0378665)
    (0.425913,1.16755) +- (0,0.0262021)
  };
  \addplot[red, thick, solid, no marks, forget plot] coordinates {(0.256822,4.24844) (0.287307,3.10273) (0.316845,2.40381) (0.373156,1.58702) (0.425913,1.16755)};
  \addplot[red, thick, dashed, no marks, forget plot] coordinates {(0.128495,26.6136) (0.142369,20.9665) (0.159478,14.6645) (0.192928,8.8097) (0.225369,5.72402) (0.256822,4.24844)};
  \addplot[red, thick, densely dotted, no marks, forget plot] coordinates {(0.0699738,84.6385) (0.0810868,81.8888) (0.0993503,45.5921) (0.128495,26.6136)};
  \addplot[OliveGreen, mark=*, only marks, mark options={solid}, forget plot,
    error bars/.cd, y dir=both, y explicit]
  coordinates {
    (0.0993503,35.8045) +- (0,7.98984)
    (0.117295,25.4549) +- (0,3.49665)
    (0.128495,20.3922) +- (0,2.55163)
    (0.142369,13.3358) +- (0,1.06263)
    (0.159478,7.74167) +- (0,0.572877)
    (0.192928,3.22959) +- (0,0.234809)
    (0.225369,2.20485) +- (0,0.0513895)
    (0.256822,1.49445) +- (0,0.036195)
    (0.287307,1.16019) +- (0,0.0273304)
    (0.316845,0.960591) +- (0,0.0224336)
    (0.373156,0.79653) +- (0,0.0171393)
    (0.425913,0.748861) +- (0,0.0158013)
  };
  \addplot[OliveGreen, thick, solid, no marks, forget plot] coordinates {(0.256822,1.49445) (0.287307,1.16019) (0.316845,0.960591) (0.373156,0.79653) (0.425913,0.748861)};
  \addplot[OliveGreen, thick, dashed, no marks, forget plot] coordinates {(0.128495,20.3922) (0.142369,13.3358) (0.159478,7.74167) (0.192928,3.22959) (0.225369,2.20485) (0.256822,1.49445)};
  \addplot[OliveGreen, thick, densely dotted, no marks, forget plot] coordinates {(0.0993503,35.8045) (0.117295,25.4549) (0.128495,20.3922)};
  \addplot[orange, mark=square*, only marks, mark options={solid}, forget plot,
    error bars/.cd, y dir=both, y explicit]
  coordinates {
    (0.128495,39.1429) +- (0,14.9824)
    (0.142369,26.4808) +- (0,4.96758)
    (0.159478,13.1033) +- (0,1.19962)
    (0.192928,8.8337) +- (0,0.684646)
    (0.225369,4.20373) +- (0,0.0984793)
    (0.256822,2.7597) +- (0,0.0640864)
    (0.287307,1.94845) +- (0,0.0469289)
    (0.316845,1.44891) +- (0,0.0352211)
    (0.373156,1.029) +- (0,0.0235598)
    (0.425913,0.853362) +- (0,0.018284)
  };
  \addplot[orange, thick, solid, no marks, forget plot] coordinates {(0.256822,2.7597) (0.287307,1.94845) (0.316845,1.44891) (0.373156,1.029) (0.425913,0.853362)};
  \addplot[orange, thick, dashed, no marks, forget plot] coordinates {(0.128495,39.1429) (0.142369,26.4808) (0.159478,13.1033) (0.192928,8.8337) (0.225369,4.20373) (0.256822,2.7597)};
  \addplot[Purple, mark=triangle*, only marks, mark options={solid}, forget plot,
    error bars/.cd, y dir=both, y explicit]
  coordinates {
    (0.192928,35.7278) +- (0,7.0359)
    (0.225369,14.115) +- (0,0.286426)
    (0.256822,8.35735) +- (0,0.14299)
    (0.287307,5.39055) +- (0,0.0896458)
    (0.316845,3.58847) +- (0,0.0623683)
    (0.373156,2.14769) +- (0,0.0363315)
    (0.425913,1.41721) +- (0,0.0235815)
  };
  \addplot[Purple, thick, solid, no marks, forget plot] coordinates {(0.256822,8.35735) (0.287307,5.39055) (0.316845,3.58847) (0.373156,2.14769) (0.425913,1.41721)};
  \addplot[Purple, thick, dashed, no marks, forget plot] coordinates {(0.192928,35.7278) (0.225369,14.115) (0.256822,8.35735)};
  \addplot[blue, line width=2.2pt, opacity=0.3, forget plot] coordinates {(0.04,468.75) (0.04062,454.51) (0.04125,440.71) (0.04189,427.33) (0.04254,414.35) (0.04321,401.77) (0.04388,389.57) (0.04456,377.73) (0.04525,366.26) (0.04595,355.14) (0.04667,344.35) (0.04739,333.9) (0.04813,323.76) (0.04888,313.92) (0.04964,304.39) (0.05041,295.15) (0.05119,286.18) (0.05199,277.49) (0.0528,269.07) (0.05362,260.89) (0.05445,252.97) (0.0553,245.29) (0.05616,237.84) (0.05703,230.62) (0.05791,223.61) (0.05881,216.82) (0.05973,210.24) (0.06066,203.85) (0.0616,197.66) (0.06256,191.66) (0.06353,185.84) (0.06451,180.19) (0.06552,174.72) (0.06654,169.42) (0.06757,164.27) (0.06862,159.28) (0.06969,154.45) (0.07077,149.75) (0.07187,145.21) (0.07299,140.8) (0.07412,136.52) (0.07527,132.38) (0.07644,128.36) (0.07763,124.46) (0.07883,120.68) (0.08006,117.01) (0.0813,113.46) (0.08257,110.01) (0.08385,106.67) (0.08515,103.43) (0.08648,100.29) (0.08782,97.246) (0.08918,94.293) (0.09057,91.429) (0.09198,88.652) (0.09341,85.96) (0.09486,83.35) (0.09633,80.818) (0.09783,78.364) (0.09935,75.984)};
  \addplot[red, line width=2.2pt, opacity=0.3, forget plot] coordinates {(0.04,446.43) (0.04062,432.87) (0.04125,419.73) (0.04189,406.98) (0.04254,394.62) (0.04321,382.63) (0.04388,371.01) (0.04456,359.75) (0.04525,348.82) (0.04595,338.23) (0.04667,327.96) (0.04739,318) (0.04813,308.34) (0.04888,298.98) (0.04964,289.9) (0.05041,281.09) (0.05119,272.56) (0.05199,264.28) (0.0528,256.25) (0.05362,248.47) (0.05445,240.92) (0.0553,233.61) (0.05616,226.51) (0.05703,219.63) (0.05791,212.96) (0.05881,206.5) (0.05973,200.23) (0.06066,194.15) (0.0616,188.25) (0.06256,182.53) (0.06353,176.99) (0.06451,171.61) (0.06552,166.4) (0.06654,161.35) (0.06757,156.45) (0.06862,151.7) (0.06969,147.09) (0.07077,142.62) (0.07187,138.29) (0.07299,134.09) (0.07412,130.02) (0.07527,126.07) (0.07644,122.24) (0.07763,118.53) (0.07883,114.93) (0.08006,111.44) (0.0813,108.06) (0.08257,104.77) (0.08385,101.59) (0.08515,98.507) (0.08648,95.516) (0.08782,92.615) (0.08918,89.803) (0.09057,87.075) (0.09198,84.431) (0.09341,81.867) (0.09486,79.381) (0.09633,76.97) (0.09783,74.632) (0.09935,72.366)};
  \addplot[OliveGreen, line width=2.2pt, opacity=0.3, forget plot] coordinates {(0.04,1824.6) (0.0408,1688.6) (0.04161,1562.8) (0.04245,1446.4) (0.04329,1338.8) (0.04416,1239.1) (0.04504,1147) (0.04594,1061.7) (0.04686,982.82) (0.04779,909.83) (0.04875,842.29) (0.04972,779.79) (0.05072,721.97) (0.05173,668.45) (0.05276,618.94) (0.05382,573.11) (0.05489,530.7) (0.05599,491.45) (0.05711,455.13) (0.05825,421.51) (0.05941,390.39) (0.0606,361.59) (0.06181,334.93) (0.06304,310.25) (0.0643,287.41) (0.06559,266.26) (0.0669,246.68) (0.06823,228.55) (0.0696,211.77) (0.07099,196.23) (0.0724,181.84) (0.07385,168.52) (0.07533,156.19) (0.07683,144.76) (0.07837,134.18) (0.07993,124.38) (0.08153,115.31) (0.08316,106.9) (0.08482,99.115) (0.08651,91.903) (0.08824,85.222) (0.09,79.032) (0.0918,73.298) (0.09364,67.985) (0.09551,63.062) (0.09741,58.5) (0.09936,54.273) (0.1013,50.356) (0.1034,46.725) (0.1054,43.36) (0.1075,40.241) (0.1097,37.35) (0.1119,34.67) (0.1141,32.185) (0.1164,29.881) (0.1187,27.745) (0.1211,25.764) (0.1235,23.928) (0.126,22.224) (0.1285,20.644)};
  \addplot[orange, line width=2.2pt, opacity=0.3, forget plot] coordinates {(0.04,3944) (0.04095,3594.6) (0.04192,3276.3) (0.04291,2986.2) (0.04393,2721.9) (0.04497,2481.1) (0.04604,2261.6) (0.04713,2061.6) (0.04825,1879.3) (0.0494,1713.2) (0.05057,1561.8) (0.05177,1423.9) (0.05299,1298.1) (0.05425,1183.6) (0.05554,1079.1) (0.05685,983.93) (0.0582,897.17) (0.05958,818.09) (0.061,746) (0.06244,680.29) (0.06392,620.4) (0.06544,565.8) (0.06699,516.02) (0.06858,470.65) (0.07021,429.28) (0.07187,391.57) (0.07358,357.18) (0.07532,325.83) (0.07711,297.25) (0.07894,271.18) (0.08081,247.41) (0.08273,225.74) (0.08469,205.98) (0.0867,187.96) (0.08875,171.52) (0.09086,156.53) (0.09302,142.86) (0.09522,130.39) (0.09748,119.01) (0.09979,108.64) (0.1022,99.172) (0.1046,90.538) (0.1071,82.661) (0.1096,75.474) (0.1122,68.918) (0.1149,62.935) (0.1176,57.476) (0.1204,52.495) (0.1232,47.949) (0.1262,43.8) (0.1291,40.014) (0.1322,36.558) (0.1353,33.403) (0.1386,30.524) (0.1418,27.895) (0.1452,25.495) (0.1486,23.304) (0.1522,21.303) (0.1558,19.476) (0.1595,17.808)};
  \addplot[Purple, line width=2.2pt, opacity=0.3, forget plot] coordinates {(0.04,15151) (0.04128,13357) (0.0426,11776) (0.04397,10382) (0.04537,9152.8) (0.04683,8069.3) (0.04833,7114) (0.04987,6271.9) (0.05147,5529.4) (0.05312,4874.9) (0.05482,4297.8) (0.05657,3789.1) (0.05839,3340.6) (0.06026,2945.2) (0.06219,2596.6) (0.06418,2289.2) (0.06623,2018.3) (0.06835,1779.4) (0.07054,1568.8) (0.0728,1383.2) (0.07513,1219.5) (0.07754,1075.2) (0.08002,947.92) (0.08258,835.75) (0.08522,736.86) (0.08795,649.67) (0.09077,572.8) (0.09368,505.03) (0.09668,445.28) (0.09977,392.6) (0.103,346.15) (0.1063,305.2) (0.1097,269.1) (0.1132,237.27) (0.1168,209.2) (0.1205,184.46) (0.1244,162.64) (0.1284,143.41) (0.1325,126.45) (0.1367,111.5) (0.1411,98.313) (0.1456,86.689) (0.1503,76.44) (0.1551,67.403) (0.1601,59.435) (0.1652,52.41) (0.1705,46.215) (0.1759,40.753) (0.1816,35.937) (0.1874,31.691) (0.1934,27.946) (0.1996,24.645) (0.206,21.733) (0.2126,19.166) (0.2194,16.903) (0.2264,14.906) (0.2337,13.146) (0.2411,11.594) (0.2489,10.225) (0.2568,9.0185)};
  \addplot[black!60, densely dashdotted, forget plot] coordinates {(0.04,1) (0.469,1)};
\nextgroupplot[xmin=0.04, xmax=0.469, ymin=0.499, ymax=970, xmode=log, log basis x=10, xtick={0.05,0.1,0.15,0.2,0.3,0.4}, xticklabels={5,10,15,20,30,40}]
  \addplot[blue, mark=diamond*, only marks, mark options={solid}, forget plot,
    error bars/.cd, y dir=both, y explicit]
  coordinates {
    (0.0699738,10.2204) +- (0,0.745014)
    (0.0810868,8.5591) +- (0,0.634552)
    (0.0993503,6.53831) +- (0,0.460966)
    (0.128495,4.56225) +- (0,0.301284)
    (0.142369,3.82074) +- (0,0.23755)
    (0.159478,3.22591) +- (0,0.19668)
    (0.192928,2.90371) +- (0,0.180771)
    (0.225369,2.20833) +- (0,0.0505734)
    (0.256822,1.8843) +- (0,0.0417866)
    (0.287307,1.61247) +- (0,0.0349888)
    (0.316845,1.4495) +- (0,0.0299627)
    (0.373156,1.16759) +- (0,0.0236287)
  };
  \addplot[blue, thick, solid, no marks, forget plot] coordinates {(0.256822,1.8843) (0.287307,1.61247) (0.316845,1.4495) (0.373156,1.16759)};
  \addplot[blue, thick, dashed, no marks, forget plot] coordinates {(0.128495,4.56225) (0.142369,3.82074) (0.159478,3.22591) (0.192928,2.90371) (0.225369,2.20833) (0.256822,1.8843)};
  \addplot[blue, thick, densely dotted, no marks, forget plot] coordinates {(0.0699738,10.2204) (0.0810868,8.5591) (0.0993503,6.53831) (0.128495,4.56225)};
  \addplot[red, mark=diamond, only marks, mark options={solid}, forget plot,
    error bars/.cd, y dir=both, y explicit]
  coordinates {
    (0.0699738,129.385) +- (0,14.5632)
    (0.0810868,93.4089) +- (0,10.3702)
    (0.0993503,44.7671) +- (0,4.72091)
    (0.128495,26.174) +- (0,2.01382)
    (0.142369,20.9645) +- (0,1.52044)
    (0.159478,16.323) +- (0,1.17806)
    (0.192928,10.233) +- (0,0.746821)
    (0.225369,6.63483) +- (0,0.156204)
    (0.256822,4.61975) +- (0,0.113073)
    (0.287307,3.24475) +- (0,0.0769843)
    (0.316845,2.49637) +- (0,0.0599763)
    (0.373156,1.70924) +- (0,0.0388269)
    (0.425913,1.32815) +- (0,0.029311)
  };
  \addplot[red, thick, solid, no marks, forget plot] coordinates {(0.256822,4.61975) (0.287307,3.24475) (0.316845,2.49637) (0.373156,1.70924) (0.425913,1.32815)};
  \addplot[red, thick, dashed, no marks, forget plot] coordinates {(0.128495,26.174) (0.142369,20.9645) (0.159478,16.323) (0.192928,10.233) (0.225369,6.63483) (0.256822,4.61975)};
  \addplot[red, thick, densely dotted, no marks, forget plot] coordinates {(0.0699738,129.385) (0.0810868,93.4089) (0.0993503,44.7671) (0.128495,26.174)};
  \addplot[OliveGreen, mark=*, only marks, mark options={solid}, forget plot,
    error bars/.cd, y dir=both, y explicit]
  coordinates {
    (0.0699738,8.90052) +- (0,0.679926)
    (0.0810868,7.42572) +- (0,0.530809)
    (0.0993503,5.26152) +- (0,0.358998)
    (0.128495,3.21083) +- (0,0.185771)
    (0.142369,2.61093) +- (0,0.154267)
    (0.159478,2.11977) +- (0,0.107922)
    (0.192928,1.56984) +- (0,0.0654425)
    (0.225369,1.30845) +- (0,0.0281353)
  };
  \addplot[OliveGreen, thick, dashed, no marks, forget plot] coordinates {(0.128495,3.21083) (0.142369,2.61093) (0.159478,2.11977) (0.192928,1.56984) (0.225369,1.30845)};
  \addplot[OliveGreen, thick, densely dotted, no marks, forget plot] coordinates {(0.0699738,8.90052) (0.0810868,7.42572) (0.0993503,5.26152) (0.128495,3.21083)};
  \addplot[orange, mark=square*, only marks, mark options={solid}, forget plot,
    error bars/.cd, y dir=both, y explicit]
  coordinates {
    (0.0699738,16.4618) +- (0,1.27005)
    (0.0810868,12.6977) +- (0,0.98534)
    (0.0993503,9.29587) +- (0,0.685416)
    (0.128495,4.61849) +- (0,0.318833)
    (0.142369,4.18923) +- (0,0.277017)
    (0.159478,3.58722) +- (0,0.222869)
    (0.192928,2.3849) +- (0,0.131424)
    (0.225369,1.76766) +- (0,0.0391522)
    (0.256822,1.46429) +- (0,0.0315874)
    (0.287307,1.24594) +- (0,0.0246054)
  };
  \addplot[orange, thick, solid, no marks, forget plot] coordinates {(0.256822,1.46429) (0.287307,1.24594)};
  \addplot[orange, thick, dashed, no marks, forget plot] coordinates {(0.128495,4.61849) (0.142369,4.18923) (0.159478,3.58722) (0.192928,2.3849) (0.225369,1.76766) (0.256822,1.46429)};
  \addplot[orange, thick, densely dotted, no marks, forget plot] coordinates {(0.0699738,16.4618) (0.0810868,12.6977) (0.0993503,9.29587) (0.128495,4.61849)};
  \addplot[Purple, mark=triangle*, only marks, mark options={solid}, forget plot,
    error bars/.cd, y dir=both, y explicit]
  coordinates {
    (0.0699738,35.592) +- (0,2.83224)
    (0.0810868,27.8555) +- (0,2.21443)
    (0.0993503,16.5337) +- (0,1.28118)
    (0.128495,10.2617) +- (0,0.725366)
    (0.142369,7.41319) +- (0,0.549193)
    (0.159478,6.56488) +- (0,0.471124)
    (0.192928,4.40087) +- (0,0.29844)
    (0.225369,3.49265) +- (0,0.0823242)
    (0.256822,2.63447) +- (0,0.0608688)
    (0.287307,2.14263) +- (0,0.048676)
    (0.316845,1.75994) +- (0,0.0389148)
    (0.373156,1.28683) +- (0,0.0272173)
  };
  \addplot[Purple, thick, solid, no marks, forget plot] coordinates {(0.256822,2.63447) (0.287307,2.14263) (0.316845,1.75994) (0.373156,1.28683)};
  \addplot[Purple, thick, dashed, no marks, forget plot] coordinates {(0.128495,10.2617) (0.142369,7.41319) (0.159478,6.56488) (0.192928,4.40087) (0.225369,3.49265) (0.256822,2.63447)};
  \addplot[Purple, thick, densely dotted, no marks, forget plot] coordinates {(0.0699738,35.592) (0.0810868,27.8555) (0.0993503,16.5337) (0.128495,10.2617)};
  \addplot[blue, line width=2.2pt, opacity=0.3, forget plot] coordinates {(0.04,18.75) (0.04062,18.463) (0.04125,18.181) (0.04189,17.902) (0.04254,17.628) (0.04321,17.359) (0.04388,17.093) (0.04456,16.832) (0.04525,16.574) (0.04595,16.32) (0.04667,16.071) (0.04739,15.825) (0.04813,15.583) (0.04888,15.344) (0.04964,15.109) (0.05041,14.878) (0.05119,14.651) (0.05199,14.426) (0.0528,14.206) (0.05362,13.988) (0.05445,13.774) (0.0553,13.563) (0.05616,13.356) (0.05703,13.152) (0.05791,12.95) (0.05881,12.752) (0.05973,12.557) (0.06066,12.365) (0.0616,12.176) (0.06256,11.989) (0.06353,11.806) (0.06451,11.625) (0.06552,11.447) (0.06654,11.272) (0.06757,11.1) (0.06862,10.93) (0.06969,10.763) (0.07077,10.598) (0.07187,10.436) (0.07299,10.276) (0.07412,10.119) (0.07527,9.964) (0.07644,9.8115) (0.07763,9.6614) (0.07883,9.5136) (0.08006,9.368) (0.0813,9.2247) (0.08257,9.0835) (0.08385,8.9445) (0.08515,8.8076) (0.08648,8.6729) (0.08782,8.5402) (0.08918,8.4095) (0.09057,8.2808) (0.09198,8.1541) (0.09341,8.0293) (0.09486,7.9065) (0.09633,7.7855) (0.09783,7.6664) (0.09935,7.549)};
  \addplot[red, line width=2.2pt, opacity=0.3, forget plot] coordinates {(0.04,468.75) (0.04062,454.51) (0.04125,440.71) (0.04189,427.33) (0.04254,414.35) (0.04321,401.77) (0.04388,389.57) (0.04456,377.73) (0.04525,366.26) (0.04595,355.14) (0.04667,344.35) (0.04739,333.9) (0.04813,323.76) (0.04888,313.92) (0.04964,304.39) (0.05041,295.15) (0.05119,286.18) (0.05199,277.49) (0.0528,269.07) (0.05362,260.89) (0.05445,252.97) (0.0553,245.29) (0.05616,237.84) (0.05703,230.62) (0.05791,223.61) (0.05881,216.82) (0.05973,210.24) (0.06066,203.85) (0.0616,197.66) (0.06256,191.66) (0.06353,185.84) (0.06451,180.19) (0.06552,174.72) (0.06654,169.42) (0.06757,164.27) (0.06862,159.28) (0.06969,154.45) (0.07077,149.75) (0.07187,145.21) (0.07299,140.8) (0.07412,136.52) (0.07527,132.38) (0.07644,128.36) (0.07763,124.46) (0.07883,120.68) (0.08006,117.01) (0.0813,113.46) (0.08257,110.01) (0.08385,106.67) (0.08515,103.43) (0.08648,100.29) (0.08782,97.246) (0.08918,94.293) (0.09057,91.429) (0.09198,88.652) (0.09341,85.96) (0.09486,83.35) (0.09633,80.818) (0.09783,78.364) (0.09935,75.984)};
  \addplot[OliveGreen, line width=2.2pt, opacity=0.3, forget plot] coordinates {(0.04,29.424) (0.04062,28.559) (0.04125,27.719) (0.04189,26.904) (0.04254,26.114) (0.04321,25.348) (0.04388,24.604) (0.04456,23.883) (0.04525,23.183) (0.04595,22.504) (0.04667,21.845) (0.04739,21.206) (0.04813,20.586) (0.04888,19.985) (0.04964,19.401) (0.05041,18.835) (0.05119,18.286) (0.05199,17.753) (0.0528,17.236) (0.05362,16.734) (0.05445,16.248) (0.0553,15.775) (0.05616,15.317) (0.05703,14.873) (0.05791,14.441) (0.05881,14.023) (0.05973,13.617) (0.06066,13.223) (0.0616,12.84) (0.06256,12.469) (0.06353,12.109) (0.06451,11.76) (0.06552,11.421) (0.06654,11.092) (0.06757,10.773) (0.06862,10.463) (0.06969,10.163) (0.07077,9.871) (0.07187,9.588) (0.07299,9.3134) (0.07412,9.0469) (0.07527,8.7882) (0.07644,8.5372) (0.07763,8.2936) (0.07883,8.0572) (0.08006,7.8278) (0.0813,7.6051) (0.08257,7.389) (0.08385,7.1792) (0.08515,6.9757) (0.08648,6.7781) (0.08782,6.5864) (0.08918,6.4002) (0.09057,6.2196) (0.09198,6.0443) (0.09341,5.8741) (0.09486,5.7089) (0.09633,5.5485) (0.09783,5.3929) (0.09935,5.2418)};
  \addplot[orange, line width=2.2pt, opacity=0.3, forget plot] coordinates {(0.04,51.02) (0.04062,49.493) (0.04125,48.012) (0.04189,46.575) (0.04254,45.182) (0.04321,43.83) (0.04388,42.52) (0.04456,41.249) (0.04525,40.016) (0.04595,38.821) (0.04667,37.661) (0.04739,36.536) (0.04813,35.446) (0.04888,34.388) (0.04964,33.362) (0.05041,32.367) (0.05119,31.402) (0.05199,30.465) (0.0528,29.557) (0.05362,28.677) (0.05445,27.823) (0.0553,26.994) (0.05616,26.191) (0.05703,25.411) (0.05791,24.656) (0.05881,23.922) (0.05973,23.211) (0.06066,22.522) (0.0616,21.852) (0.06256,21.204) (0.06353,20.574) (0.06451,19.964) (0.06552,19.371) (0.06654,18.797) (0.06757,18.24) (0.06862,17.7) (0.06969,17.175) (0.07077,16.667) (0.07187,16.174) (0.07299,15.695) (0.07412,15.231) (0.07527,14.781) (0.07644,14.344) (0.07763,13.921) (0.07883,13.51) (0.08006,13.111) (0.0813,12.725) (0.08257,12.35) (0.08385,11.986) (0.08515,11.633) (0.08648,11.291) (0.08782,10.959) (0.08918,10.636) (0.09057,10.324) (0.09198,10.021) (0.09341,9.7266) (0.09486,9.4413) (0.09633,9.1645) (0.09783,8.8959) (0.09935,8.6354)};
  \addplot[Purple, line width=2.2pt, opacity=0.3, forget plot] coordinates {(0.04,114.55) (0.04062,111.08) (0.04125,107.7) (0.04189,104.43) (0.04254,101.26) (0.04321,98.19) (0.04388,95.209) (0.04456,92.318) (0.04525,89.516) (0.04595,86.798) (0.04667,84.163) (0.04739,81.608) (0.04813,79.131) (0.04888,76.729) (0.04964,74.399) (0.05041,72.141) (0.05119,69.951) (0.05199,67.827) (0.0528,65.769) (0.05362,63.772) (0.05445,61.836) (0.0553,59.959) (0.05616,58.139) (0.05703,56.374) (0.05791,54.663) (0.05881,53.004) (0.05973,51.395) (0.06066,49.835) (0.0616,48.322) (0.06256,46.855) (0.06353,45.433) (0.06451,44.054) (0.06552,42.717) (0.06654,41.42) (0.06757,40.163) (0.06862,38.944) (0.06969,37.762) (0.07077,36.616) (0.07187,35.505) (0.07299,34.427) (0.07412,33.382) (0.07527,32.369) (0.07644,31.387) (0.07763,30.434) (0.07883,29.51) (0.08006,28.615) (0.0813,27.746) (0.08257,26.904) (0.08385,26.088) (0.08515,25.296) (0.08648,24.529) (0.08782,23.784) (0.08918,23.062) (0.09057,22.363) (0.09198,21.684) (0.09341,21.026) (0.09486,20.388) (0.09633,19.769) (0.09783,19.169) (0.09935,18.588)};
  \addplot[black!60, densely dashdotted, forget plot] coordinates {(0.04,1) (0.469,1)};
\end{groupplot}
\node[anchor=north] at ([xshift=0.175cm, yshift=-0.9cm]group c1r3.south east) {\pgfplotslegendfromname{legend_subthreshold_lambda_erasure_grid_pm_corr_mergeZZ_obsZZ-XX}};
\end{tikzpicture}

%% file: sections/8_discussion.tex
\section{Discussion}\label{sec:discussion}

CBQC and MBQC have evolved as distinct paradigms, each with its own set of advantages and challenges. In particular, they have been shown to be suitable for different physical platforms and error correction protocols have been designed around this. Circuit-based error correction has been largely developed around matter-based qubits, while measurement-based error correction has been largely developed around photonic qubits, emphasised by the development of FBQC as a linear-optical measurement-based error correction protocol. The results presented in this work are a starting point towards bridging the gap between these two paradigms, demonstrating that hybrid protocols can be designed to leverage the strengths of both paradigms in the context of distributed matter-photonic architectures. In particular, we have shown that logical gates using surface code lattice surgery can be implemented with high loss thresholds and augmented distances using matter-based quantum emitters and fusion networks. This is true both for the hybrid protocols we have constructed and for circuit-based implementations where Bell pair preparation is modelled using linear-optical fusions.

Under pure elementary-outcome erasure, \textsc{Bell}, \textsc{BellZigZag}, and $3d$-\textsc{Chain} share the asymptotic $50\%$ threshold. Truncating the hybrid chains lowers this threshold to approximately $38\%$ for $8$-\textsc{Chain} and $32\%$ for $5$-\textsc{Chain}. These thresholds should not be directly compared with existing thresholds for distributed lattice surgery based on purely phenomenological- or circuit-noise models, since they model a different error process. Interface erasure thresholds are higher than interface depolarising thresholds, though interface erasure rates are higher due to fusion failure and loss. Post-selecting on successful heralds will result in much lower error rates on Bell pairs, though reduce generation rates. An exact comparison of the trade-offs between erasure-based decoding of network errors and post-selection requires more precise hardware assumptions and remains an avenue for future work.

Adding resource-state errors reduces the erasure thresholds approximately linearly in the error rate, as seen in previous work~\cite{bartolucciFusionbasedQuantumComputation2023}. For local circuit noise, erasure thresholds decrease slowly as low levels of local noise are added and more rapidly as $p_c$ approaches the local-noise threshold. This can be attributed to the fact that these noise sources affect different parts of the system, making it less common for error chains to combine across regions.

Fusion boosting becomes more useful in the presence of local noise and resource-state errors. Boosting can improve photon-loss thresholds for all protocols when local noise is high and can be used to probe far-subthreshold regimes when local noise is low. The subthreshold results show that the interface distance is relevant only once other error mechanisms are sufficiently weak. The change in finite-distance suppression as local noise decreases is consistent with the asymptotic merge-observable advantage predicted by the augmented interface distance, though no hybrid protocol achieves the best protection of the perpendicular observable. For any process in which the protection of the perpendicular observable is more important, the presented hybrid protocols will show no advantage. Where roughly equal protection of both observables is required, it may be possible to even out the suppression by tailoring the failure bases of fusion measurements to create error biases, as explored in Ref.~\cite{bombinIncreasingErrorTolerance2023}. Additionally, since we have investigated a specific hybrid construction, there is a large design space still to explore and alternative interface structures may lead to improved performance.

The principal local-noise limitation of the hybrid interfaces arises from the deterministic boundary fusions between the circuit and resource-state regions. Errors in these operations can introduce half-distance error chains on one of the two observables. In the sampled low-noise regime, these mechanisms make only a small contribution and do not obscure the erasure suppression of the linear-chain interfaces; as local noise increases, they contribute to masking this advantage. This limitation belongs to the boundary fusion implementation considered here rather than to hybrid protocols in general. Flagged measurement gadgets~\cite{chamberlandFlagFaulttolerantError2018} could be used to catch these errors, or fault-equivalent/distance-preserving transformations~\cite{rodatzFaultToleranceConstruction2026,rodatzFloquetifyingStabiliserCodes2024} could be used to construct alternative local boundary realisations, such as those proposed in Refs.~\cite{schmidtResourceadaptiveDistributedFault2026,thakurSyndromeExtractionFramework2026}.

The simulations assume that the required resource states are supplied deterministically when the interface is measured. We do not model explicit resource state generation, nor do we consider hardware-realistic noise models for emitters. These omissions are especially significant for larger chain protocols. The results therefore establish an interface-distance and logical-suppression advantage within the stated noise models; they do not establish a reduction in emitter count, preparation time, or total computational overhead. A lower-level treatment could build on the graphical dataflow framework of de Felice et al.~\cite{defeliceDataflowProgrammingFramework2026}, which combines ZX diagrams with explicit linear-optical components.

These results provide a promising starting point for hybrid error correction in matter--photonic architectures. There are many avenues for future work. We have only considered a single code family in this work, but the methods should in principle be applicable to any other CSS codes. We have also only directly investigated simple lattice surgery merge-and-split operations using two logical qubits. However, the framework could equally be applied to more complex lattice operations, such as fan-outs and multi-target $CX$ gates. This opens the door to compilation routines which optimise the use of both measurement-based and circuit-based operations to reduce the overall resource overhead of a computation.

%% file: sections/9_conclusion.tex
\section{Conclusion}

We have proposed a method for constructing hybrid error correction protocols that combine circuit-based and measurement-based error correction based on the ZX calculus. We have shown that this method can be used to construct hybrid logical gates using surface-code lattice surgery and analysed the resulting interface detector error models. The untruncated hybrid interface retains the $50\%$ elementary-outcome erasure threshold of the Bell-pair baselines while increasing the merge-record interface distance from $d+1$ to $2d+1$; truncating the resource states trades a lower threshold for smaller resources. Fusion boosting levels can be used to improve loss thresholds as local circuit noise levels increase. The subthreshold simulations reveal a crossover between two regimes. At $p_c=p_r=10^{-3}$, local errors largely mask the interface-distance differences: $3d$-\textsc{Chain} gives the strongest suppression of the merge observable though fails to improve over \textsc{BellZigZag} for the perpendicular observable. At lower local noise, the linear-chain suppression factors rise more rapidly as erasure is reduced, exposing the augmented interface distance.